\documentclass{emulateapj}
\usepackage{color}
\usepackage{graphicx}
\font\boldsym=cmmib10
\def \bea {\begin{eqnarray}}
\def \ena {\end{eqnarray}}                  

\def \beqa {\begin{eqnarray}}
\def \eeqa {\end{eqnarray}}
\def \bee {\begin{equation}}
\def \ene {\end{equation}}
\def    \ba     {\bf  a}

\def \mcs {\mbox{cos}^{2}}
\def \mss {\mbox{sin}^{2}}

\def	\tH	{{\tau_\H}}
\def    \simlt  {\lower.5ex\hbox{$\; \buildrel < \over \sim \;$}}
\def    \simgt  {\lower.5ex\hbox{$\; \buildrel > \over \sim \;$}}

\def	\Angstrom	{\,{\rm \AA}}		% Angstrom
\def	\ba	{{\bf a}}
\def	\be	{{\bf e}}

\def 	\bE	{{\bf E}}
\def	\beq	{\begin{equation}}

\def	\bJ	{{\bf J}}

\def    \bmu    {{\hbox{\boldsym\char'026}}}	%bold \mu
\def    \bomega {{\hbox{\boldsym\char'041}}}	%bold \omega

\def    \kB     {k_{\rm B}}

\def	\cm	{\,{\rm cm}}

\def	\D	{{\rm D}}

\def	\eeq	{\end{equation}}

\def	\erg	{\,{\rm ergs}}

\def	\gtsim	{\simgt}
\def	\H	{{\rm H}}
\def	\He	{{\rm He}}

\def	\IR	{{\rm IR}}

\def	\ltsim	{\simlt}
\def	\micron	{\mu{\rm m}}

\def	\nH	{n_{\rm H}}

\def	\s	{\,{\rm s}}

\def	\xhat	{\hat{\bf x}}

\def	\yhat	{\hat{\bf y}}

\def	\zhat	{\hat{\bf z}}

\def	\NC	{N_{\rm C}}

\newlength{\figwidth}
\newlength{\figwidthw}
\newlength{\figwidthww}
\newlength{\figwidthd}
\countdef\decade=200
\advance\decade by \year
\countdef\hours=201
\advance\hours by \time
\divide\hours by 60
\countdef\mins=202
\advance\mins by \hours
\multiply\mins by 60
\multiply\hours by 100
\countdef\miltime=203
\advance\miltime by \hours
\advance\miltime by \time
\advance\miltime by -\mins

\begin{document}
%\title{
%------------- enable for labeling preprint ---------------------------
%        \vspace*{-3.0em}
%        {\normalsize\rm To be submitted to {\it The Astrophysical Journal}}\\ 
%        {\normalsize\rm \todayd: DRAFT \vers}\\
%        \vspace*{1.0em}

%\shorttitle{improved model of spinning dust}
%\shortauthors{Hoang, Lazarian, & Draine}
\title{Improving the model of emission from spinning dust: effects of
  grain wobbling and transient spin-up} 
\author{Thiem Hoang\altaffilmark{1}, B. T. Draine\altaffilmark{2} and 
A. Lazarian\altaffilmark{1}}

\altaffiltext{1}{Astronomy Department, University of Wisconsin, Madison, WI 53706}
\altaffiltext{2}{Department of Astrophysical Sciences, Princeton University, Princeton, NJ 08544}

\begin{abstract}
Observations continue to support the interpretation of
the anomalous microwave foreground as
electric dipole radiation from spinning dust grains as
proposed by Draine \& Lazarian (1998ab).
In this paper we present a refinement of the original
model by improving the treatment of a number of physical
effects. First, we consider a {\it disk-like} grain rotating with
angular velocity at an arbitrary angle with respect to the grain symmetry axis
(i.e., grain wobbling) and derive the rotational damping and
excitation coefficients arising from infrared emission,
plasma-grain interactions and electric dipole emission. 
The angular velocity distribution function and the electric dipole emission spectrum
for disk-like grains are calculated using the Langevin equation,
for cases both with and without fast internal relaxation. Our results show that
for fast internal relaxation, the peak emissivity of spinning dust,
compared to earlier studies, increases by a factor of $\sim $ 2 for the Warm 
Neutral Medium (WNM), the Warm Ionized Medium (WIM), the Cold
Neutral Medium (CNM) and the Photodissociation Region (PDR), 
and by a factor $\sim$ 4 for Reflection Nebulae (RN). The frequency at the 
emission peak also increases by factors $\sim$1.4 to $\sim$2 for these media.
Without internal relaxation, the increase of  emissivity is comparable, 
but the emission spectrum is more extended to higher
frequency. The increased emission results from the non-sphericity of grain shape
and from the anisotropy in damping and excitation along directions
parallel and perpendicular to the grain symmetry axis. Second, we
provide a detailed numerical study including transient spin-up of
grains by single-ion collisions. The range of grain size in
which single-ion collisions are important is identified. The 
impulses broaden the emission spectrum and increase the
peak emissivity for the CNM, WNM and WIM, although the increases are not as
large as those due to the grain wobbling. In addition, we present an improved treatment of
rotational excitation and damping by infrared emission.
\end{abstract}


\section{\label{sec:intro}
         Introduction}

Diffuse Galactic microwave emission in the 10 -- 100 GHz frequency range
carries important information on the fundamental properties
of the interstellar medium, but it also interferes with Cosmic
Microwave Background (CMB) experiments (see Bouchet et al.\ 1999,
Tegmark et al.\ 2000, Efstathiou 2003).

It used to be thought that there were only three major components of
the diffuse microwave Galactic foreground: synchrotron emission,
free-free radiation from plasma (thermal bremsstrahlung) and thermal
emission from dust. 
However, in the range of frequency from 10 to
100 GHz an anomalous microwave foreground 
which was difficult to reconcile with the
components above was first reported by Kogut et al.\ (1996a, 1996b).

de Oliveira-Costa et al.\ (2002) gave this emission the 
nickname ``Foreground X'', reflecting its mysterious
nature.  
This component is spatially correlated with 100 $\mu$m
thermal emission from dust, but its intensity is much higher than one
would expect by extrapolating the thermal dust emission
spectrum to the microwave range.  Draine \& Lazarian (1998a,b) proposed
that this foreground was electric dipole radiation from ultrasmall spinning
dust grains.  Although such emission from
spinning dust had been discussed previously (see Erickson 1957,
Ferrara \& Dettmar 1994), DL98a were the first to include the variety
of excitation and damping processes that are relevant for very small
grains.
As time went on
alternative models for the enigmatic foreground have appeared
to be inconsistent with observations\footnote{For instance, the
  dust-correlated synchrotron emission suggested by Bennett
  et al.\ (2003) has now been ruled out
  (see de Oliveira-Costa et al.\ 1999, Finkbeiner, Langston, \&
  Minter 2004, Boughn \& Pober 2007, Gold et al.\ 2009, 2010).}, 
while the predictions of the
spinning dust model have thus far been confirmed. As a result, 
spinning dust is now the
principal explanation for the mysterious ``Foreground X''.

Although the model in Draine \& Lazarian (1998ab) provided
quantitative predictions consistent with observational data, the current
state of precision measurement of the foregrounds calls for refinement
of the model, using a better description of the complex grain
dynamics and modifying some of the original assumptions.

Recent studies showed that the correspondence of the DL98 model to
observations can be improved by adjusting the parameters of the
model. For instance, 
Dobler et al.\ (2009) used the Wilkinson
Microwave Anisotropy Probe (WMAP) 5 year data to
show that a broad bump with frequency
at $\sim$ 40GHz correlated with H$\alpha$, a tracer of the warm ionized
medium (WIM).
They showed that this
bump is consistent with predictions from a DL98 model modified so that
grains have a 
 characteristic dipole moment of $3.5$ D at grain size 1 nm,
 and the gas number density of the WIM is $n_{\H}=0.15 \cm^{-3}$ 
(cf. $n_{\H}=0.1 \cm^{-3}$ in the DL98 model).

On the theoretical front, Ali-Ha\"imound et al.\ (2009) improved
the accuracy of predictions of the model of emission from spinning
dust using the Fokker-Planck equation for the angular velocity ${\bomega}$.
 The authors quantified the
deviations of the grain angular velocity distribution function from
the Maxwellian approximation 
that had been used by DL98b for the sake of simplicity.
With the other assumptions being identical to DL98b, their
findings are not much different from DL98b's predictions.  However,
both the DL98 model and the refined model by Ali-Ha\"imoud et
al. (2009) disregarded the non-sphericity of grains and the anisotropy
in the damping and excitation processes.
Obviously, this assumption is inexact for non-spherical
grains or when there exists any anisotropy in the damping and
excitation processes. This present paper is intended to go deeper into
studies of grain dynamics for a disk-like grain geometry, relaxing
more of the simplifying assumptions in the original DL98b treatment.

The main thrust of our present work is to provide a better description
of several physical processes which have not been addressed in their
complexity either in DL98b or the papers that followed. In particular:
(i) the effects on
electric dipole emission arising from the wobbling of the axis of
major inertia of the {\it disk-like} grain around the angular momentum
due to internal relaxation and the anisotropy of grain
rotational damping and excitation (not yet been treated in the
literature, so far as we are aware); and (ii) transient spin-up of very small
grains due to single-ion collisions. 

First of all, the former
process, i.e., imperfect internal alignment in the non-spherical grain, 
can alter the frequency at which 
the electric dipole emits. In the DL98 model, the emission frequency 
is identical to the angular frequency $\omega/2\pi$.
However, if the dipole moment is fixed in the grain body,
then the complex motion of the grain axes around the angular momentum $\bJ$
will result in emission at frequencies different from its angular frequency. 

In addition, imperfect alignment is essential for many
astrophysical processes, e.g., for grain alignment (see Lazarian 2007
and Lazarian \& Hoang 2009 for recent reviews). 
In our quest to understand
the rotational dynamics of grains which do not rotate about
 their axis of major inertia we capitalize on improved understanding
 of internal randomization arising from thermal fluctuations within the
 grain (Lazarian 1994, Lazarian \& Roberge 1997, 
Weingartner 2009, Hoang \& Lazarian 2009). Disalignment of the grain's 
principal axis from the direction of the angular momentum $\bJ$ will 
cause the angular velocity to increase, leading to increased electric 
dipole emission at high frequency. Our paper provides a quantitative 
description of the effect on the spinning dust emissivity.

The latter process -- collisions with ions -- 
is important for small grains
where the angular momentum of an impinging ion can be larger than the
pre-collision grain angular momentum, resulting in a rotational excitation
spike. In DL98b it was noted that the effect is expected to increase
the spinning dust emissivity, but no quantitative
description was given.

While the treatment of high impulse ion collisions is easily performed within
our Langevin code, the treatment of grain wobbling for non-spherical
grains requires a careful and somewhat tedious modification of our
treatment of the angular momentum diffusion and damping in the DL98 model. In
particular, we have to consider separately parallel and perpendicular
contributions to grain damping and excitation. In addition, we provide
an improved treatment of infrared emission from spinning dust grains.

The structure of the paper is as follows. In \S 2, we present elements
of the DL98 model and our modifications to that model. In \S 3 we
provide detailed calculations for rotational damping and 
excitations arising from plasma-grain interactions and
electric dipole emission for the disk-like grain geometry.
 Refined calculations for the effects of infrared emission are presented in \S
4 and 5. In \S 6, we study the effect of the grain precession on the 
electric dipole emission spectrum and identify its frequency modes. In \S 7 and 8 we 
present our numerical techniques for finding
the distribution functions of grain angular velocity and electric dipole emission,
 and benchmark calculations. In \S 9, we present our results for emissivities from
spinning dust for various idealized environments, and clarify the role
of grain shape and differential rotational damping and excitation
processes to the increase of peak emissivity and frequency. Discussion
and summary are presented in \S 10 and 11, respectively.

\section{Revisiting the DL98 model}

\subsection{Elements of the DL98 model}

Here we present the grain model and our notation.  DL98 pointed out that
the abundant polycyclic aromatic hydrocarbon (PAH) particles required to
explain the observed infrared emission provide a population of particles
that must be spinning and emitting electric dipole radiation in rotational
transitions.  The smaller PAH particles are expected to be planar.
The grain size $a$ is
defined as the radius of a sphere of equivalent volume.  Grains are
assumed to be disk-like with height $L$ and radius $R$ for $a<a_{2}$ and
 spherical  for $a \ge a_{2}$. $a_{2}=6\times 10^{-8}$ cm is chosen in DL98b. 
The surface equivalent radius $a_{\rm s}$ is defined to be the the radius of the
sphere with the same surface area of the grain. The excitation
equivalent radius $a_{\rm x}$ is defined to be the radius of the sphere
with the same $\int r^{2}dS$ with $r$ being the distance from the
surface element $dS$ to the center of mass of the grain (see Appendix
A).

The electric dipole moment $\mu$ of a grain arises from the
intrinsic electric dipole moment associated with asymmetric molecules
or substructures, and from the asymmetric distribution of any excess charge
present. The latter is shown to be less important.

A grain acquires charge through collisions with ions and electrons in
gas, and through photoemission. Assuming that the charging and
photoemission are in equilibrium (i.e., ionization equilibrium), the
distribution of grain charge $f(Z)$ for a given grain size can be
obtained by solving the ionization equilibrium equations (see Draine
\& Sutin 1987, Weingartner \& Draine 2001b).

A grain in the gas experiences collisions with atoms and ions,
plasma-grain interactions, infrared emission and electric dipole
emission. All these processes result in damping and excitation of
grain rotation. DL98b assumed that the angular velocity $\bomega$ is
perfectly aligned with the grain axis of major inertia $\ba_{1}$, and
isotropically oriented in space, and derived the damping and
excitation coefficients assuming perfect internal
alignment. The dimensionless damping and excitation coefficients, $F$ and $G$
are defined as 
\bea F_j=-\frac{\tau_{\H}}{\omega_{j}}\frac{d\omega_{j}}{dt},\label{eq_F}
\\
G_j=\frac{\tau_{\H}}{2\kB T_{\rm gas}}\frac{I_\| d\omega_{j}^{2}}{dt},
\label{eq_G}
\ena
where $j$=n, i, p and IR denote collisions of the grain with neutral,
ion, plasma-grain interactions and infrared emission,
$(1/2)I_{\|}d\omega_{j}^{2}/dt$ is the increase of kinetic energy of rotation along
one axis due to the excitation process $j$; $\tau_{\rm H}$ is the
damping time of the grain in a purely H I gas of temperature
$T_{\rm gas}$, and $I_\|$ is the moment of inertia  
along the grain
symmetry axis. For an uncharged grain in a gas of purely atomic hydrogen, 
$F_{\H}=G_{\H}=1$.

The emissivity per H
 due to 
the electric dipole emission of spinning dust with angular velocity $\omega$ is
\beq
\label{eq:emissivity}
{j_\nu\over n_\H} = 
{1\over 4\pi}{1\over n_\H}
\int_{a_{\rm min}}^{a_{\rm max}} da {dn\over da} 
4\pi \omega^2 f_\omega 2\pi 
\left(2\mu_{\perp}^{2}\omega^{4} \over 3c^{3}\right)~~~,
\label{emiss}
\eeq
where $n_{\rm H}$ is the density of H nuclei, $f_\omega$ is the distribution
function for the angular velocity $\omega$, $\mu_{\perp}$
is the electric dipole moment
perpendicular to the rotation axis, and 
$dn/da$ is the grain size distribution
function with $a$ in the range from $a_{\rm min}$ to
$a_{\rm max}$. Here we take the grain size distribution from
Draine \& Li (2007),
and consider only carbonaceous grains. In the
DL98 model, for the sake of simplicity, $f_\omega$ was assumed to be
a Maxwellian distribution.

\subsection{Our improvements of the DL98 model}

Both the DL98 model and the refined model by Al-Ha\"imoud
et al.\ (2009) assumed that very small grains have disk-like shape and
 large grains are spherical, but they ignored the non-sphericity of 
grain shape when calculating the rotational distribution function 
and emissivity. In the present
paper, we modify the DL98 model as follows.

First, we consider the rotation of a disk-like grain 
 with angular velocity $\bomega$ not perfectly aligned 
with its symmetry axis (i.e. grain wobbling or imperfect alignment), relaxing 
the assumption of the perfect internal alignment of $\bomega$ with the symmetry axis
 in the DL98 model. 
The rotational damping and
excitation coefficients parallel and perpendicular to the grain
symmetry axis resulting from grain infrared emission, plasma-grain
interactions and electric dipole emission are then derived.

Second, we will identify the frequency modes of dipole emission as a result
of complex motion of the grain axes with respect to a fixed angular momentum $\bJ$. 

Third, we will find the
exact distribution functions for 
angular velocity $f_\omega$ and electric dipole emission frequency 
$f_{\nu}$ for {\it non-spherical} grains using numerical simulations
of the Langevin equation (LE), instead of assuming the Maxwellian
distribution as in the DL98 model or using the Fokker-Planck equation
(FP) as in Ali-Ha\"imoud et al.\ (2009). Using the LE approach, we
investigate the effect of grain wobbling on the emission spectrum of
spinning dust. The effect of internal thermal fluctuations, which
results in deviation of the grain symmetry axis from the angular momentum $\bJ$
(Lazarian 1994; Lazarian \& Roberge 1997), is also studied.

Finally, the transient rotational spin-up due to single-ion collisions
with very small grains
 is numerically studied using the LE, in which 
single-ion collisions are treated as Poisson-distributed discrete events.

To see how the grain wobbling can modify results from the DL98 model,
let us consider the simple case of 3D rotation for an axisymmetric
grain with principal moments of inertia $I_{1}>I_{2}=I_{3}$ along
principal axes $\ba_{1}$, $\ba_{2}$ and $\ba_{3}$,
respectively. Denote $I_{\|}=I_{1}$ and
$I_{\perp}=I_{2}=I_{3}$. The ratio of moments of inertia is defined as 
$h=I_{\|}/I_{\perp}$. \footnote{The eigenvalues of the moment of
inertia tensor of the disk of density $\rho$, radius $R$, and height $L$ are
$I_{\|}=\pi\rho R^{4}L/2, I_{\perp}=(\rho \pi R^{4}L/12)[3+(L/R)^{2}]$.}
For the moment,
we ignore the damping due to the electric dipole emission.
Assuming that the rotation along three principal axes are independent, the mean
square angular velocity for this case is 
\bea 
\langle
\omega^{2}\rangle=\langle \omega_{\|}^{2}\rangle+2\langle
\omega_{\perp}^2\rangle, 
\ena 
where $\langle\omega_{\perp}^2\rangle=
\langle\omega_{2}^2\rangle=\langle\omega_{3}^2\rangle$. The mean square angular
velocities along the parallel and perpendicular direction to the
symmetry axis are given by (see DL98b) 
\bea 
\langle
\omega_{\|,\perp}^{2}\rangle=\frac{G_{\|,\perp}}{F_{\|,\perp}}
\frac{\kB T_{\rm gas}}{I_{\|,\perp}},
\ena 
where $F_{\|,\perp}$ and $G_{\|,\perp}$ are total damping and
excitation coefficients defined by equations (\ref{eq_F}) and
(\ref{eq_G}) corresponding to the rotation parallel and perpendicular
to the grain symmetry axis.  After some manipulations, we get 
\bea
\langle \omega^{2}\rangle=3\langle
\omega_{\|}^{2}\rangle+\frac{G_{\|}}{F_{\|}}
\frac{2\kB T_{\rm gas}}{I_{\|}}\left(\frac{\alpha_{\|}}{\alpha_{\perp}}-1\right),
\ena 
where
\bea
\alpha_{\|}=\frac{I_{\|}F_{\|}}{G_{\|}},~~
\alpha_{\perp}=\frac{I_{\perp}F_{\perp}}{G_{\perp}}.
\label{alpha}
\ena 
Therefore, the increase of mean square angular velocity becomes
\bea \frac{\langle
  \omega^{2}\rangle-3\langle\omega_{\|}^{2}\rangle}
  {3\langle\omega_{\|}^{2}\rangle}
\equiv
\frac{\langle \omega^{2}\rangle-\langle\omega^{2}\rangle_{\rm DL98}}
     {\langle\omega^{2}\rangle_{\rm DL98}}
=
\frac{2}{3}\left(\frac{\alpha_{\|}}{\alpha_{\perp}}-1\right),
\label{dome2}
\ena 
where
$\langle\omega^{2}\rangle_{\rm DL98}=
3\langle\omega_{\|}^{2}\rangle=\left(G_{\|}/F_{\|}\right)\times
\left(3\kB T_{\rm gas}/I_{\|}\right)$.

It can be seen that the excess of mean square angular velocity of the
grain depends only on the ``anisotropy ratio''
\beq
\eta=\frac{\alpha_{\|}}{\alpha_{\perp}}=h\times\left(\frac{F_{\|}G_{\perp}}{F_{\perp}G_{\|}}\right) .
\label{eq:eta}
\eeq
The anisotropy can arise from the
non-sphericity of grain shape (i.e, $h\ne 1$) and from
the differential damping and excitation along the direction parallel
and perpendicular to the symmetry axis (i.e., $F_{\|}\ne F_{\perp},~
G_{\|}\ne G_{\perp}$). For the disk-like grain, $\eta>1$,
the increase of $\langle \omega^{2}\rangle$ is given by equation
(\ref{dome2}), and therefore we expect an increase of the frequency at the
emission peak (hereafter peak frequency). Since the power radiated by a
rotating grain is a nonlinear function of $\omega$, we expect a
substantial increase in the emissivity of spinning dust. We will study
the correlation of the increase of peak frequency to the anisotropy in
\S~9.3.

Infrequent
hits by single-ion collisions are able to transiently spin-up the
grain, producing spikes of angular velocity $\omega$. 
As a result, we expect both
the peak frequency and total emissivity increase when the transient
spin-up is taken into account. This issue we address in \S 8.

\subsection{Idealized Environments}

Table \ref{ISM} presents physical parameters for idealized environments where
$n_{\rm H}$ is the 
hydrogen number density, $T_{\rm gas}$ and $T_{\rm d}$ are gas and dust
temperature, $\chi=u_{\rm rad}/u_{\rm ISRF}$ is the ratio of radiation energy density
$u_{\rm rad}$ to the mean radiation density for the diffuse interstellar medium 
$u_{\rm ISRF}$ (see Mathis, Mezger, \& Panagia 1983), 
$n({\rm H}_{2}),~ n({\rm H}^{+}),~ n({\rm M}^{+})$
are the molecular hydrogen density, ion hydrogen density and ionized metal
density, respectively. Physical parameters
for the CNM, WNM, WIM, RN are similar to those in DL98b. The
parameters for the PDR are taken to be similar to those inferred for
the Orion Bar (see Allers et al.\ 2005).
\begin{table}
\caption{Idealized Environments For Interstellar Matter}\label{ISM}
\begin{tabular}{llllll} \hline\hline\\
\multicolumn{1}{c}{\it Parameters} & \multicolumn{1}{c}{CNM}& 
{WNM} &WIM &RN &PDR\\[1mm]
\hline\\
$n_{\rm H}$~(cm$^{-3}$) &30 &0.4 &0.1 &$10^{3}$ &$10^{5}$ \\[1mm]
$T_{\rm gas}$~(K)& 100 & 6000 &8000 &100 &1000\\[1mm]
$T_{\rm d}$~(K)& 20& 20 &20 &40 &80\\[1mm]
$\chi$ &1 &1 &1 &1000 &30000\\[1mm]
$x_{\rm H}$ &0.0012 &0.1 &0.99 &0.001 &0.0001\\[1mm]
$x_{\rm M}$ &0.0003 &0.0003 &0.001 &0.0002 &0.0002\\[1mm]
$y=2n({\rm H}_{2})/n_{\rm H}$&{$0.$} &0. &0. &0.01 &0.01\\[1mm]
\\[1mm]
\hline\hline\\
\end{tabular}
\end{table}

\section{Rotational Damping and Excitation for Imperfect Alignment}
Below we present calculations of rotational damping and excitation for
a disk-like (or cylindrical) grain with principal moments of inertia
$I_{\|}$ and $I_{\perp}$. The disk has radius $R$ and height $L$.
 The angular velocity $\bomega$ is at an arbitrary angle with the grain
symmetry axis $\ba_{1}$. When the angular velocity $\bomega$ is not
aligned with $\ba_{1}$, the rotation of the grain consists of the
rotation about $\ba_{1}$ with angular velocity $\omega_{\|}$ and the
rotation about an axis perpendicular to $\ba_{1}$ with angular
velocity $\omega_{\perp}$. Parallel and perpendicular components of
dimensionless damping and excitation coefficients are defined as in
equations (\ref{eq_F}) and (\ref{eq_G}), but $I_{\|}, \omega, \tau_{\rm H}$
are replaced by $I_{\|}, \omega_{\|}$ and $\tau_{\rm
  H,\|}$ and $I_{\perp}, \omega_{\perp}$ and $\tau_{\rm
  H,\perp}$.

\subsection{Collisional Damping and Excitation}

For a disk-like grain, the damping times for rotation parallel and
perpendicular to the grain symmetry axis due to gas collisions are
given in Appendix B. For a pure H gas of density $n_{\rm H}$, they
read 
\bea 
\tau_{\rm H,\|}&\approx&4.12\times 10^{10}
 a_{-7}\hat{\rho}{T}_{\rm 2}^{-1/2}\left(\frac{30 \mbox{
     cm}^{-3}}{n_{\rm H}}\right)\Gamma_{\|} {~\mbox
   s},\label{tgas1}\\ \tau_{\rm H,\perp}&\approx&4.58\times 10^{9}
 a_{-7}\hat{\rho}T_{\rm 2}^{-1/2}\left(\frac{30 \mbox{
     cm}^{-3}}{n_{\rm H}}\right)\Gamma_{\perp} {~\mbox
   s},~~~~\label{tgas2}
\\
%%btd 091213  I changed r to a -- see also appendix B where I changed
%%            definitions of \Gamma factors
\Gamma_{\|}&\equiv&\frac{8}{9}\left(\frac{6L}{R}\right)^{2/3}
\frac{1}{({2L}/{R}+1)},
\\ 
\Gamma_{\perp}&\equiv&\left(\frac{4}{3}\right)^{1/3}
\left(\frac{L}{R}\right)^{2/3}\left[3+({L}/{R})^{2}\right]\times
 \frac{1}{g_{\perp}}, 
\ena 
where $\hat{\rho}\equiv\rho/2{~\rm g\,cm^{-3}}$,
$T_2\equiv T_{\rm gas}/100$ K, $a_{-7}\equiv a/10^{-7}$ cm, and
%%btd 091213  change q_\perp to g_\perp for consistency by App. B
\bea g_{\perp}=
\frac{1}{6}\left(L\over R\right)^{3}+
\frac{L}{R}+\frac{1}{2}\left(\frac{L}{R}\right)^{2}+\frac{1}{2}.
\ena 
For dimensionless
collisional damping and excitation coefficients, 
$F_{\{\rm n,\rm i\},\|}=F_{\{\rm n,\rm i\},\perp}$ and 
$G_{\{\rm n,\rm i\},\|}=G_{\{\rm n,\rm i\},\perp}$  because they are
normalized over those of purely H gas.

It is convenient to define components of thermal angular velocity
 parallel and perpendicular to the grain symmetry axis at the gas 
temperature $T_{\rm gas}$:
\bea
\omega_{\rm T,\|}=\left(\frac{2\kB T_{\rm gas}}{I_{\|}}\right)^{1/2},~
\omega_{\rm T,\perp}=\left(\frac{2\kB T_{\rm gas}}{I_{\perp}}\right)^{1/2}.
~~~~\label{omega_T}
\ena

\subsection{Electric Dipole Damping}

The electric dipole moment $\bmu$ of the grain is assumed to be fixed
in the grain body system, and given by
\bea
\bmu=\mu_{1}\ba_{1}+\mu_{2}\ba_{2}+\mu_{3}\ba_{3},\label{mueq}
\ena 
where $\mu_{1},\mu_{2}$ and $\mu_{3}$ are components of $\bmu$ along 
the grain principal axes $\ba_{1}, \ba_{2}$ and $\ba_{3}$, respectively.

The rotation of the electric dipole
along $\ba_{1}$ results in the damping for $\omega_{\|}$, and its
rotation along the axes perpendicular to $\ba_{1}$ results in the
damping for $\omega_{\perp}$. The decrease of parallel 
and perpendicular components of
angular velocity due to electric dipole emission are given by (see Appendix B3)
\bea
\frac{I_{\|}d\omega_{\|}}{dt}=-\frac{I_{\|}^{2}\omega_{\|}^{3}}{3\kB T_{\rm
    gas}}\frac{1}{\tau_{\rm
    ed,\|}},\\ \frac{I_{\perp}d\omega_{\perp}}{dt}
= -\frac{I_{\perp}^{2}\omega_{\perp}^{3}}{3\kB T_{\rm
    gas}}\frac{1}{\tau_{\rm ed,\perp}}, 
\ena
where the damping times
for rotation along and perpendicular to $\ba_{1}$ are given by 
\bea
\tau_{\rm
  ed,\|}=\frac{3I_{\|,\perp}^{2}c^{3}}{6\mu_{\perp}^{2}\kB T_{\rm
    gas}},~~\tau_{\rm
  ed,\perp}=\frac{3I_{\|,\perp}^{2}c^{3}}{6(\mu_{\|}^{2}+\mu_{\perp}^{2}/2)\kB T_{\rm
    gas}},~~~\label{eq_taued}
\ena 
and $\mu_{\|}^{2}=\mu_{1}^{2}$ and $\mu_{2}^{2}=\mu_{3}^{2}=\mu_{\perp}^{2}/2$ have been
 assumed.

Following DL98b, the dipole moment is
given by 
\bea
\mu^{2}=23\left[\left(\frac{a_{\rm x}}{a}\right)^2 \langle
  Z^2\rangle+3.8\left(\frac{\beta}{0.4~\D}\right)^2 a_{-7}\right]a_{-7}^{2}
~\mbox{Debye}^{2},~~~
\label{mu2}
\ena
where $\langle Z^{2}\rangle$ is the
mean square grain charge, $\beta$ is the dipole moment per atom of the
grain, and $a_{-7}=a/10^{-7}$cm.  Plugging equation (\ref{mu2}) into
(\ref{eq_taued}) with the assumption of uniform distribution of $\bmu$
 along the grain principal axes, and using moments of inertia for the disk, we obtain
\bea
\tau_{\rm ed,\|}&=&\frac{3I_{\|}^{2}c^{3}}{4\mu^{2}\kB T_{\rm
    gas}},\nonumber
\\
&=&
1.6\times10^{11}
\hat{\rho}^{2}a_{-7}^{8}\times\frac{(R/L)^{4/3}}
     {[(a_{\rm x}/a)^2 \langle Z^2\rangle+3.8(\beta/0.4\D)^2a_{-7}]T_2}~\s,\nonumber\\
\label{tauedpar}
\ena
and
\bea
\tau_{\rm ed,\perp}
&=&
\frac{3I_{\perp}^{2}c^{3}}{4\mu^{2}\kB T_{\rm gas}},\nonumber
\\
&=&
4.0\times10^{10}
\hat{\rho}^{2} a_{-7}^8\times\frac{(R/L)^{4/3}
     \left(1+\frac{1}{3}(L/R)^2\right)^2}
     {[(a_{\rm x}/a)^2 \langle
Z^2\rangle+3.8(\beta/0.4\D)^2 a_{-7}]T_2}~\s.\nonumber\\
\label{tauedper}
\ena
For grains larger than $a_{2}$, $\tau_{\rm ed,\|}=\tau_{\rm ed,\perp}$.

\subsection{Damping and Excitation by the Plasma}
The problem of plasma-grain interactions for perfect internal
alignment was studied in DL98b, and refined by Ali-Ha\"imoud et
al. (2009). Here we consider the damping and excitation by the plasma
for the case of imperfect internal alignment. We assume for simplicity that 
$\bmu$ is directed along $\ba_{2}$ axis. We consider only neutral
grain ($Z_{\rm g}=0$, as in DL98b), and assume a spherical grain with a cylindrical
excitation equivalent radius $a_{\rm cx}$ (e.g. Ali-Ha\"imoud et
al. 2009).

For an ionized gas with ion density $n_{\rm i}$, ion mass $m_{\rm i}$, and
charge $Z_{\rm i}$, the dimensionless excitation coefficients parallel and
perpendicular to the grain symmetry axis from plasma drag are given
by 
\bea 
G_{\rm p,\|,\perp}=\frac{n_{\rm i}}{n_{\rm
    H}}\left(\frac{m_{\rm i}}{m_{\rm
    H}}\right)^{1/2}\left(\frac{Z_{\rm i}e\mu}{\kB T_{\rm
    gas}a_{\rm cx}^{2}}\right)^{2}{3\over
  2}\int_{0}^{\infty}ue^{-u^{2}}du~g_{\|,\perp},\nonumber\\
\label{gp}
\ena
where $e$ is the elementary charge, $u\equiv v/v_{\rm T}$, with $v$
being the velocity and $v_{\rm T}^2\equiv2\kB T_{\rm gas}/m_{\rm i}$,
and
\bea 
g_{\|,\perp}=\int_{b_{\rm max}/a_{\rm cx}}^{\infty}\frac{dl}{l}\mathcal
I_{\|,\perp}\left(\frac{u\Omega_{\|,\perp} }{l}\right),~\\
\Omega_{\|,\perp}=\left(\frac{m_{\rm i}a_{\rm cx}^{2}}{2\kB T_{\rm
      gas}}\right)^{1/2}\omega_{\|,\perp},\nonumber
\ena
where
$\mathcal I_{\|,\perp}\left({u\Omega_{\|,\perp} }/{l}\right)$ are given by equations (\ref{ipar}) and
(\ref{iper}) (see Appendix C2 for the definition of $b_{\rm max}$ and
detailed calculations).
\begin{figure}
\includegraphics[width=0.5\textwidth]{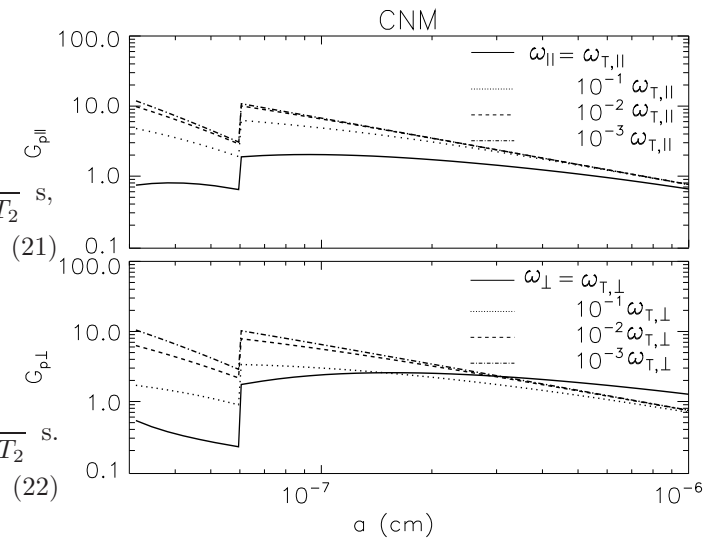}
\caption{$G_{{\rm p},\|}$ and $G_{{\rm p},\perp}$ as function of $a$ of a neutral
  grain for different values of $\omega_{\|}/\omega_{\rm T,\|}$ and
 $\omega_{\perp}/\omega_{\rm T,\perp}$  in the CNM. Electric
  dipole moment $\beta_{0}=0.4$ D is adopted.}
\label{Gpar}
 \end{figure}
Using the Fluctuation-Dissipation theorem (see Lazarian 1995; Lazarian
\& Roberge 1997 for the application to grain diffusion coefficients),
we obtain $F_{\rm p,\|}=G_{\rm p,\|}$, and $F_{\rm p,\perp}=G_{\rm
  p,\perp}$.

Figure \ref{Gpar} presents $G_{\rm p,\|}$ and $G_{\rm p,\perp}$ for
different values $\omega_{\|}/\omega_{\rm T,\|}$ and
 $\omega_{\perp}/\omega_{\rm T,\perp}$ for CNM. It can
be seen that $G_{\rm p,\|}$ and $G_{\rm p,\perp}$ increase as the
grain rotates slower, and for $\omega_{\|,\perp} <10^{-3}$, they converge to the
value for non-rotating grain, i.e, $\omega_{\|,\perp}=0$.

\section{\label{sec:PAH photoabs}
         Photoexcitation of PAHs}

Small PAHs are planar molecules, consisting of $sp^2$-bonded C atoms in 
a two-dimensional hexagonal structure, with peripheric H atoms.
PAHs absorb strongly in the ultraviolet.
At wavelengths $3000 \ltsim \lambda \ltsim 1500\Angstrom$, photoabsorption is
primarily due to $\pi\rightarrow\pi^*$ electronic transitions, 
excited by electric fields parallel to the plane defined by the C atoms.
PAHs -- especially PAH ions -- 
are also able to absorb at longer wavelengths, although the absorption
is much weaker. The electronic transitions responsible for the absorption
are not certain,
but we will here assume that this absorption also involves transitions with
electric dipole matrix elements parallel to the C atom plane.

Suppose the interstellar radiation field to consist of a unidirectional
component providing a fraction $\gamma$ of the local energy density of
starlight, plus an isotropic 
component.
Let $\hat{\bf n}_\star$ 
be a unit vector parallel to the direction of propagation
of the unidirectional component of the starlight.
Let the dimensionless factor 
\beq
U \equiv \frac{u_\star}{u_{\rm MMP}},
\eeq
where $u_\star$ is the starlight energy density,
and $u_{\rm MMP}=6.85\times10^{-14}\erg\cm^{-3}$ is the
estimate of Mathis, Mezger \& Panagia (1983, hereafter MMP)
for the local starlight energy density.
Let $\dot{N}_{abs,0}$ be the photon absorption rate for the PAH if
illuminated by isotropic starlight with $U=1$.
Assuming the unidirectional and isotropic components to have similar spectra,
the PAH photoexcitation rate is (Sironi \& Draine 2009)
\beqa \label{eq:dotNabs}
\dot{N}_{abs} &=& U\Psi(\theta)\dot{N}_{abs,0}~,
\\
\Psi(\theta) &\equiv&
\left[(1-\gamma) + 
\gamma\frac{3}{4}\left(1+\cos^2\theta\right)\right],
\\
\cos\theta&\equiv& \hat{\bf n}_\star\cdot\ba_{1},
\eeqa
where $\ba_{1}$ is a unit vector normal to the C atom plane.

\section{\label{sec:PAH IR}
         Revised Rotational Excitation and Damping by IR Emission}

DL98b discuss the various processes contributing
to the rotational excitation and deexcitation of very small dust grain.
The dominant processes include direct collisions of atoms and ions with
the dust grain, torques on the dust grain due to electric fields
produced by passing ions (``plasma drag''), and exchange of angular momentum
with the electromagnetic field via absorption and emission of photons. 
Because a dust grain radiates many more photons than it absorbs, the emission
processes are much more important than the change of angular momentum
resulting from photon absorptions.

If the grain is rotating, then there will be a tendency for the
infrared emission to, on average, reduce the grain angular momentum (
Martin 1972).
This process was included in the study by DL98b.
As noted by 
Ali-Ha\"imoud et al.\ (2009), 
the DL98b expression for the rotational damping torque 
from infrared emission
was too small by a factor of two owing to an algebraic error.
Here we make a more refined estimate of rates for excitation and
deexcitation of PAH rotation due to infrared emission.
We employ the PAH model of Draine \& Li (2007, hereafter DL07); this grain
model reproduces various observations of infrared emission, and therefore
should provide an improved estimate of rotational excitation and
damping from infrared emission.  The present treatment also takes into
account anisotropy in the infrared emission from planar PAHs.

Table 2 lists the character assumed for each of the vibrational modes
in the DL07 PAH model.

\begin{table}[htb]
\begin{center}
\caption{\label{tab:mode type}
         Optically-Active Vibrational Modes}
\begin{tabular}{l c l}
\hline
$\lambda(\micron)$ & in-plane fraction & identification$^*$\cr
\hline
3.30    & 1 & C-H stretch\cr
5.27	& 0 & oop C-H bend overtone?\cr
5.70	& 0 & oop C-H bend overtone?\cr
6.22	& 1 & C-C stretch\cr
6.69	& 1 & C-C stretch\cr
7.417	& 1 & C-C stretch\cr
7.598	& 1 & C-C stretch\cr
7.850   & 1 & C-C stretch\cr
8.33    & 1 & ip C-H bend\cr
8.61    & 1 & ip C-H bend\cr
10.68   & 0 & oop mono C-H bend\cr
11.23	& 0 & oop mono C-H bend\cr
11.33	& 0 & oop mono C-H bend\cr
11.99   & 0 & oop duo C-H bend\cr
12.62   & 0 & oop trio C-H bend\cr
12.69   & 0 & oop trio C-H bend\cr
13.48   & 0 & oop quartet C-H bend\cr
14.19   & 0 & oop quartet C-H bend\cr
15.90   & 2/3 & ?\cr
16.45	& 2/3 & ?\cr
17.04	& 2/3 & ?\cr
17.375	& 2/3 & ?\cr
17.87   & 2/3 & ?\cr
18.92   & 2/3 & ?\cr
15.     & 2/3 & FIR continuum?\cr
\hline
\multicolumn{3}{l}{$^*$~ip = in-plane mode; oop = out-of-plane mode}
\end{tabular}
\end{center}
\end{table}

Having in mind PAHs, we consider a grain with dynamical symmetry,
with eigenvalues $I_1 > I_2=I_3$ of the moment of inertia tensor. 
Let $\ba_{1}, \ba_{2}$ and $\ba_{3}$ be unit vectors parallel 
to the principal axes corresponding to moments of inertia
$I_1, I_{2}$ and $I_{3}$, respectively.
The vibrational modes will be assumed to have oscillations that are
either parallel to $\ba_{1}$
(``out-of-plane'') or perpendicular to $\ba_{1}$ (``in-plane'').
Let the angular momentum vector have components $J_{\|}$ parallel to $\ba_{1}$,
and $J_\perp$ perpendicular to $\ba_{1}$ (i.e., in the $\ba_{2}$-$\ba_{3}$
plane).

%------------------------ begin figure 1 ---------------------------
\begin{figure}
\begin{center}
\includegraphics[angle=0,width=0.5\textwidth]{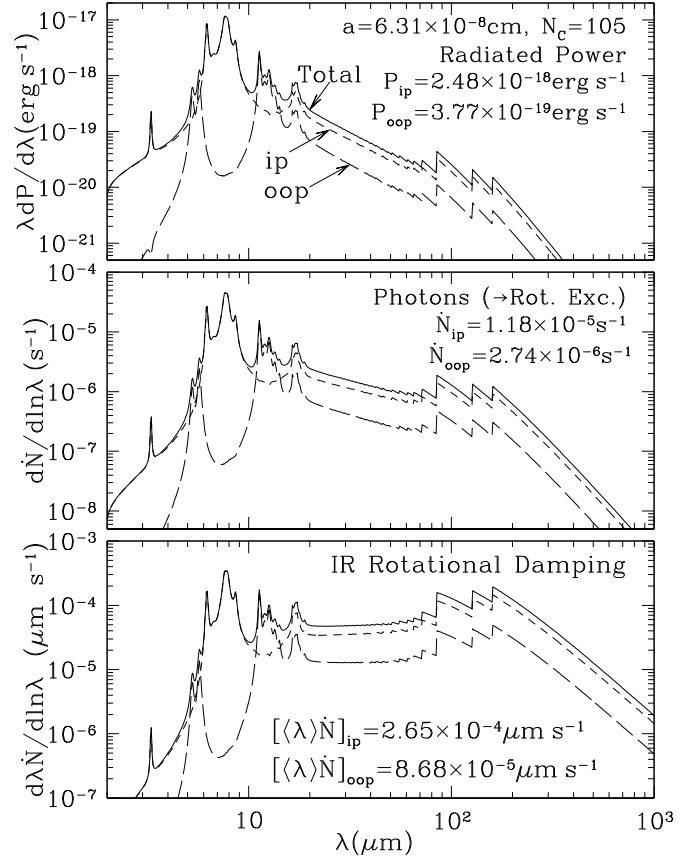}
\caption{\label{fig:spec}
Moments of the infrared emission from a PAH ion with
effective radius $a=6.31\Angstrom$ ($N_C=105$ carbon atoms), heated by
the average interstellar radiation field ($U=1$).
The radiated power (top panel) and
photon emission rate (middle panel) are 
dominated by $5\ltsim\lambda\ltsim15\micron$
emssion, but the rotational damping coefficient (bottom panel)
has an appreciable contribution from emission out to 
$\lambda\approx 200\micron$.
The sawtooth discontinuities in the spectra at $\lambda\gtsim50\micron$
arise from approximations in the calculational method (see text).
}
\end{center}
\end{figure}
%------------------------- end figure 1 ----------------------------

Suppose that the grain is absorbing photons at a rate
$\dot{N}_{abs}$, and radiating photons with a power per unit
frequency $P_\nu$. For $U<10^4$, small PAHs 
are in the single-photon-heating regime, with 
virtually all of the energy of an absorbed starlight photon being radiated
before the next photon absorption occurs.
This allows us to assume a universal time-averaged emission spectrum 
for each PAH, with the emission intensity proportional to the photon
absorption rate:
\beq
P_{m,\nu} \approx U\Psi(\theta) P_{m,\nu,0}~,
\eeq
where $P_{m,\nu,0}$ is the time-averaged power per unit frequency radiated by
in-plane ($m=ip$) or out-of-plane ($m=oop$) vibrations of a single PAH molecule
or ion
heated by isotropic radiation with $U=1$.
Figure \ref{fig:spec} shows $\nu P_{m,\nu,0}$ for a PAH with $\NC=105$ C atoms,
where $P_{m,\nu,0}$ is obtained by solving the equations of statistical
equilibrium for a PAH subject to stochastic heating
(Draine \& Li 2001, Li \& Draine 2001b), using the ``thermal-discrete''
method, where the emission from a PAH at a given energy is estimated using
a thermal approximation, but the downward transitions are treated as
discrete transitions to a lower energy level.
The model employs a heat capacity based on a realistic spectrum of
vibrational modes (Draine \& Li 2001), and updated
vibrational band strengths (Draine \& Li 2007) which, for a suitable
PAH size distribution, closely reproduce observed IR spectra.
The sawtooth discontinuities in the model spectrum arises from
approximations in the treatment.  The long-wavelength emission from a real
PAH would presumably be dominated by a finite number of emission lines, but the
continuous spectrum obtained here is expected to provide a reasonable
representation for the distribution of emitted power over wavelength.

% $\propto U\Psi(\theta)$, as in eq.\ (\ref{eq:dotNabs}).

The photon emission rate
\beq
\dot{N}_m \approx U\Psi \dot{N}_{m,0},
\eeq
where
\beq \label{eq:Ndot integral}
\dot{N}_{m,0}=\int_0^\infty \frac{1}{h\nu} P_{m,\nu,0} d\nu.
\eeq
The middle panel of Figure \ref{fig:spec} 
shows the integrand in eq.\ (\ref{eq:Ndot integral})
for the time-averaged 
rate of photon emission $\dot{N}_m$
by in-plane and out-of-plane modes, for PAH neutrals and ions.

The rotational damping also depends on the photon-weighted wavelength,
\beq \label{eq:<lambda>}
\langle\lambda\rangle_m \equiv \frac{1}{\dot{N}_m}
\int_0^\infty \frac{\lambda}{h\nu}P_{m,\nu}d\nu,
\eeq
for $m=ip$ or $m=oop$.
The lower panel of
Figure \ref{fig:spec} shows the integrand in eq.\ (\ref{eq:<lambda>}).
For the PAH emission properties assumed by Draine \& Li (2007), the small
fraction of the total power radiated at $\lambda\gtsim100\micron$ contributes
substantially to the integral in eq.\ (\ref{eq:<lambda>}), but does not
dominate the integral.

%------------------------- begin figure 2 -------------------------------
\begin{figure}
\begin{center}
\includegraphics[angle=0,width=0.5\textwidth]{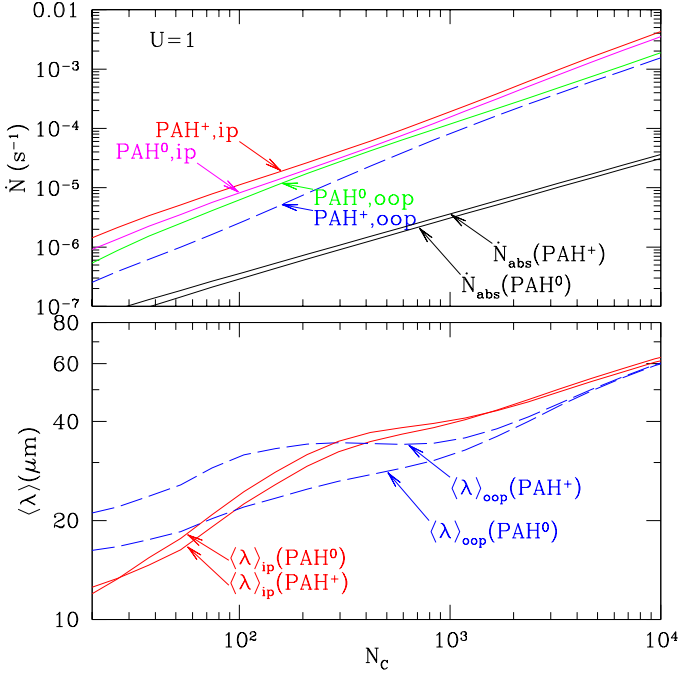}
\caption{\label{fig:rotxdx vs a}
Rate of photon emission $\dot{N}$ (upper panel) and $\langle\lambda\rangle$ 
(lower panel)
as a function of $N_C$, the number of C atoms in the PAH, for PAHs in the
local interstellar radiation field.
}
\end{center}
\end{figure}
%---------------------------- end figure 2 ----------------------------

Figure \ref{fig:rotxdx vs a} shows how $\dot{N}_m$ and
$\langle\lambda\rangle$ depend on the the number $\NC$ of C atoms in the PAH. 

%------------------------------ begin figure 3 ---------------------------------
\begin{figure}
\begin{center}
\includegraphics[angle=0,width=0.5\textwidth]{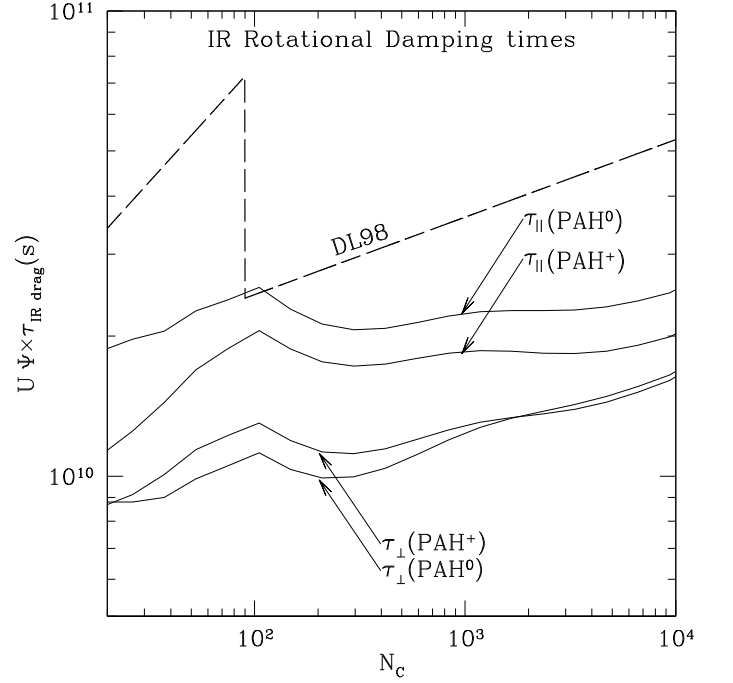}
\caption{\label{fig:taudrag}
Time scale for damping of angular momentum parallel to the symmetry
axis ($J_{\|}$) and perpendicular to the symmetry axis ($J_\perp$).
}
\end{center}
\end{figure}
%-------------------------------------------------------------------------------

Let $J$ be the grain angular momentum.
We assume that the grain has no intrinsic helicity, so that a nonrotating
grain is equally likely to emit left- or right-circularly polarized photons.
If the grain is initially non-rotating, IR emission will
result in\footnote{The factor 2 in front of $\hbar^{2}$ is pointed out by
 Silsbee et al. (2010).}
\beqa \label{eq:d/dt(Jz2)}
\langle \frac{d}{dt} J_{\|}^2\rangle 
&=& 
\frac{1}{2}U\Psi(\theta)\left(\dot{N}_{abs,0}+\dot{N}_{ip,0}\right) 2\hbar^2,
\\ \label{eq:d/dt(Jperp2)}
\langle \frac{d}{dt} J_\perp^2\rangle 
&=& 
\frac{1}{2}U\Psi(\theta)
\left[\dot{N}_{oop,0}+\frac{1}{2}
\left(\dot{N}_{abs,0}+\dot{N}_{ip,0}\right)\right]
2\hbar^2.~~~~~
\eeqa

The infrared emission will, on average, remove angular momentum
from the grain, with rates
(see Appendix B2)
\beqa
\left(\frac{dJ_{\|}}{dt}\right)_{\rm IR} 
&=& 
-\frac{3\hbar}{2\pi I_{\|} c}U\Psi(\theta)
\dot{N}_{ip,0}\langle\lambda\rangle_{ip}
J_{\|},~~~~\\
\left(\frac{dJ_\perp}{dt}\right)_{\rm IR}
&=& 
-\frac{3\hbar}{2\pi I_{\perp} c}U\Psi(\theta) 
\left[ 
\frac{1}{2}\dot{N}_{ip,0}\langle\lambda\rangle_{ip}
+
\dot{N}_{oop,0}\langle\lambda\rangle_{oop}
\right] 
J_\perp.~~~~
\eeqa

The characteristic time scales for rotational damping by IR emission are
\beqa
\tau_{\rm IR,\|} &\equiv&\frac{-J_{\|}}{\left(dJ_{\|}/dt\right)_{\rm IR}} =
\frac{1}{U\Psi(\theta)}\frac{2\pi I_{\|} c}{3\hbar}
        \left[\dot{N}_{ip,0}\langle\lambda\rangle_{ip}\right]^{-1},~~~
\\
\tau_{\rm IR,\perp} &\equiv& \frac{-J_\perp}{\left(dJ_\perp/dt\right)_{\rm IR}}=
\frac{1}{U\Psi(\theta)}\frac{2\pi I_{\perp} c}{3\hbar}\nonumber\\
&&\times\left[(1/2)\dot{N}_{ip,0}\langle\lambda\rangle_{ip}
+ \dot{N}_{oop,0}\langle\lambda\rangle_{oop}\right]^{-1}.~~~
\eeqa

The rotational damping times $\tau_{\rm IR,\|}$ and 
$\tau_{\rm IR,\perp}$ are shown in Figure \ref{fig:taudrag} for PAHs in the
average interstellar radiation field.  We show separate damping times
for PAH$^0$ and PAH$^+$, but the timescale for changes in PAH
ionization are short compared to the time scale for changes in angular
momentum, so one should take an appropriate weighted mean: 
\beqa
\tau_{\rm IR,\|}^{-1} &=& f_0 \tau_{\rm IR,\|}^{-1}({\rm PAH}^0) +
(1-f_0)\tau_{\rm IR,\|}^{-1}({\rm PAH}^+), \\ \tau_{\rm IR,\perp}^{-1}
&=& f_0 \tau_{\rm IR,\perp}^{-1}({\rm PAH}^0) + (1-f_0)\tau_{\rm
  IR,\perp}^{-1}({\rm PAH}^+), 
\eeqa
where $f_0$ is the neutral
fraction for PAHs of that size.

Also shown is the rotational damping time estimated by DL98b for
graphitic grains.  Our new estimates for $\tau_{\rm IR,\|}$ and
$\tau_{\rm IR,\perp}$ are smaller than the DL98b estimate by factors
of 1--6, depending on size, charge state, and the orientation of
$\ba_{1}$ relative to the angular momentum $\bJ$.  A factor of 2 is
attributable to an algebraic error in DL98b (see Ali-Ha\"imoud et
al. 2009); the remaining differences are due to use of realistic PAH
emission properties rather than the power-law heat capacity and
power-law opacity used by DL98b, as well as different assumptions
regarding the grain moment of inertia.

%-----------------------------------------

Using equations (\ref{eq_F}) and (\ref{eq_G}), the parallel and
perpendicular components of the dimensionless coefficients of damping
and excitation from infrared emission are then given by 
\bea
 F_{\rm
  IR,\|}&=&\frac{\tau_{\rm H,\|}}{\tau_{\rm IR,\|}},~~~F_{\rm
  IR,\perp}=\frac{\tau_{\rm H,\perp}}{\tau_{\rm
    IR,\perp}},\label{fir_par}\\
 G_{\rm
  IR,\|}&=&\langle\frac{dJ_{\|}^{2}}{dt} \rangle\frac{\tau_{\rm
    H,\|}}{2I_{\|}\kB T_{\rm gas}}\nonumber\\
 &=&\frac{1}{2}U\Psi(\theta)\frac{(\dot{N}_{abs,0}+\dot{N}_{ip,0})2\hbar^{2}
}{2I_{\|}\kB T_{\rm gas}}\tau_{\rm H,\|},~~~\label{gir_par}\\
 G_{\rm
  IR,\perp}
&=&\langle\frac{dJ_{\perp}^{2}}{dt} \rangle\frac{\tau_{\rm
    H,\perp}}{2I_{\perp}\kB T_{\rm gas}}\nonumber
\\ 
&=&
\frac{1}{2}U\Psi(\theta)\frac{\left[\dot{N}_{oop,0}+\frac{1}{2}
(\dot{N}_{abs,0}+\dot{N}_{ip,0})\right]2\hbar^{2}}
{2I_{\perp}\kB T_{\rm gas}}\tau_{\rm H,\perp},
~~~~\label{gir_per} 
\ena 
where $I_{\|}$ and $I_{\perp}$ are
given by equation (\ref{Iper}).

\begin{figure}
\includegraphics[width=0.5\textwidth]{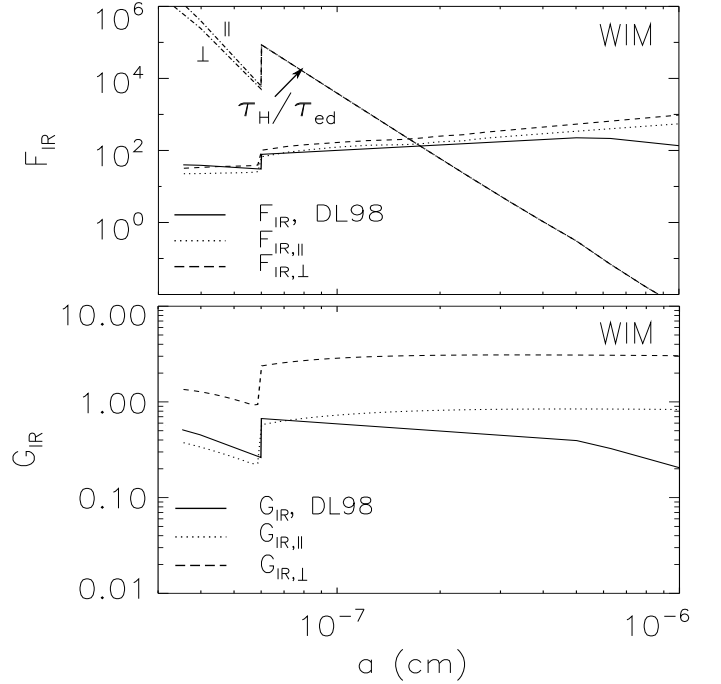}
\caption{Comparison of our new damping and excitation coefficients
  arising from infrared emission, $F_{\rm IR}, G_{\rm IR}$ with those
  in DL98b for WIM with $f_{0}\approx 0$. The ratio of damping times
  $\tau_{\rm H}/\tau_{\rm ed}$ for rotation parallel and perpendicular to the
  grain symmetry axis also shown.}
\label{FG_IRWIM}
 \end{figure}

\begin{figure}
\includegraphics[width=0.5\textwidth]{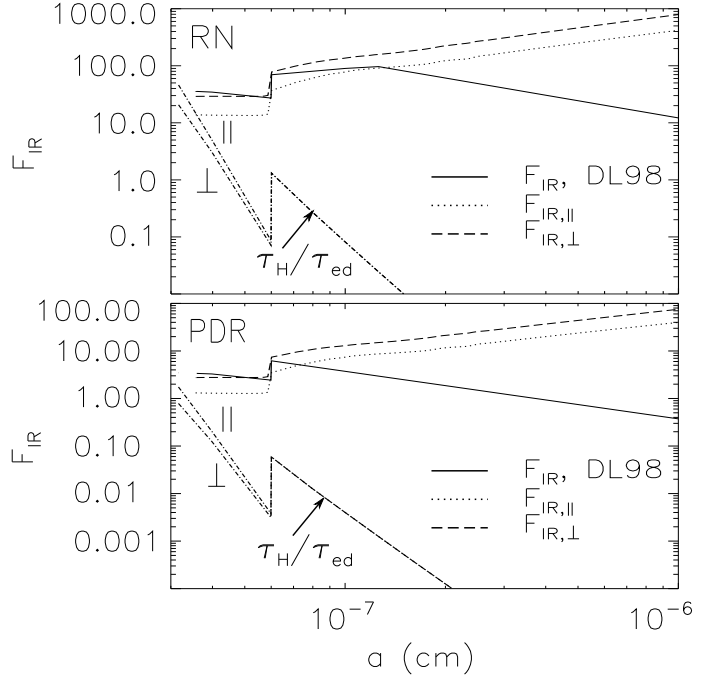}
\caption{Similar to figure \ref{FG_IRWIM} but for RN and PDR. Here the
  neutral fraction of PAH $f_{0}=1$.}
\label{fir_new}
 \end{figure}

The new coefficients for rotational excitation and damping by 
infrared emission derived in this
section are compared with those from the DL98b model in Figures
\ref{FG_IRWIM} and \ref{fir_new} for WIM, RN and PDR. There the
electric dipole damping rate for parallel and perpendicular direction
is also shown. For spherical grains (i.e. $a > 6\times 10^{-8}$cm), the
parallel and perpendicular components of $F_{\rm IR}$ and $G_{\rm IR}$ are not
the same as the electric dipole damping rate, i.e., there exists the
anisotropy in the damping and excitation. Such a high anisotropy of
the damping and excitation from infrared emission can result in
important effects of the dynamics of spinning dust grains.

\section{Effect of grain precession on the dipole emission}
Earlier studies (DL98ab, Ali-Haimoud et al. 2009) assumed that the frequency 
of dipole emission from rotating
grain is the same as the rotational frequency  $\omega/2\pi$.
Since the electric dipole is fixed in the body system, the rotation and 
precession of the grain with respect to fixed $\bJ$ can induce
some modification for the dipole emission frequency from 
its rotational frequency.

\begin{figure}
\includegraphics[width=0.4\textwidth]{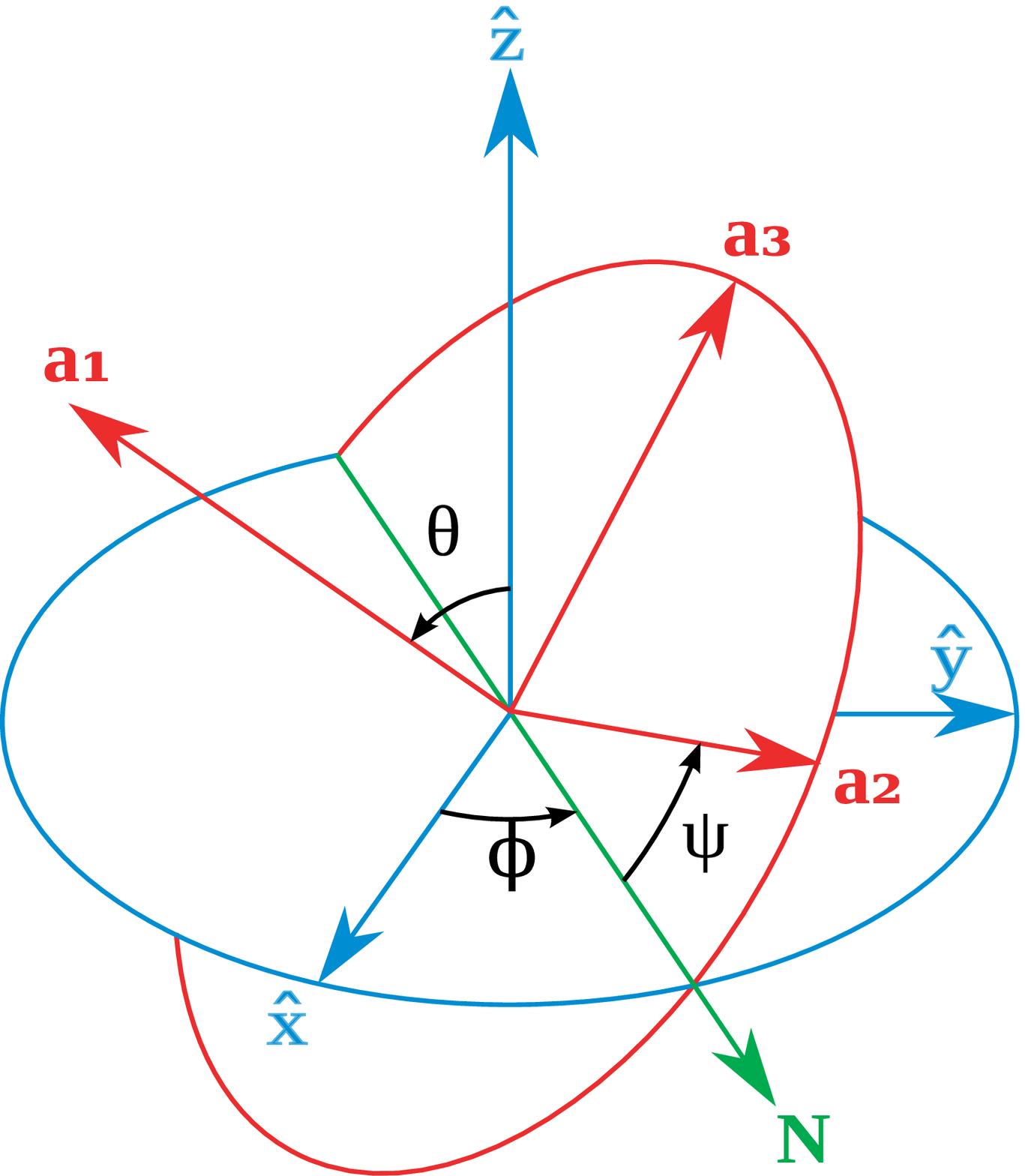}
\caption{Torque-free motion of the grain in an external system with 
$\zhat \| \bJ$ and $\xhat,\yhat\perp \bJ$ described by three Euler angles 
$\theta, \phi$ and $\psi$.}
\label{preces}
 \end{figure}

Let us consider the simple case of a disk-like grain where the dipole lies in
the plane $\ba_{1}\ba_{2}$: 
\bea
\bmu=\mu_{\|}\ba_{1}+\mu_{\perp}\ba_{2}
\ena
in the grain body system.
The torque-free motion of this grain 
with angular momentum $\bJ$ and ratio of inertia moments $h$ 
consists of the precession of the symmetry axis $\ba_{1}$ about $\bJ$ with 
constant angle $\theta$ and rate $\dot\phi$, and the rotation of grain itself 
about the symmetry axis with the rate $\dot\psi$ where $\phi, \psi$ and 
$\theta$ are Euler angles (see Figure \ref{preces}). The precession and 
rotation rates  are respectively given by (Landau \& Lifshitz 1976)
\bea
\dot\phi=\frac{J}{I_{\perp}},~~~\dot{\psi}=(1-h)\frac{J}{I_{\|}}\cos\theta,
~~.\label{omep}
\ena

Precession and rotation of the grain with respect to $\bJ$ results in an 
acceleration for the dipole moment:
\bea
\ddot{\bmu}=\mu_{\|}\ddot{\ba}_{1}+\mu_{\perp}\ddot{\ba}_{2},\label{dotmu}
\ena
where $\ddot{\ba}_{1}$ and $\ddot{\ba}_{2}$ are functions of time and 
given in Appendix F. 

The instantaneous emission power of the dipole moment is defined by 
\bea
P_{\rm ed}(J,\theta,t)=\frac{2}{3c^{3}}|\ddot\bmu|^{2}.\label{power}
\ena

The emission 
power of the grain
 is obtained by averaging (\ref{power}) over the torque-free motion of the grain, 
i.e., over angles $\psi$ and $\phi$ in the 
range from $0$ to $2\pi$. The final result for the case $\mu_{\|}=0$ and 
$\mu_{\perp}=|\bmu|$ is 
\bea
\frac{P_{\rm ed}(J,\theta)}{2\mu_{\perp}^{2}/3c^{3}}&=&\frac{1+\mcs\theta}{2}
(\dot\phi+\dot\psi)^{4}\nonumber\\
&&-2(\dot\phi^{3}\dot\psi+\dot\phi\dot\psi^{3})
(1-\cos\theta)^2+\frac{1}{2}\dot\psi^{4}\sin^{2}\theta.~~~~\label{ped}
\ena

\begin{figure}
\includegraphics[width=0.5\textwidth]{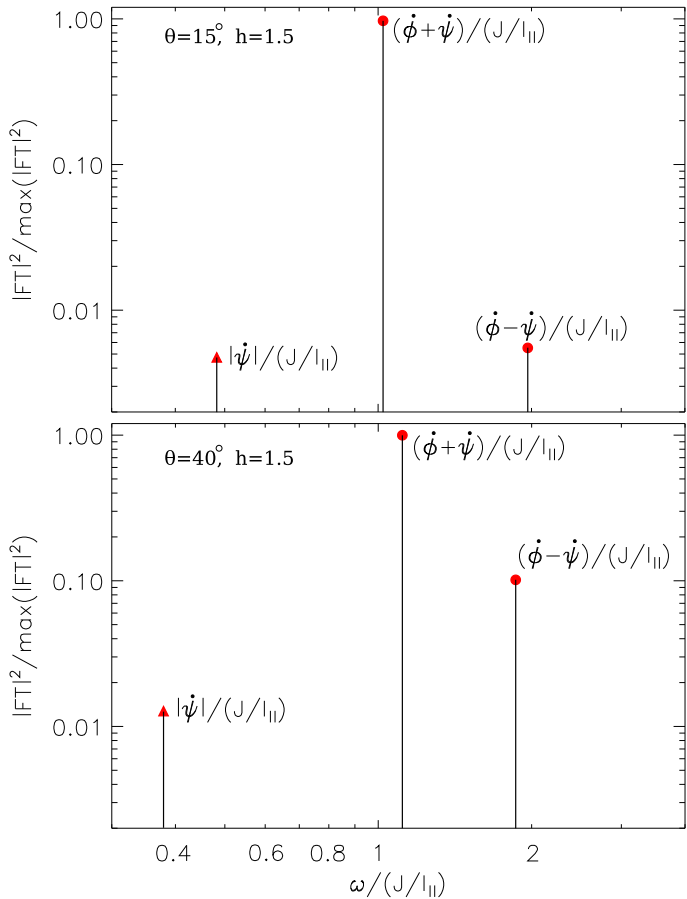}
\caption{Normalized emission spectrum for a disk-like grain ($h=1.5$) with
   $\mu_\parallel=0$, for two values of the angle $\theta$ between the
   principal axis ${\bf a}_1$ and the angular momentum ${\bf J}$.  The
   two components of $|{\rm FT}(\ddot{\mu}_x)|^2$ (or $|{\rm FT}(\ddot{\mu}_y)|^2$) are
   indicated by circles, while the component of $|{\rm FT}(\ddot{\mu}_z)|^2$ is indicated by a
   triangle. }
\label{fft}
\end{figure}

To find the frequency of dipole emission, we perform the Fourier transform 
for the components of dipole acceleration $\ddot{\bmu}$. 
Normalized square amplitude of the Fourier transform for the components 
$\ddot\mu_{x}$ (or $\ddot\mu_{y}$) and $\ddot\mu_{z}$ are 
shown in Figure  \ref{fft} for $\theta=15$ and $40^{\circ}$ for a disk-like grain 
with $h=1.5$. We can see that the emission spectrum from the electric dipole 
corresponds to three frequency modes, $\dot\psi/2\pi$, resulting from the 
component $\ddot\mu_{z}$, and $(\dot\phi\pm\dot\psi)/(2\pi)$, arising from
 the component $\ddot{\mu}_{x}$ (or$\ddot{\mu}_{y}$).
The emission power from the former mode is negligible compared to that 
from the later modes, which have frequencies given by
\bea
\nu=\frac{\dot\phi\pm\dot\psi}{2\pi}=
\frac{J}{I_{\|}}\frac{h\mp(h-1)|\cos\theta|}{2\pi},\label{nued}
\ena
in which the major mode $(\dot{\phi}+\dot{\psi})/2\pi$ for $\theta<90^{\circ}$ 
(or $(\dot\phi-\dot\psi)/2\pi$ for $\theta>90^{\circ}$) is much stronger 
than the second mode $(\dot\phi-\dot\psi)/2\pi$ ($(\dot\phi+\dot\psi)/2\pi$).
 When $\theta$ increases, the amplitude of higher frequency mode increases, 
and that mode becomes important as $\theta$ approaches $90^{\circ}$ 
(see Fig. \ref{fft}).  
However, for the fast internal  relaxation, the angle $\theta$ 
fluctuates with small amplitude about $\bJ$, and the lower frequency mode 
is dominant. Therefore, we are interested only in the emission of this 
lower frequency mode in the present paper.
The ratio of emission frequency of the dominant mode to the rotation 
frequency $\omega/2\pi$ is given by
\bea \label{eq:nu(J,theta)}
\frac{\nu}{\omega/2\pi}=
\frac{\dot\phi-|\dot\psi|}{\omega}=\frac{h-(h-1)|\cos\theta|}
{\sqrt{\mcs\theta+h^{2}\mss\theta}}\le 1.\label{nuratio}
\ena
It turns out that the dominant emission frequency is {\it smaller} than the rotation 
frequency, and it is very close to the rotation frequency for very small 
$\theta$. 
The emissivity at the frequency 
$\nu$ per H is given by
\beq
{j_\nu\over n_\H} = 
{1\over 4\pi}{1\over n_\H}
\int_{a_{\rm min}}^{a_{\rm max}} da {dn\over da} 
j_{\nu,a}~,\label{jnued}
\eeq
where
\bea
{j_{\nu,a}}= P_{\rm ed}(J,\theta) f_{\nu}(J,\theta),\label{jnup1}
\ena
and $f_{\nu}(J,\theta)d\nu$ is the probability of the dipole emission
in $[\nu,\nu+d\nu]$.

If we assume that all of the emission is at the dominant frequency
given by (\ref{eq:nu(J,theta)}), then 
equation (\ref{jnup1}) 
can be rewritten in terms of integrals over $J$ and $\theta$ by 
introducing a Delta function:
\bea
{j_{\nu,a}}&=&
\int\int d\theta dJ P_{\rm ed}(J,\theta)\delta(\nu'-\nu) f_{J}
f_{\theta}.~~~~~~~\label{jnup2}
\ena
where $\nu'=\nu'(J,\theta)$ is the emission frequency corresponding to $J$
and $\theta$, $f_{J}dJ$ is the probability of the grain having
angular momentum in $[J,J+dJ]$, and
$f_{\theta}d\theta$ is the probability of the
angle being in $[\theta,\theta+d\theta]$.

It is possible to calculate $j_{\nu,a}$ using 
(\ref{jnup2}).
 However, we can assume that the emission power is approximately 
given by a power law of the emission frequency 
\bea
P_{\rm ed}(\nu)=\frac{2}{3c^{3}}\frac{2}{3}\mu_{\perp}^{2}(2\pi\nu)^{4},\label{ped_nu}
\ena
We found that the emissivity obtained using (\ref{jnued}) and 
(\ref{jnup1}) with this approximation does not differ more than $10\%$ 
from more rigorous calculations
 using (\ref{jnup2}). Therefore, in the following, we calculate the emissivity using 
equations (\ref{jnued}-\ref{jnup1}) with the approximated emission power 
$P_{\rm ed}(\nu)$ given 
by equation (\ref{ped_nu}).

\section{Numerical Methods}

The most important element in calculating the emissivity 
from spinning dust grains is to find the probability 
distribution $f_\omega$ and $f_\nu$ for rotation resulting in emission at frequency $\nu$
%distribution $f_\nu(\nu)$ for rotation resulting in emission at frequency $\nu$
for a given 
grain size $a$ and a given electric dipole
moment $\mu$. Here we present a numerical approach using the Langevin
equation to obtain those distribution functions.

\subsection{The Langevin Equation}

The Langevin equation is a well-known equation for describing
stochastic dynamics (see Gardner 1983). In terms of its assumptions
and applicability it is equivalent to the well known Fokker-Planck
equation. In particular, the equation can handle only situations when
the increments of the variable are smaller than the variable. Thus,
the Langevin equation cannot be applied directly for treating the high
impulse collisions of spinning dust grains with ions. Therefore in what
follows we use a hybrid approach, describing the grain dynamics for
most of the computational time using the Langevin equation and
describing the infrequent (but high impact) collisions with ions as
impulsive events.

The angular momenta of the impacting ions are drawn from distribution
functions that include focusing by the attractive potential when
approaching neutral or negatively-charged grains. 
For describing grain
dynamics the Langevin equation was first suggested in Roberge, DeGraff
\& Flaherty (1993) and found more applications in later works on
grain alignment (see Roberge \& Lazarian 1999, Hoang \& Lazarian 2008,
2009).

\subsubsection{1D rotation}

In an idealized situation when the grain rotates only about its axis
of major inertia $\ba_{1}$, namely 1D rotation, the rotation of the
grain is determined by the value of angular velocity
$\omega\equiv\omega_{\|}$, and distribution function 
$f_{\nu}=2\pi f_\omega$.\footnote{This situation is similar that in
  the DL98 model and in the improved treatment by Ali-Ha\"imoud et
  al. (2009).} Note that for this case, $\tau_{\rm H}\equiv
\tau_{\rm H,\|}$, and $\omega_{\rm T}\equiv \omega_{\rm T,\|}$, and notations $\|$
and $\perp$ are omitted in this section.

For a given grain size, the evolution of $\omega$ can be described by
a stochastic differential (Langevin) equation. In dimensionless units
of $\omega'=\omega/\omega_{\rm T}$ and $t'=t/\tau_{\rm H}$, the Langevin
equation reads 
\bea 
{d\omega'}=-\left[F_{\rm
    tot}\omega'+\frac{2\omega'^{3}\tau_{\rm H}}{3\tau_{\rm ed}}\right] dt'
+\sqrt{{G_{\rm tot}}\over \tau_{\rm H}}dq,\label{le} \ena where
$dt'=dt/\tau_{\rm H}$ is the time step, $dq$ is a random variable with
variance $\langle dq^{2}\rangle=dt$, and the total damping $F_{\rm
  tot}$ and diffusion coefficients $G_{\rm tot}$ are given by \bea
F_{\rm tot}=F_{\rm i}+F_{\rm n}+F_{\rm IR}+F_{\rm p},\\ G_{\rm
  tot}=G_{\rm i}+G_{\rm n}+G_{\rm IR}+G_{\rm p},\label{FG} 
\ena 
where
$F_{j}$ and $G_{j}$ with $j$=i, n, IR, p are damping and excitation
coefficients corresponding to ion-grain, neutral-grain interactions,
infrared emission and plasma drag.

To solve equation (\ref{le}) for $\omega'$, we numerically integrate
it for $N$ time steps with total time $T$. After each time step $dt$,
the value $\omega'$ is updated via equation \bea
\omega'_{i+1}=\omega'_{i}-\left[F_{\rm
    tot}\omega'_{i}+\frac{2\omega_{i}^{'3}\tau_{\rm H}}{3\tau_{\rm
      ed}}\right] dt'+\sqrt{{G_{\rm tot}}\over \tau_{\rm
    H}}dq,\label{ome_ip1} \ena for $i$ running from $1$ to $N$. The
initial value $\omega'_{0}$ is a random variable in the range [0,1]
generated from a uniform distribution function.\footnote{The initial value is
unimportant as it is ``forgotten'' if the simulation is sufficiently long.}
The obtained solution
$\omega'_{i}$ is binned and used to find the distribution
function $f_\omega^{(1)}$. We note that $f_\omega^{(1)}$
is for the rotation along one axis only. By assuming 
isotropic distribution of the vector $\bomega$, the distribution
function $f_\omega\propto \omega^2f_\omega^{(1)}$ with normalization
condition $\int_{0}^{\infty}f_\omega d\omega=1$. 
Results for $f_\omega$ are presented
in \S 7.3.

\subsubsection{Grain Wobbling: 3D rotation}
When the angular momentum is not completely aligned with the grain
axis of major inertia (i.e., 3D rotation), its orientation within the
grain body has to be accounted for in finding $f_\omega$ and $f_\nu$.

Problems of internal alignment, i.e. to what extent the angular
momentum $J$ and the grain axis of major inertia are
aligned have been always at the focus of the quantitative treatment of
grain alignment (see Lazarian 2007, for a review). The initial works
(e.g. Jones \& Spitzer 1967) usually assumed a Maxwellian
distribution for $J$. Then Purcell (1979) noticed
that the internal relaxation of energy within a wobbling interstellar
grain may be fast (i.e, the timescale of internal relaxation is
smaller compared to the gas damping time)\footnote{Purcell (1979)
  introduced a new magneto-mechanical effect, which he termed Barnett
  relaxation, which he showed to be able to align 0.1 $\mu$m grains
  over a time scale of the order of a year. Later another
  magneto-mechanical effect, termed nuclear relaxation, was introduced
  in Lazarian \& Draine (1999). Nuclear relaxation is
  much faster than Barnett relaxation for 0.1 $\mu$m
  grains, but is not applicable to the tiny spinning grains 
  considered here.}. 
Due to the fast internal relaxation demonstrated in
Purcell (1979) for a period of time the works on grain dynamics
assumed that the grain axis of major inertia must
always be directed along the grain angular momentum, as this alignment
minimizes the energy of the rotating grain. However, Lazarian (1994)
noticed that this cannot be true for thermally rotating grains and
restored the notion of grain wobbling as an essential element of grain
dynamics.

In the 3D rotation case, we consider the evolution of $\bJ$ in the lab
coordinate
system instead of $\bomega$. For a general grain shape, $\bJ$ is completely
determined by three
components $J_{x,y,z}$ in the inertia coordinate system $\be_{1}\be_{2}\be_{3}$
  (see Appendix E).
 The evolution of $\bJ$ in time is also described by three Langevin equations
\bea
dJ_{i}=A_{i}dt+\sqrt{B_{ii}}dq_{i},\mbox{~for~} i=\mbox{~x,~y,~z},\label{le3d}
\ena
where $dq_{i}$ is the random Gaussian variables with $\langle dq_{i}^{2}\rangle=dt$, 
$A_{i}=\langle {\Delta J_{i}}/{\Delta t}\rangle$ 
and $B_{ii}=\langle \left({\Delta J_{i}}\right)^{2}/{\Delta t}\rangle$ are damping and
diffusion coefficients defined in the inertial coordinate system.
In dimensionless units $\bJ'=\bJ/I_{\|}\omega_{\rm T,\|}$,
$t'=t/\tau_{\rm H,\|}$, equation (\ref{le3d}) becomes 
\bea
dJ'_{i}=A'_{i}dt'+\sqrt{B'_{ii}}dq'_{i},\label{le3d2}
\ena
where $\langle dq_{i}^{'2}\rangle=dt'$, and
\bea
A'_{i}&=&-J'_{i}{F_{\rm tot,\|}}\left({\mbox{cos}^{2}\theta+
\gamma_{\rm H} \mbox{sin}^{2}\theta}\right)\nonumber\\
&&-\frac{2}{3}\frac{J_{i}^{'3}}{\tau'_{\rm ed}}\left({\mbox{cos}^{4}\theta+
\gamma_{\rm ed}\mbox{sin}^{4}\theta}+\frac{h^{3}+3h}{2}\sin^{2}\theta\cos^{2}\theta\right),~~~~\\
B'_{ii}&=&\frac{B_{ii}}{2I_{\|}\kB T_{\rm gas}}\tau_{\rm H,\|},
\ena
where 
\bea
\tau'_{\rm ed}&=&\frac{\tau_{\rm ed,\|}}{\tau_{\rm H,\|}},~~~
\gamma_{\rm H}=\frac{F_{\rm tot,\perp}\tau_{\rm H,\|}}{F_{\rm tot,\|}\tau_{\rm H,\perp}},~~
\gamma_{\rm ed}=\frac{I_{\|}\tau_{\rm ed,\|}}{I_{\perp}\tau_{\rm ed,\perp}},
\ena
and $\theta$ is the angle between $\ba_{1}$ and $\bJ$, $F_{\rm tot,\|}, 
F_{\rm tot,\perp}$ are total damping coefficients parallel and perpendicular 
to the grain symmetry axis, and $B_{ii}$ are given in Appendix E. 

\subsection{Effect of Internal Thermal Fluctuations}
 
Very little is known about internal relaxation in microscopic spinning
grains. Thus, in what follows, for our description of grain wobbling
we consider two extreme models. The first model assumes fast
internal relaxation, with the distribution of 
deviations of the axis of grain maximal
inertia from the direction of $\bJ$ determined by the 
dimensionless ratio $J^2/I_\parallel \kB T_{\rm d}$,
where $T_{\rm d}$
is the temperature of the grain body (Lazarian
\& Roberge 1997). The second model assumes no internal
relaxation, with a Maxwellian distribution function for the angle between
$\ba_{1}$ with $\bJ$.

For a given value of angular momentum, and in the presence of fast
internal relaxation, the angle $\theta$ between the grain axis of
major inertia $\ba_{1}$ and $\bJ$ follows the local thermal
equilibrium distribution function 
\bea f_{\rm LTE}(\theta)&=&A \mbox{
  exp}\left(-\frac{J^{2}}{2I_{\|}\kB T_{\rm d}}(1+[h-1]
\mbox{sin}^{2}\theta)\right),\nonumber\\ &=&A
\mbox{
  exp}\left(-J^{'2}\frac{T_{\rm gas}}{T_{\rm d}}(1+[h-1]
\mbox{sin}^{2}\theta)\right),\label{fle}
\ena 
where $h=I_{\|}/I_{\perp}, J'={J}/{I_{\|}\omega_{\rm T,\|}}$,
and $A$ is the normalization constant such that $\int_{0}^{\pi} f_{\rm
  LTE}(\theta)2\pi\sin\theta d\theta=1$ (Lazarian \& Roberge 1997). 

When the timescale of internal relaxation is much longer than that of
gas damping time, i.e., without internal relaxation, the angle
distribution function is simply Maxwellian: 
\bea 
f_{\rm Mw}(\theta)=
\frac{h}{4\pi}
\frac{1}{\left(\mbox{cos}^{2}\theta+h\mbox{sin}^{2}\theta\right)^{3/2}}
~~~,
\label{fmax}
\ena 
where $\int_{0}^{\pi} f_{\rm Mw}(\theta)2\pi\sin\theta
d\theta=1$ (see Jones \& Spitzer 1967; Lazarian \& Roberge 1997).

To follow the evolution of $\bJ$ in time in the presence of
internal relaxation, we use the following algorithm. First, we generate
initial values at $t=0$ for
$J'$, angle $\xi$ between $\bJ$ and $\be_{1}$ and
azimuthal angle $\chi$ from uniform distribution functions. Thus,
$J_{z}'(t=0)=J' \cos\xi, J_{x}'(t=0)=J'\sin\xi\cos\chi$ and
$J_{y}'(t=0)=J'\sin\xi\sin\chi$. Then we solve
equations (\ref{le3d2}) with the time step $dt$ to get
$J'_{x}, J'_{y}$ and $J'_{z}$ and then $J'$ at the moment $t+dt$. 
Due to the
internal relaxation, the angle $\theta$ at that moment is
unknown. Therefore, at the end of each timestep, $\theta$ is assumed
to be a random angle generated from the distribution function
(\ref{fle}, fast internal relaxation) or (\ref{fmax}, without internal
relaxation). With $\theta$ known, we can obtain the value of angular 
velocity $\omega_{i}=\sqrt{\omega_{\|}^{2}+\omega_{\perp}^{2}}$  with 
its components $\omega_{\|}=(J/I_{\|}) \cos\theta$, and
$\omega_{\perp}=(J/I_{\perp})\sin\theta$. The dipole emission frequency 
of the dominant mode  $\nu_{i}$ is calculated by (\ref{nued}).
This process is repeated for $N$
timesteps. The obtained value of angular velocity and dipole emission 
frequency at the end of each time step $\omega_{i}$ and $\nu_{i}$
are binned and will be used to find the distribution function 
$f_\omega$ and $f_{\nu}$.

\subsection{Benchmark Calculations}\label{sec:bench}

To benchmark our Langevin code, we first run simulations for the
perfect internal alignment case to find $f_\omega=(1/2\pi)f_\nu$ 
to compare with the analytical distribution function obtained from the
Fokker-Planck (FP)
equation (Ali-Ha\"imoud et al.\ 2009). Physical parameters for
different phases are given in Table 1. For very small PAHs, the dust 
temperature fluctuates due to transient heating by single photon. 
Therefore, we assume a smaller dust temperature $T_{\rm d}=10$K for the 
CNM, WNM and WIM, and $T_{\rm d}=20$ and $40$K for the RN and PDR, 
respectively, for smallest grains with  $a<7\times 10^{-8}$cm. Rotational
 damping and excitation coefficients in Figure (\ref{fg_wim}) are adopted
 for the benchmarks. We note that the results of LE simulations depend on the
choice of right time step. Very small timesteps are inefficient,
while large timesteps can introduce error. We
define the timestep by $dt=\epsilon~ {\rm
  min}[1/F_{\rm tot},1/G_{\rm tot},\tau_{\rm H}/\tau_{\rm ed}]$. Our tests
show that $\epsilon=10^{-1}$ is a good choice. We run the LE
simulations for $N=10^{7}$ steps.

\begin{figure}
\includegraphics[width=0.48\textwidth]{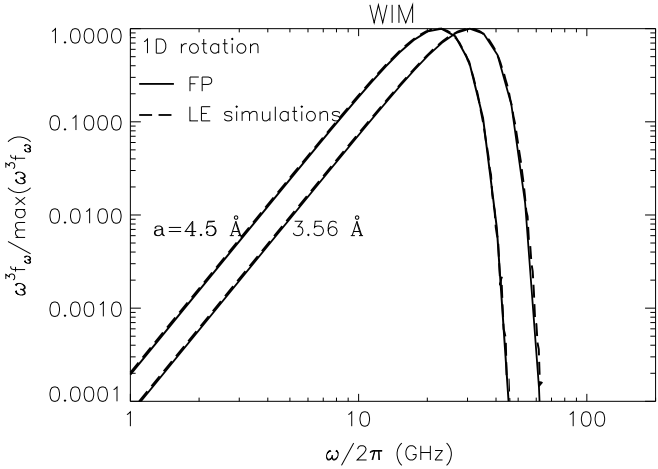}
\caption{Comparison of normalized $\omega^{3}f_{\omega}$ obtained from analytical
  solution of the FP equation and from numerical simulations of the
  Langevin equation for grain sizes $a=3.56$ and $4.5$ \AA. }
\label{f1}
 \end{figure}

Figure \ref{f1} shows normalized $\omega^{3}f_\omega$ for a grain
with electric dipole moment parameter $\beta=\beta_{0}$ and sizes
$a=3.56$ and $4.5$ \AA~ obtained from the FP equation (solid lines)
and the LE simulations (dashed lines) for the WIM.\footnote{Hereafter,
  the distribution function $f_\omega$ for the case $\beta=\beta_{0}$
  is presented.} It can be seen that the LE simulations are in
excellent agreement with analytical solutions of the FP equation.

\begin{figure}
\includegraphics[width=0.5\textwidth]{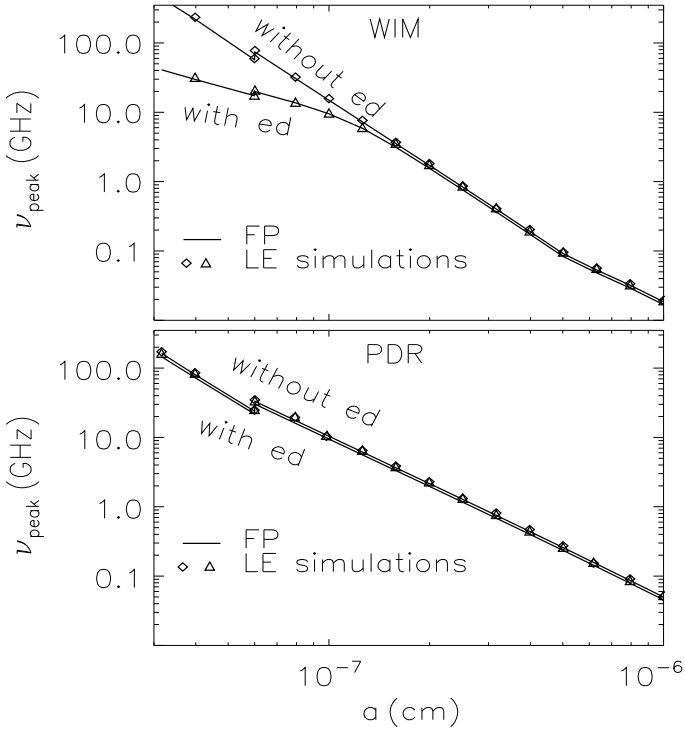}
\caption{Comparison of peak frequency with and without electric dipole
  damping (ed) obtained from simulations of LE and FP for WIM 
({\it upper panel}) and  PDR ({\it lower panel}).  }
\label{f1b}
 \end{figure}

Another important parameter we need to benchmark is the peak frequency
 as a function of grain size. 
For the perfect alignment case, the peak frequency is given by equation (160) in
Ali-Ha\"imoud et al.\ (2009). Using the distribution $f_\omega$ from
LE simulations, we can obtain the peak frequency by maximizing the
function $\omega^{4}f_\omega$. Results from these two
approaches are presented in Figure \ref{f1b} for the WIM and PDR for
both cases with and without electric dipole damping. In all cases,
results from LE simulations coincide with those from the FP. For the
PDR, the results with and without electric dipole damping are not much
different because the electric dipole damping is negligible.

\section{Impulsive Excitation by Single-Ion Collisions}
DL98b showed that for grains smaller than $7\times 10^{-8}$cm, the
angular impulse due to an individual ion-grain collision may be
comparable to the angular momentum of the grain. Thus, infrequent hits
of ions can result in transient rotational excitation for small grains.

Let $\tau_{\rm icoll}^{-1}$ be the mean rate of ion collisions 
with the grain, given by
\bea \tau_{\rm icoll}^{-1}&=&f(Z_{\rm g}=0) n_{\rm i}\pi
a^{2}\left(\frac{8\kB T_{\rm gas}}{m_{\rm i}\pi}\right)^{1/2}
\left[1+\frac{\sqrt{\pi}}{2}\Phi \right]+
\nonumber
\\ &&\sum_{Z_{\rm g}\ne
  0}f(Z_{\rm g}) n_{\rm i}\pi
a^{2}\left(\frac{8\kB T_{\rm gas}}{m_{\rm i}\pi}\right)^{1/2}g
\left(\frac{Z_{\rm g}Z_{\rm i}e^{2}}{a\kB T_{\rm gas}}\right),~~~~~\label{Ri}
\ena 
where $\Phi=\left({2Z_{\rm i}^{2}e^{2}}/{a\kB T_{\rm gas}}\right)^{1/2}$, $ g(x)=
1-x$ for $x<0$ and $g(x)=e ^{-x}$ for $x>0$, and
$f(Z_{\rm g})$ is the grain charge distribution function (see Appendix D).
The probability of the next collision occurring
in $[t,t+dt]$ is 
\bea dP=
\tau_{\rm icoll}^{-1}\exp\left(-t/\tau_{\rm icoll}\right)dt.
\ena
The rms angular momentum per ion collision $\langle \delta J^{2}\rangle$ 
is inferred by dividing the total rms angular momentum to the collision rate, 
and its final formula is given in Appendix D. 

The upper panel in Figure \ref{rate} presents the collision rate, 
$\tau_{\rm icoll}^{-1}$
compared to the rate of electric dipole damping for the grain with
$\beta=0.4$D, $\tau_{\rm ed}^{-1}$, for the CNM, WNM and WIM. The effect
of single-ion collisions is determined by the critical size $a_{\rm cri}$
corresponding to $\tau_{\rm icoll}^{-1}=\tau_{\rm ed}^{-1}$. For the WIM, it
can be seen that for grains with $a\le a_{\rm cri}=8.6\times 10^{-8}$cm, the
electric dipole damping time
is shorter than the time between two ion collisions. The active range
of single-ion collisions for the WNM is $a\le a_{\rm cri}= 1.2\times
10^{-7}$cm. The critical size is smaller for the CNM with $a_{\rm cri}=3.8\times
10^{-8}$cm. Although, the single-ion collisions are important for the CNM, 
WNM and WIM, they are not important for the RN and PDR.

\begin{figure}
\includegraphics[width=0.5\textwidth]{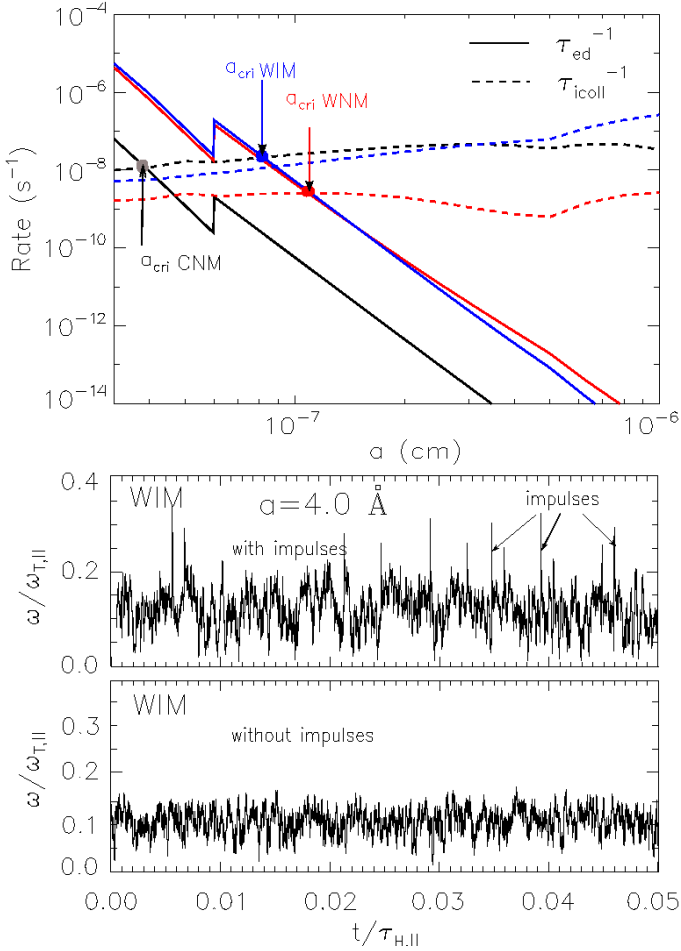}
\caption{{\it Upper panel: }Comparison of ion collision rate $\tau_{\rm icoll}^{-1}$ with the
  electric dipole damping rate $\tau_{\rm ed}^{-1}$ of grains with
  electric dipole moment parameter 
  $\beta=0.4$D for the CNM, WNM and WIM. The filled circle symbols denote
the critical size $a_{\rm cri}$ when
  $\tau_{\rm icoll}^{-1}=\tau_{\rm ed}^{-1}$. {\it Lower panel: }The  evolution of the angular 
velocity $\omega/\omega_{\rm T,\|}$ as a
  function of $t/\tau_{\rm H,\|}$ for the case with and 
without impulses from
  single-ion collisions in the WIM for $a=4$\AA. }
\label{rate}
 \end{figure}

To account for single-ion collisions, we first use the Poisson
distribution to generate the time intervals between successive collisions
$\Delta \tau_{{\rm icoll},j}$, which is a random variable in this case for a
given $\tau_{\rm icoll}$. Thus, the $nth$ collision occurs at the moment 
$t_{{\rm icoll},n}=\sum_{j=1}^n\Delta \tau_{{\rm icoll},j}$. 
We then run the integration for the LE (i.e., eq.
\ref{le3d2}) over time to find $J$ and $\omega$. The $nth$
collision deposits an
angular momentum $\Delta {\bf J}_n$
that is assumed to be randomly-oriented, with magnitude drawn
from the distribution function with the rms $\langle \delta J^{2}\rangle$
 appropriate for the grain charge and radius.
The impinging ions are taken to be protons for the WNM and WIM, and a mix of
protons and C$^+$ ions, with fractional abundances $x_{\rm H}$ and $x_{\rm M}$ 
from Table 1, for the CNM. For the WIM and WNM in which the focusing effect 
due to  negatively-charged grain is negligible, we can assume that the grain 
charge is given by the mean charge $\langle Z\rangle$. But for the CNM in 
which the focusing effect is important, we consider the grain in various 
charge states $Z$, and find the corresponding 
$f_\omega$ and $f_{\nu}$.  

To identify the role of single-ion collision on the distribution
function, we first consider the perfect alignment case and solving 1D
LE (Equation (\ref{le})). The resulting $\omega$ for the WIM is shown in
the lower panel of Figure \ref{rate} for a grain size $a=4$\AA. It can be
seen that due to the electric dipole damping, the grain rotates
subthermally most of the time with $\omega < \delta \omega$. Ionic
impulses result in the transient rotational spin-up followed by the
continuous damping by electric dipole emission.

\begin{figure}
\includegraphics[width=0.48\textwidth]{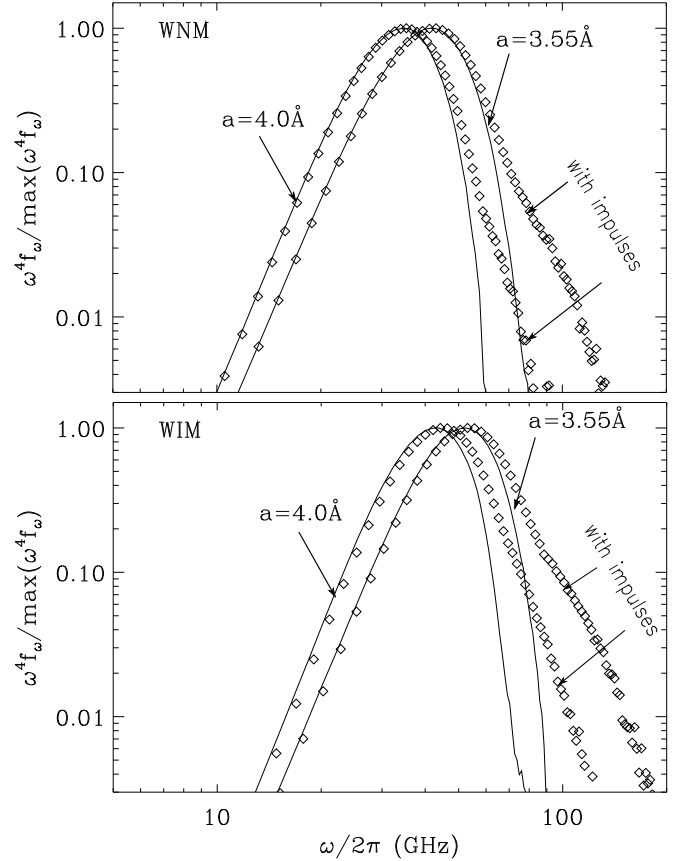}
\caption{Normalized $\omega^4f_\omega$ in the
  presence of impulses from single-ion collisions for the WNM ({\it
    upper panel}) and the WIM ({\it lower panel}) for two typical
  grain sizes. Impulses broaden the emission spectra to higher frequency
 (see diamond symbols).}
\label{f2}
 \end{figure}

Distribution functions obtained from LE simulations in the presence 
of impulses and the grain wobbling
are compared with those when only the grain wobbling is considered
(i.e., without impulses) in Figure \ref{f2} for the WNM (upper
panel) and the WIM (lower panel) and for two grain sizes $3.55$ and
$4.0$\AA. It can be seen that the ionic impulses extend the
distribution function to higher frequency, but change only slightly the peak 
frequency of the spectrum. This effect is more important for smaller grains.

\section{Emissivity}
Here we calculate the emissivity at the frequency $\nu$ using equation (\ref{jnued})
 with $f_{\nu}$ obtained from the LE simulations and the emission power 
$P_{\rm ed}(\nu)$ from equation (\ref{ped_nu}). To compare with earlier studies, 
we also calculate the emissivity  using equation (\ref{emiss}) with 
the Maxwellian distribution function $f_\omega$ from DL98b and the
analytical distribution function from the FP equation (FP). 

The emissivity is calculated assuming that $25\%$ of the grains have 
 the electric dipole moment
parameter $\beta=2\beta_{0}$, $50\%$ have $\beta=\beta_{0}$ and $25\%$ have 
$\beta=0.5\beta_{0}$ with $\beta_{0}=0.4$D. The electric dipole moment $\bmu$
is assumed to be isotropically distributed along the grain principal axes.
 We adopt the models of grain size distribution from Draine \&
Li (2007) with
the total to selective extinction $R_{\rm V}=3.1$ and the total carbon
abundance per hydrogen nucleus $b_{\rm C}=5.5\times 10^{-5}$ for diffuse
environments CNM, WNM and WIM, and $R_{\rm V}=5.5,~ b_{\rm C}= 2.8\times 10^{-5}$
for the RN and PDR for carbonaceous grains with
$a_{\rm min}=3.55$ \AA~and $a_{\rm max}=100$ \AA.

\subsection{Grain Wobbling: Imperfect Internal Alignment}

Here we present our new calculations of emissivity accounting for
imperfect internal alignment for the idealized environments listed in
Table 1. We run simulations of LE (eq. \ref{le3d2}) to find the 
distribution functions $f_{\nu}$ for
128 grain sizes in the range from $a_{\rm min}$ to $a_{\rm max}$. For plasma
drag and infrared emission, we use our new damping and excitation
coefficients from Figures (\ref{Gpar}) and (\ref{FG_IRWIM}). For
collisional damping and excitation, we use the results in DL98b as
shown in Figure (\ref{fg_wim}).

\subsubsection{Distribution function $f_\omega$}
For a more instructive comparison of our results with earlier studies, 
here we present the distribution function of rotational frequency 
$f_\omega$ instead of the distribution function of dipole emission 
frequency $f_\nu$, which is slightly different from the former one 
for the grains of the planar geometry. 

\begin{figure}
\includegraphics[width=0.48\textwidth]{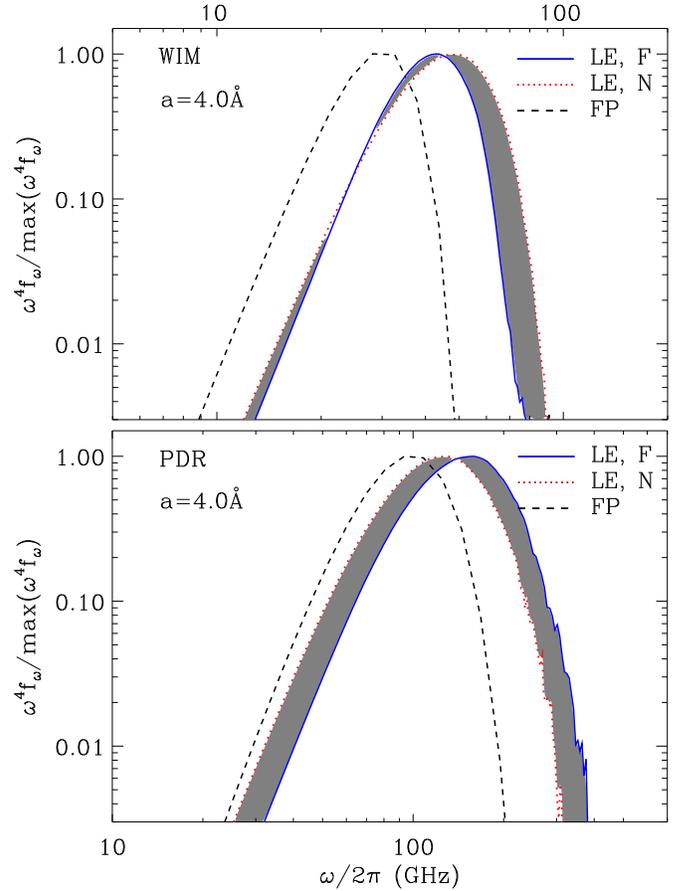}
\caption{Comparison of normalized $\omega^{4}f_\omega$ for the case
  of perfect internal alignment using the FP method (dashed line, FP) in 
Ali-Ha\"imoud et al. (2009) with that in the presence 
for grain wobbling obtained from simulations of LE, both with fast internal 
relaxation (LE, F) and without internal relaxation (LE, N). Shading areas denote 
the transition from fast internal relaxation 
(solid line) to without internal
  relaxation (dot line). For both the WIM and PDR, the peak
  frequency of distribution function increases, and the spectrum
  shifts to higher frequency in the later cases.  }
\label{fome_imper}
\end{figure}

Figure \ref{fome_imper} shows normalized $\omega^4f_\omega$
for the case of grain wobbling with and
without fast internal relaxation, compared to that obtained using the FP
method in Ali-Ha\"imoud et al.\ (2009) for a grain size $a=4$ \AA~ for
the WIM and PDR. 
To find the distribution function for the FP case, we use the parallel
damping and excitation coefficients for the distribution function given
by equation (34) in Ali-Ha\"imoud et al. (2009) because the grain is assumed 
to rotate along the symmetry axis.

It
can be seen that in both environments, the grain wobbling results in a
higher peak frequency, and the entire distribution functions are
shifted to higher frequencies. Furthermore, the distribution function
for the case without internal relaxation is extended to higher
frequency than that with fast internal relaxation because the higher
rotation frequency in the later case corresponds to higher internal
dissipation rate of energy (see dotted line in the lower panel).

\subsubsection{Emissivity}
We calculate the emissivity per H using the distribution function
$f_\nu$ found in the previous subsection for different
environments in the presence of fast internal relaxation and without
internal relaxation.

\begin{figure}
\includegraphics[width=0.48\textwidth]{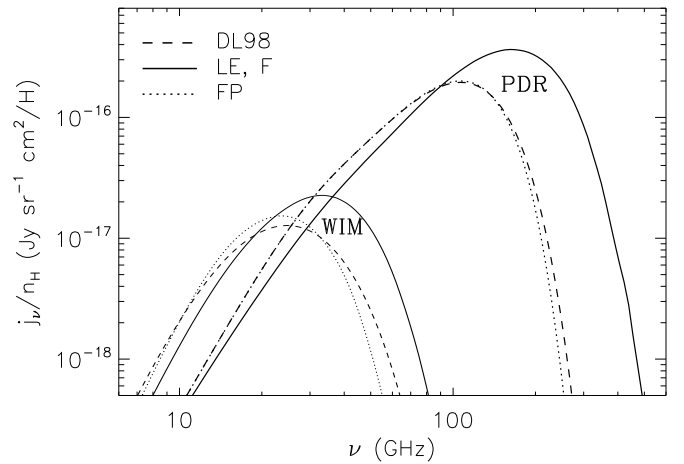}
\caption{Comparison of emissivity spectrum for the grain wobbling
  (solid line) with those from the 1D rotation case (dot and dashed
  lines) and the fast internal relaxation (LE, F) is taken into account. The
  peak frequency of WIM increases from $\sim 23$ to $\sim 33$GHz and its 
increases from $\sim 106$ to $\sim 161$ GHz
  for PDR, relative to that from the FP equation method in 
Ali-Ha\"imoud et al. (2009). Emissivity increased by a factor of $\sim 1.5$
 and $1.8$   for WIM and PDR, respectively.}
\label{jnu_imper}
\end{figure}

In Figure \ref{jnu_imper}, we compare our results in the presence of
fast internal relaxation with those obtained using methods in DL98b
and Ali-Ha\"imoud et al.\ (2009) for perfect internal alignment for the
WIM and RN. The parallel damping and excitation coefficients are taken
in use to find the distribution function for the later cases.

First, it can be seen that the grain wobbling results in increases of
the peak frequency. Compared to results from the FP equation method in 
Ali-Ha\"moud et al. (2009), we see that the peak frequency increases 
from $\sim 23$ GHz to $\sim 33$GHz for the WIM and from $\sim 106$ to
 $\sim 161$ GHz for PDR. Also, the peak emissivity of spinning dust 
in the later case is increased by factors of $\sim 1.5 $ for the WIM 
and $\sim 1.8$ for PDR, respectively.

\begin{figure}
\includegraphics[width=0.48\textwidth]{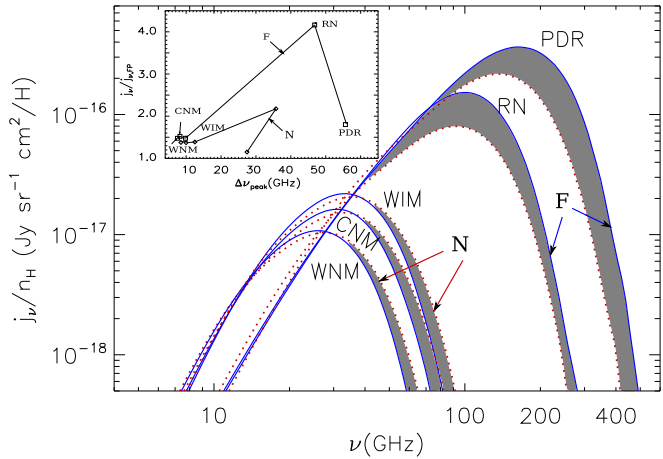}
\caption{Emissivity per H in the case of grain wobbling for various
  environments. Gray shaded areas represent the transition between fast internal
 relaxation (F, solid lines) to no internal relaxation (N, dot lines). 
Subplot shows $j_{\nu}/j_{\nu,\rm FP}$ with $j_{\nu,\rm FP}$ being the peak emissivity
  from the method of FP equation, as a function of the increase of peak frequency
  $\Delta \nu_{\rm peak}$ for fast internal relaxation (F) and no internal
 relaxation (N).}
\label{jnu_freq}
\end{figure}

The emissivity per H for various idealized environments are summarized in
Figure \ref{jnu_freq} with and without fast internal relaxation,
respectively. The solid lines denote the emission spectra for the
case of fast internal relaxation. The shaded areas represent the
expected spectra when the internal relaxation varies from fast to very
slow.  The subplot represents $j_{\nu}/j_{\nu,\rm FP}$ with $j_{\nu}$
and $j_{\nu,\rm FP}$ being the emissivity at the peak of the
emission spectrum for the imperfect alignment case and that from the FP equation 
method, as a function of the increase in the peak frequency $\Delta
\nu=\nu_{\rm peak}-\nu_{\rm peak,\rm FP}$ of emission spectrum.

Figure \ref{jnu_freq} shows that the increase of emissivity and peak frequency
 for the grain wobbling case is important. The peak frequency is increased 
from $8$ to $55$GHz (i.e., by factors of 1.2 to 1.8) different environments
 (see subplot in Figure \ref{jnu_freq}).
The peak emissivity is increased by
factors 1.47 (1.42), 1.5 (1.4), 1.52 (1.37), 1.8 (1.2)  for WNM, WIM, CNM 
and PDR, respectively, corresponding to the cases with fast internal 
relaxation (without internal relaxation), except for the RN where the peak
emissivity enhancement can reach 
%\btdnote{need new value}
a factor of $4.1 (2.2)$ (see the subplot). In addition, the enhancement 
of peak frequency for the case without internal relaxation is smaller 
than that for fast internal relaxation. 
That can be explained by the fact that some emissivity from its peak can
 be transferred to higher frequency, resulting more extended spectra 
 present  in the case without internal relaxation (dot line in Figure 
\ref{jnu_freq}).

The highest increase of peak emissivity present in the RN seems
difficult to explain. We note that for this medium, the infrared
damping and excitation is the most important process. Therefore, in
addition to the anisotropy due to the non-spherical geometry, the
anisotropy in the damping and excitation due to IR emission results in
such a substantial increase of peak emissivity and peak frequency (see
equation \ref{dome2}). The correlation of the increase of peak
frequency and emissivity with the anisotropy for the grain wobbling
case will be investigated in \S 9.3.

\subsection{Emissivity in the Presence of Single-Ion Collisions}

To see the effect of single-ion collision on the emissivity of
spinning dust, consider first the case of perfect alignment. We are
interested only in the CNM, WNM and WIM where the single-ion collisions are
important. 

\begin{figure}
\includegraphics[width=0.48\textwidth]{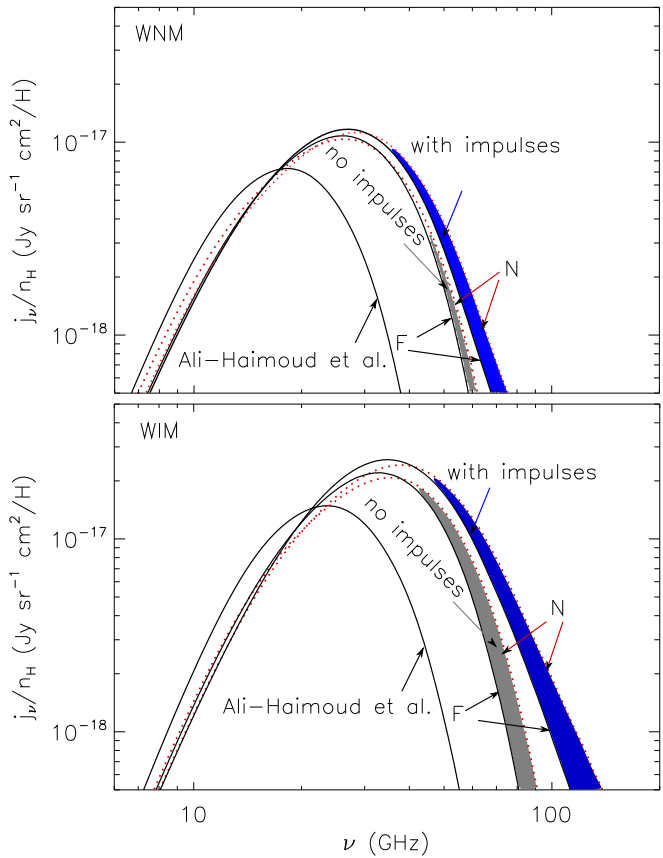}
\caption{Emissivity per H obtained for the WNM (upper panel) and WIM
  (lower panel) without ionic impulses using the FP, and with impulses
  using our LE simulations for grain wobbling. The spectra are
  efficiently broadened as a result of impulses. Fast internal
  relaxation is more efficient at highest frequency, resulting in
  narrower spectra compared to those for the case without internal
  relaxation (red color line).}
\label{jnu_imp}
\end{figure}

We run the LE simulations for the grain wobbling taking
into account the impulses for grain size from $a_{\rm min}$ to $a_{\rm cri}$
for these environments. We consider both the fast internal relaxation
and without internal relaxation. The resultant emissivity is compared
with that from the case without impulses in Figure \ref{jnu_imp} for the WNM
and WIM.

It can be seen that the ionic impulses broaden efficiently the
emission spectra to the higher frequency,  but they change slightly the 
peak frequency. For the WIM, the net effect of impulses and grain
wobbling enhances the peak emissivity by a factor of $1.73$, and the peak
frequency is increased from $\sim 23$ to $\sim 35$ GHz. The
corresponding increase in peak emissivity for the WNM is a factor of $1.58$,
and the peak frequency is increased from $\sim 19$ to $\sim 27$ GHz. 
By subtracting the effect of grain wobbling, we see that the impulses 
can increase the peak emissivity by $\sim 23\%$ and $11\%$ for the WIM 
and WNM, respectively, but they increase slightly the peak frequency
 (see Figures \ref{jnu_freq} and \ref{jnu_imp}). 

\begin{figure}
\includegraphics[width=0.48\textwidth]{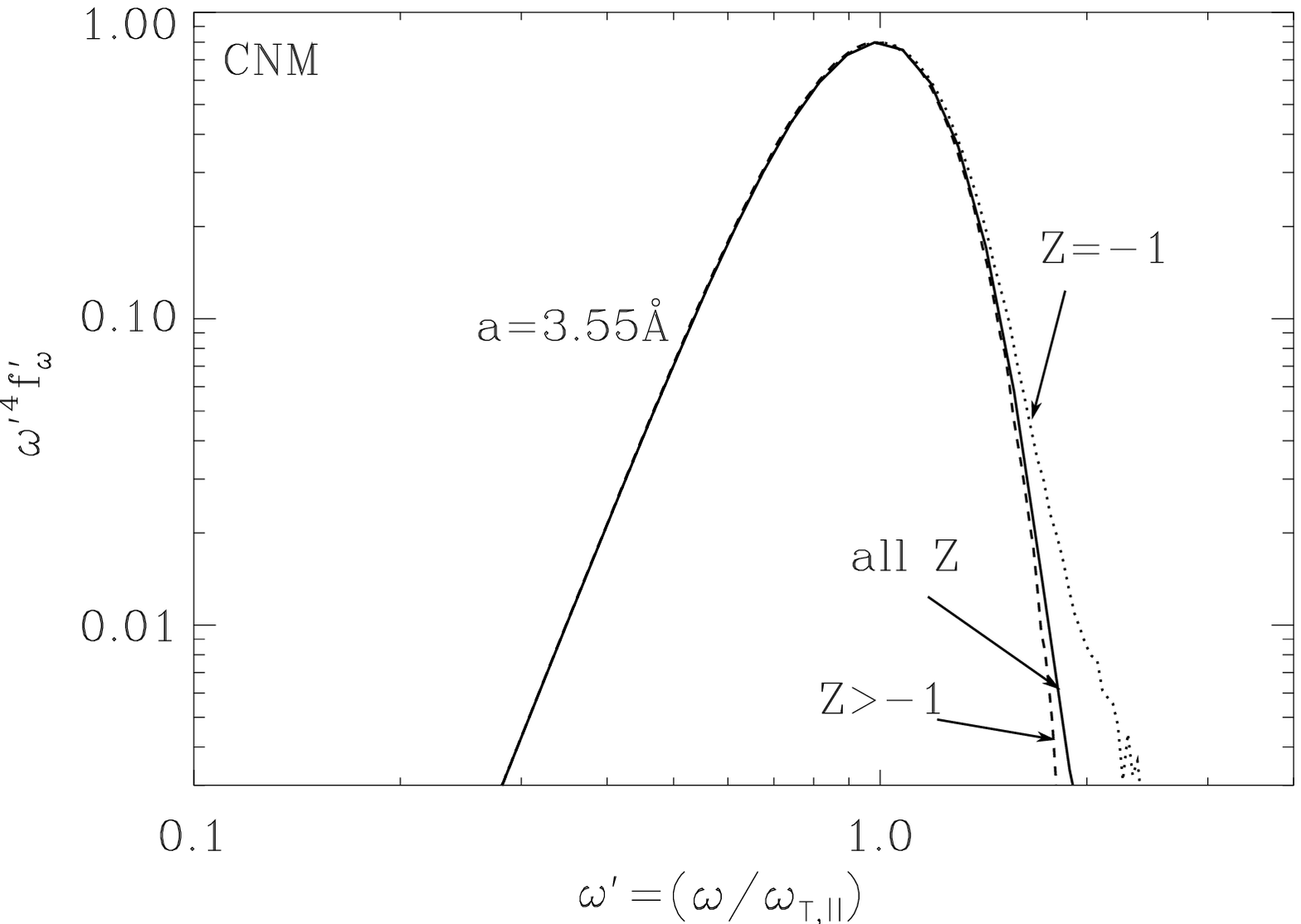}
\includegraphics[width=0.48\textwidth]{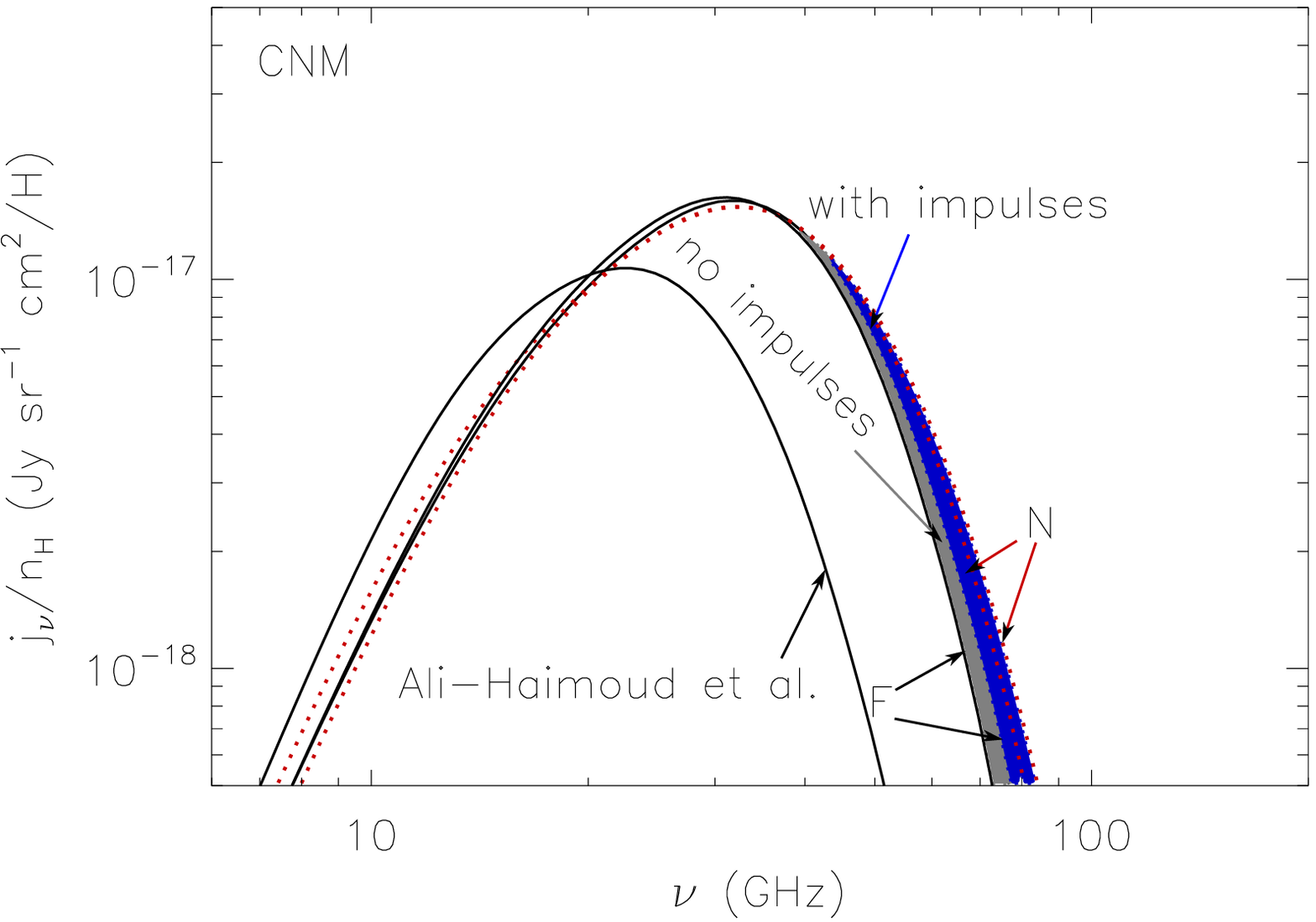}
\caption{Effect of impulses due to single-ion collisions for the CNM: normalized
 $\omega^{'4}f'_{\omega}$ with $f'_{\omega}$ is
 the normalized distribution function, as a function of 
$\omega'=\omega/\omega_{\rm T,\|}$ where $Z$ is the grain charge $Z$ (upper panel) and 
emissivity (lower panel). Gray and blue 
shaded areas correspond to the transition between fast internal relaxation (solid line) 
and without internal relaxation (dot line) for the case without and with impulses, 
respectively. The 
focusing effect for the grain with charge $Z=-1$ enhances the high-frequency tail
 (dot line in upper panel)}
\label{fj_CNM}
\end{figure}

For the CNM, the lower panel of Figure \ref{fj_CNM} shows that the effect of
 impulses on the total emissivity is 
weaker than for the WIM and WNM. The reasons for that
are as follows. First, although the focusing effect is important for 
the grain in the negatively charged state, resulting in rotational impulses at high 
frequency (see the dot line in the upper panel of Figure \ref{fj_CNM}), 
the probability of the grain in this negative charge state is relatively small. 
Second, since the effective range
 of impulses of the CNM is much narrower than that for the WIM and 
WNM (see the upper panel of Figure \ref{rate}), its emissivity, that 
is obtained by integrating over the entire grain size distribution, is obviously
less affected by impulses compared to the WIM and WNM.

We note that for a small grain in the CNM, WNM and WIM, during a short
interval of several $t_{\rm icoll}$, the grain experiences some rotational
spikes due to single-ion collisions as shown in the left lower subplot
of Figure \ref{rate}. The peak frequency and width of the spectrum
are related to the mean square value, $\langle \omega^{2}\rangle$,
which is averaged over total integration time $T$. Therefore, the overall effect of the
transient spin-up is to broaden the emission spectrum and increase the
total emissivity as seen in Figure \ref{jnu_imp}.

\subsection{Effect of the non-sphericity and anisotropy in 
            rotational damping and excitation}

To understand why the grain wobbling induces the substantial increase
of peak frequency and emissivity found in the previous sections, let
us consider a disk-like grain shape and when internal relaxation is
important.

To investigate the effect of grain wobbling for
both spherical grains ($a> a_{2}$) and disk-like grain ($a<a_{2}$),
we run LE simulations to find $f_\omega$ and $\nu_{\rm peak}$
for the case of imperfect internal alignment assuming fast
internal relaxation, for grain size from $a_{\rm min}$ to
$a_{\rm max}$ 

\begin{figure}
\includegraphics[width=0.48\textwidth]{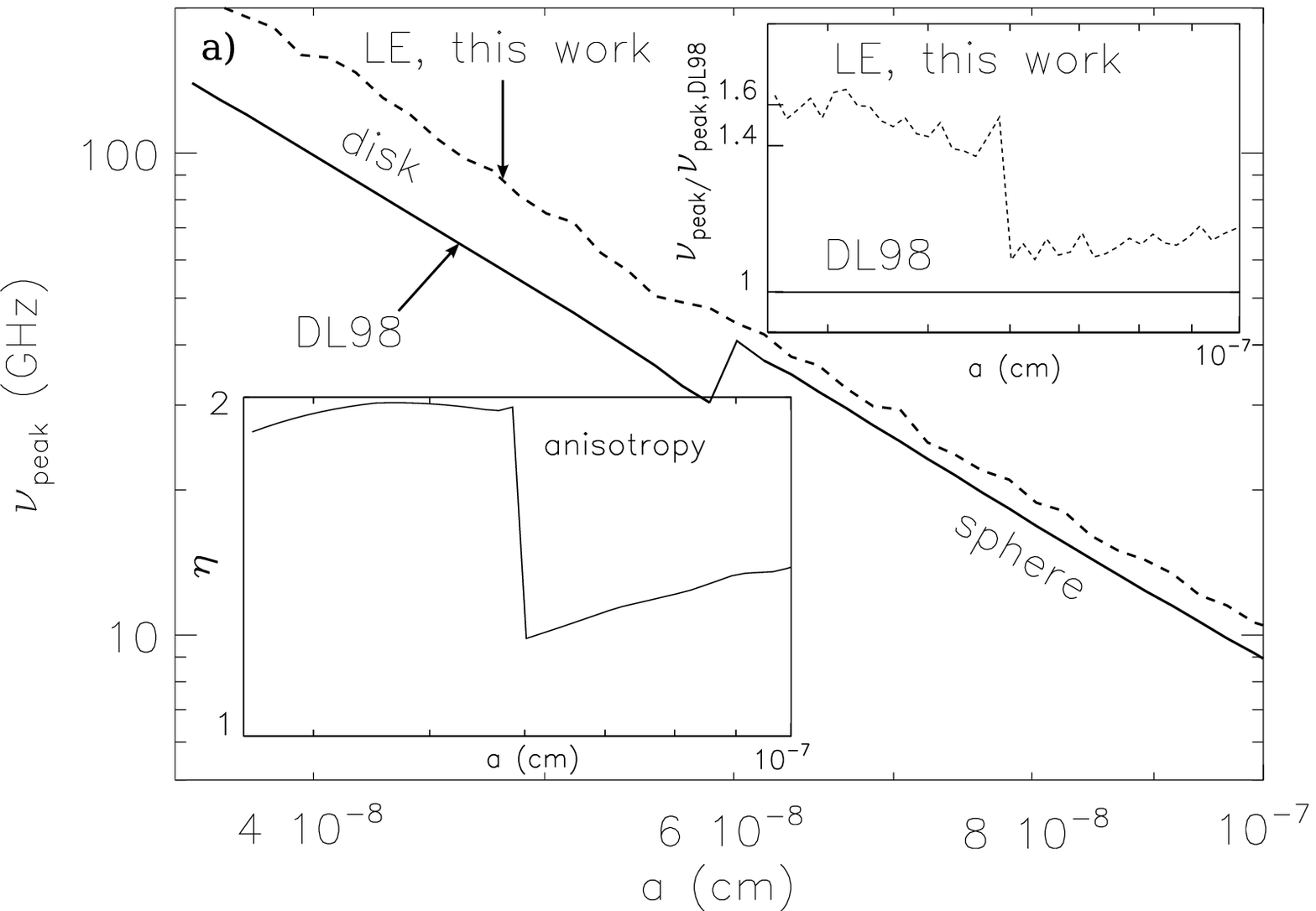}
\includegraphics[width=0.48\textwidth]{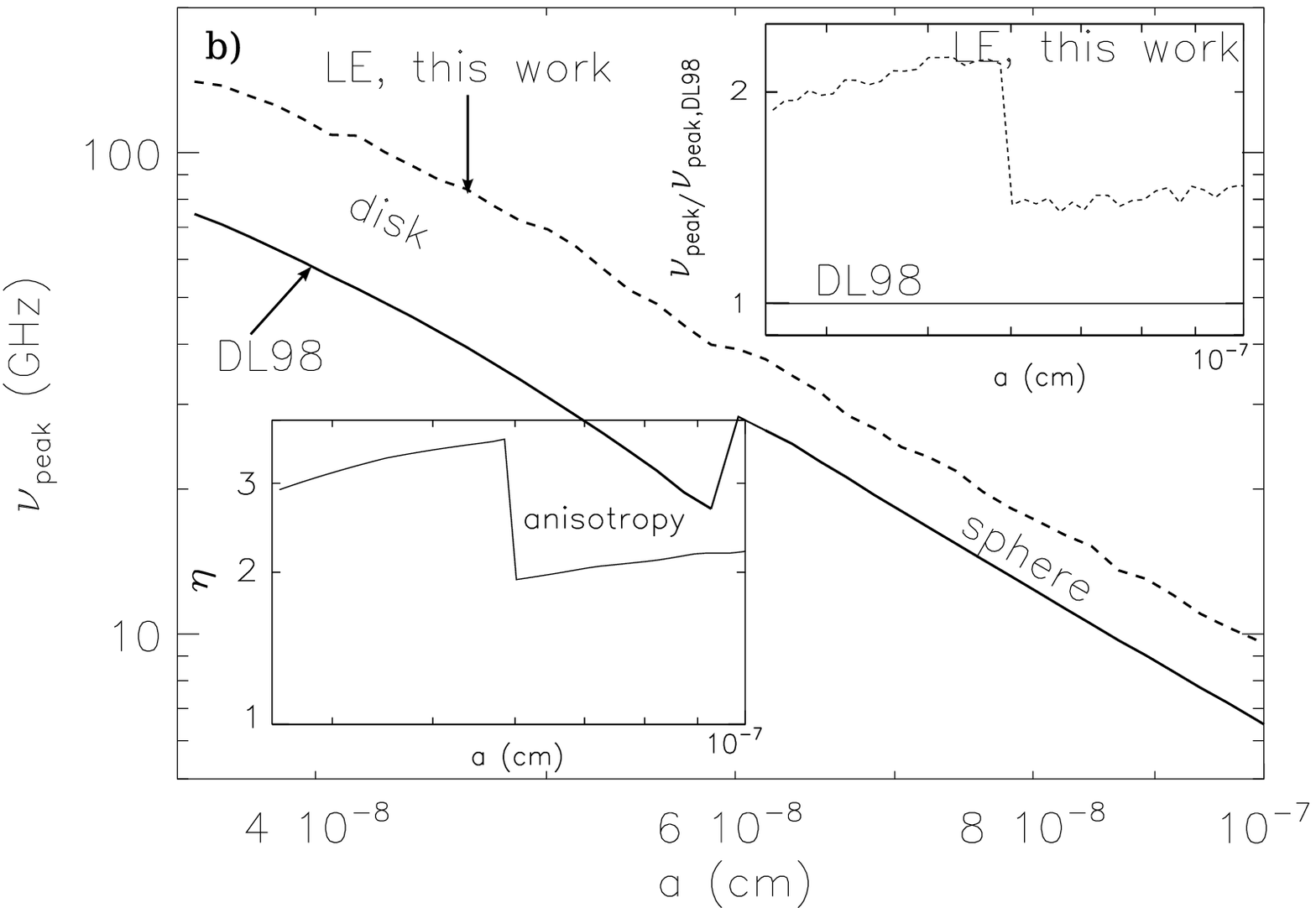}
\caption{Peak frequency as a functions of grain size for the PDR ({\it
    a}) and RN ({\it b}). Dashed and solid lines denote the result
  obtained from LE simulations $\nu_{\rm peak}$, and from the DL98 model
  $\nu_{\rm peak,\rm DL98}$, respectively. The ratio
  $\nu_{\rm peak}/\nu_{\rm peak,\rm DL98}$ versus $a$ is shown in the upper corner
  subplot, and the anisotropy $\eta=\alpha_{\|}/\alpha_{\perp}$ is
  shown in the left lower corner.}
\label{dnupeak_a}
\end{figure}

We consider here the RN and PDR for which electric dipole damping is
subdominant. Results are shown in Figure \ref{dnupeak_a} where solid
and dashed lines represent $\nu_{\rm peak}$ obtained using the DL98 model
and LE simulations, respectively. The ratio of $\nu_{\rm peak}$ to that
of DL98 model is depicted in the upper right subplot.\footnote{When the 
electric dipole damping is negligible, the distribution $f_\omega$ is
 Maxwellian, therefore $\nu_{\rm peak,\rm FP}\equiv \nu_{\rm peak, \rm DL98}$} The 
excitation anisotropy ratio
$\alpha_{\|}/\alpha_{\perp}$ (see eq.\ \ref{eq:eta})
is shown in the lower left subplot.

First of all, for the PDR, the peak frequency from LE simulations
is slightly higher than that from the DL98 model for spherical grains
($a>a_{2}=6\times 10^{-8}$cm).  For very small disk-like grains
($a<6\times 10^{-8}$cm), the peak frequency is increased by a mean
factor of $\sim 1.5$.  For the RN, one interesting feature is
that the peak frequency from LE is much larger that that from the
DL98 model for spherical grains(cf. to the PDR). We attribute the
increase to the anisotropy in the grain rotational damping and
excitation due to the infrared emission, which is the most important
process of grain excitation for the RN. 
For spherical grains in the range from $a_{2}$ to $10^{-7}$ cm, 
$\eta$ increase from $1.2$ to $1.4$ for the PDR, and from $1.8$ to $2.3$ for
the RN (see subplots in the lower left corner of Figure
\ref{dnupeak_a}). Hence, the higher anisotropy in the RN corresponds to 
the higher increase of peak frequency compared to the PDR case.

\begin{figure}
\includegraphics[width=0.48\textwidth]{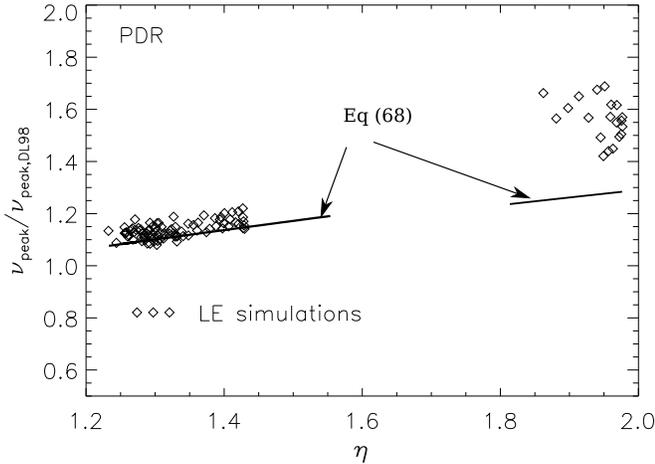}
\caption{The ratio of peak frequency $\nu_{\rm peak}/\nu_{\rm peak,\rm DL98}$ as a
  function of anisotropy $\eta$ obtained from LE compared to the
  ideal case (eq. \ref{nupeakMax}) for the PDR. The gap between
  $\eta=1.4$ and $1.8$ arises from the change in the grain geometry
  from sphere to disklike.}
\label{dnupeak_eta}
\end{figure}

To see clearly the dependence of the increase in $\nu_{\rm peak}$ in the
grain wobbling case, we plot $\nu_{\rm peak}/\nu_{\rm peak,\rm DL98}$ as a
function of $\eta$ for the PDR in Figure \ref{dnupeak_eta}. The solid
line corresponds to 
\bea
\frac{\nu_{\rm peak}}{\nu_{\rm peak,DL98}} 
\approx 
\sqrt{\frac{\langle\omega^2\rangle}{\langle\omega^2\rangle_{\rm DL98}}}
= \sqrt{\frac{2\eta+1}{3}}
\label{nupeakMax} 
\ena
where  diamonds
show $\nu_{\rm peak}$ from the LE simulations.  We see that the
ratio $\nu_{\rm peak}/\nu_{\rm peak,\rm DL98}$ increases with 
$\eta$ as expected. For small $\eta$, $\nu_{\rm peak}$ from the LE simulations
is similar to that from equation (\ref{nupeakMax}), and some
difference appears when $\eta$ increases (see Figure \ref{dnupeak_a}).
This may stem from two reasons. First, the angular velocity
distribution function $f_\omega$ for $\eta \sim 1$ is Maxwellian, but
it differs from the Maxwellian distribution when $\eta$ increases. 
Second, the assumption that $\omega_{x,y,z}$ are independent in deriving 
equation (\ref{nupeakMax}) is not valid because they have complex
 motion in the inertial coordinate system.

In other words, the anisotropy due to non-sphericity of grain shapes
{\it and} the anisotropy in the grain rotational damping and excitation both
act to increase the peak frequency of the emission.


Finally, we note that the total emissivity due to spinning dust scales
as $\omega^{4} f_\omega$, thus, the increase of $\nu_{\rm peak}$
naturally is accompanied by the increase of the peak emissivity as
observed in Figure \ref{jnu_freq}.

\section{Discussion}
\subsection{Comparison with earlier studies}

The DL98 model was the first proposal 
%is the first model attempting 
to explain the $20-40$ GHz anomalous foreground emission in terms
of rotational emission from ultra-small dust particles.  Numerous observations
(Dobler et al.\ 2009 and references therein) have confirmed the DL98
model predictions.

In the DL98 model the calculations were mostly analytical. 
The present study used a different approach, namely, the Langevin 
equation. We successfully benchmarked our calculations against the 
Fokker-Plank calculation in Ali-Ha\"imoud et al.\ (2009) for the 
perfect alignment case, and then introduced additional effects not 
included in DL98. In particular, our approach allows us to treat
discrete high impact impulses arising from single-ion collisions with
the grain. Most importantly, it enables us to study the spinning dust
emission for the case of imperfect internal alignment with and
without fast internal relaxation.

Among grain rotational damping and excitation processes, plasma drag
is shown to be important for Molecular Clouds and CNM. 
Ragot (2002) considered the interactions of the plasma with spinning dust
  from the point of view of generation of waves. We believe that for
  the case we consider, the dust-plasma interaction is accurately
  approximated by interaction of the grain with individual,
  uncorrelated, passing ions. 

DL98b noted that the typical rotational quantum number of even a small
PAH would be $J/\hbar\gtsim 10^2$, allowing the dynamics to be treated
classically.
Ysard \& Verstraete (2009) formulated the problem quantum-mechanically;
the close correspondence between their results and those of DL98b
confirms that a classical treatment is entirely adequate.\footnote{%
   In discussing some
   effects, e.g., dust-plasma interactions, DL98b did include
   quantum limitations for the angular momentum transfer, 
   i.e. that $\Delta J$ cannot be less than $\hbar$.}

While our studies confirm the general validity of the DL98 model of spinning
dust emission, they also show ways of making the model predictions
more accurate. For instance, our refined treatment of grain dynamics
provides peak emissivity predictions which differ by factors from $\sim 2$ to
$4$ from the original ones, depending on the environments. These
differences may be detectable with high precision future
observations.

The largest uncertainty within the model is the value of the dipole
electric moment. The distribution of dipole moments
adopted in DL98b was only an educated guess.
Moreover, this distribution could be affected by
interstellar processes and change from one media to another. For
instance, Dobler et al.\ (2009) showed the anomalous emission in
the 5-year WMAP data exhibits a bump at $\sim 40$ GHz. This peak
frequency is larger than the prediction by the original DL98b model
but could be explained if a modified distribution of electric dipole
moments is adopted (Dobler et al.\ 2009), but the new effects considered
here -- misaligned rotation and impulsive excitation by ions 
may also explain the observed emission spectrum.

\subsection{Effect of single-ion collisions}

DL98b pointed out that for the CNM, WNM and WIM the angular momentum
transferred in a single ion collision can be larger than the angular momentum
of the rotating grain, but neither
DL98b or the subsequent paper by Ali-Ha\"imoud
et al.\ (2009) treated the impulsive nature of the ion collisions.

In the present paper, we describe ion-grain collisions as
Poisson-distributed
discrete events.
When an ion hits the grain surface, the grain undergoes
transient spin-up and then gets damped, mainly by electric dipole damping. Due
to strong electric dipole damping, small grains spend most of their
time rotating subthermally between two rotational spikes due to
single-ion collisions. Ionic impulses are shown to be important
for grains smaller that $\sim 8.6\times 10^{-8}$, $1.2 \times
10^{-7}$ and $3.8\times 10^{-8}$cm for the WIM, WNM and CNM, respectively. 
Our quantitative simulations show that the impulses extend the high-frequency
tail of the emission spectrum but change the peak frequency only slightly.
They can enhance the peak
emissivity by $23\%$ and $11\%$ for the WIM and WNM, respectively.

\subsection{Effect of Grain Wobbling}

The DL98b model assumed perfect alignment of the grain axis of
maximal moment of inertia with the instantaneous angular momentum. 
This assumption is
valid only for strong internal relaxation {\it and} suprathermal (much
faster than thermal) rotation of grains (Lazarian 1994). The tiny
grains we deal with rotate thermally or subthermally. For such a
rotation, the grain axes are {\it not} coupled with the angular
momentum. Due to imperfect internal alignment (i.e., grain wobbling),
the anisotropy arising from grain shape, and differential damping and
excitation (e.g. due to infrared emission and plasma drag) obviously
affect the grain mean rotation rate and therefore the emissivity. 
Increased grain anisotropy, as measured by the anisotropy
ratio $\eta$ (eq.\ \ref{eq:eta}) 
produces an increase in the peak
frequency and emissivity.

As we do not know the rates for internal relaxation in
microscopic molecular size grains, we considered two cases:
very rapid internal relaxation, and very slow internal relaxation.
Using the Langevin equation, we showed that for the former case, the 
peak emissivity of spinning dust increases by a factor of $\sim 2$,
and the peak frequency of the spectrum increases by factors from 1.4 to 1.8,
depending on the environment. Grain wobbling has a larger effect in broadening
 the emission spectrum  when the internal relaxation is slow. Especially, 
for the RN when the anisotropy of damping and excitation from infrared
 emission is very high, the peak emissivity can be increased by a factor of $4$
 and peak frequency is increased by a factor of $2$.

For the WIM, the joint effect of grain wobbling and impulses due to 
single-ion collisions causes the peak frequency to shift from 
$\sim23$ to $\sim 35$ GHz. 
It is interesting to note that the new predicted peak frequency is
close to the bump with the anomalous emission found in the WMAP five year data
(Dobler et al.\ 2009). 
In their paper, using the DL98b model, they explained the
observed spectrum with a modified model of spinning dust in which the
grains have broader distribution of electric dipole moments. They
also varied the number density of the WIM to produce spectra with
higher peak frequency. These modifications do not seem to be
necessary with our improved treatment of grain dynamics.

\section{Summary}

The present paper provides a refinement
of the DL98 model of spinning dust emission. Our principal results 
are outlined below:

1. We derived the rotational damping and excitation coefficients
arising from plasma drag, infrared emission, gas drag and the electric
dipole damping for small disk-like grains with imperfect internal
alignment.

2. A nonspherical grain precessing around its angular momentum $\bJ$
radiates primarily at two frequency modes.
The lower frequency is dominant and is not 
the same as the rotational frequency $\omega/2\pi$. 

3. Using the Langevin equation,
we calculated the emissivity of grains with
disaligned rotation, both with and without fast internal
relaxation. For fast internal relaxation, allowing for disaligned
rotation, the peak emissivity of spinning dust is increased by
a factor $\sim$ 2 for the CNM, WNM, WIM and PDR; and by a factor of 
$\sim$ 4 for the RN, compared to those from the FP equation method. 
Fast internal relaxation 
also shifts the  emission spectra to higher frequencies factors of 
1.4 to 2 for these environments. Slow internal relaxation broadens 
the emission spectrum and reduces the enhancement in the peak emissivity.

4. The increase of peak emissivity and frequency
of spinning dust in the grain wobbling case compared to those 
from the DL98 model has two sources: the anisotropy of the moment of 
inertia tensor, and the anisotropy in the rotational damping and excitation
processes (e.g. infrared emission and plasma drag). Higher anisotropy
results in higher enhancement of peak frequency and emissivity.

5. We devised a method to include the high-impact ion
collisions within our statistical treatment of grain rotation. We
found that for the CNM, WNM and WIM, single-ion impulses result in
strong broadening of the emission spectrum to higher frequency and
increases the emissivity of spinning dust. Impulsive excitation is most
efficient for the WIM, but it is not important for the RN and
PDR. In all idealized environments, the transient spin-up is
subdominant compared to grain wobbling in terms of enhancement of
emissivity, but it is dominant in spectrum broadening for the CNM, WNM and
WIM.

6. For the WIM, the net effect of grain wobbling and transient spin-up
from single-ion collision increases the peak emissivity by a
factor $\sim 1.7$, and the peak frequency increases from $\sim 23$ 
to $\sim 35$ GHz. That increase in peak frequency brings the model into 
agreement with the 5-year WMAP data without modifying the electric dipole
 momentum distribution or number density of the WIM as was proposed by
 Dobler et al.\ (2009).

\acknowledgements TH and AL acknowledge the support of the Center for
Magnetic Self-Organization and the NSF grant AST 0507164. BTD acknowledges 
partial support from NSF grant AST 0406883. BTD is
grateful to Chris Hirata for helpful discussions, and to 
R.H. Lupton for availability of the SM graphics program.

\appendix

\section{A. Grain Properties}
\subsection{A1. Grain shape and size}

Here we present grain model and assumptions, notations we use
throughout the paper.  Grains are assumed to be disklike with 
radius $R$ and height $L$ for size
$a<a_{2}$ and spherical for $a\ge a_{2}$. We chose $a_{2}=6\times
10^{-8}$ cm as in DL98b.  The ``surface-equivalent'' radius $a_{\rm s}$ and
the ``excitation equivalent'' radius $a_{\rm x}$ of the grain were defined
in DL98b.

Ali-Ha\"imound et al.\ (2009) defined the radius of cylinder with the
same $\int r^2\mss\theta dS$ with $r\sin\theta$ is the distance from the
grain surface to the symmetry axis, thus
\bea 4\pi a_{\rm cx}^4 \equiv
\frac{3}{2} \oint r^2\mss\theta dS.
\label{eq:acxdef}
\ena 
The expressions for $a_{\rm s}, a_{\rm x}$ and $a_{\rm cx}$ are given in
DL98b and Ali-Ha\"imoud et al.\ (2009).  We adopt the models of grain
size distribution from Draine \& Li (2007) with the total to selective
extinction $R_{\rm V}=3.1$
and the total carbon abundance per hydrogen nucleus $b_{\rm C}=5.5\times
10^{-5}$ for diffuse environments CNM, WNM and WIM, and $R_{\rm V}=5.5,~
b_{\rm C}= 2.8\times 10^{-5}$ for the RN and PDR for
carbonaceous grains with $a_{\rm min}=3.55$ \AA~and $a_{\rm max}=100$ \AA.

\subsection{A2. Moments of Inertia}

Consider a PAH with $\NC$ carbon atoms, plus peripheral H atoms
that make a minimal contribution to the total mass.
We take the surface density of C atoms in a planar PAH to be
that in a single plane of graphite:
$\sigma_{\rm C} = 3.7\times10^{15}{\rm C}\cm^{-2}$.
We will assume that PAHs with $\NC < N_1$ consist of a monolayer,
but larger PAHs will be multilayered, with number density of
C atoms $n_{\rm C}=1.1\times10^{23}\cm^{-3}$.  
We approximate the PAH as a disk of radius $R$ and height $L$, with
\beqa
L = \frac{\sigma_{\rm C}}{n_{\rm C}}
\left[1+A \max(0,\NC^{1/3}-N_1^{1/3})\right],~~ R^2 =  \frac{\NC}{\pi n_{\rm C}L}
\eeqa
Neglecting the contribution of the H atoms, 
the moments of inertia are
\beqa
I_{\|} = \NC m_{\rm C}\frac{R^2}{2},~~I_{\perp} 
=
\NC m_{\rm C}\left(\frac{R^2}{4} + \frac{L^2}{12}\right),\label{Iper}
\eeqa
We take $N_1=10^2$ and $A=0.4$; with this choice, the PAH consists of
two layers when $\NC=364$.

\section{B. Rotational Damping\label{sec:rotational_damping}}

Gas-grain interactions, plasma-grain interactions, infrared emission,
and radio emission damp the rotation of small grains.
In this section we present the rates for these processes.

We consider a disk-like grain with temperature $T_{\rm d}$ in gas of
temperature $T_{\rm gas}$, H nucleon density $n_\H=n(\H)+2n(\H_2)$, and with
$n(\He)=0.1n_\H$. Following DL98, we ``normalize'' the various drag processes to the
drag which would be produced by ``sticky'' collisions in a pure H gas
of density $\nH$: thus for drag process $j$, we define the
dimensionless quantity $F_{j,\|,\perp}$ corresponding to the
rotational damping parallel and perpendicular to the symmetry axis
$\ba_{1}$, such that the contribution of process $j$ to the drag
torque is 
\beq
\label{eq:fdef}
I_{\|,\perp}\left({d\omega_{\|,\perp}\over dt}\right)_j 
\equiv
- \left(\frac{I_{\|,\perp}\omega_{\|,\perp}}{\tau_{\rm H,\|,\perp}}\right)
 F_{j,\|,\perp}. ~
\eeq

For the disk-like grain in a pure hydrogen gas
of density $n_{\rm H}$, $F_{j,\|,\perp}=1$, and the damping times for
rotation along and perpendicular to its symmetry axis are given by
\bea
\tau_{\rm H,\|}=\frac{3I_{\|}}{4\pi a_{\rm cx}^{4}n_{\rm H}m_{\rm
    H}(2\kB T_{\rm gas}/\pi m_{\rm H})^{1/2}},~~
\tau_{\rm
  H,\perp}=\frac{3I_{\perp}}{\pi R^{4}n_{\rm H}m_{\rm H}(2\kB T_{\rm
    gas}/\pi m_{\rm H})^{1/2}g_{\perp}},
\ena
where
\bea
g_{\perp}=\frac{1}{6}\left(L\over
R\right)^{3}+\frac{L}{R}+\frac{1}{2}\left(\frac{L}{R}\right)^{2}+\frac{1}{2}.
\ena
Using standard parameters for the interstellar diffuse medium
(ISM) and moments of inertia $I_{\|}, I_{\perp}$ for the disk, we get
%%btd 091212 I changed to use a rather than r, with redefinition of Gamma
\bea
\tau_{\rm H,\|}\approx4.12\times 10^{10}\hat{\rho}
a_{-7}{T}_{\rm 2}^{-1/2}\left(\frac{30 \mbox{ cm}^{-3}}{n_{\rm
    H}}\right)\Gamma_{\|} {~\mbox s},{~\rm and~}~\tau_{\rm H,\perp}
\approx
4.58\times 10^{9} \hat{\rho} a_{-7}T_{\rm
  2}^{-1/2}\left(\frac{30 \mbox{ cm}^{-3}}{n_{\rm
    H}}\right)\Gamma_{\perp} {~\mbox s},~~~~
\label{thper}
\ena
where
$\hat{\rho}\equiv\rho/2{~\rm gcm^{-3}}, {T}_{\rm 2}\equiv T_{\rm gas}/100$ K,
$a_{-7}\equiv a/10^{-7}$ cm, where $a\equiv(3/4)^{1/3}L^{1/3}R^{2/3}$, and
\bea
\Gamma_{\|}=
%%btd change
%\frac{8h}{3r}\frac{1}{[(\frac{2H}{r}+1)]},
\frac{8}{9}\left(\frac{6L}{R}\right)^{2/3}\frac{1}{[{2L}/{R}+1]},~~~
\Gamma_{\perp}=
%btd change
%\frac{h}{r}[3+(h/r)^{2}]\times\frac{1}{g_{\perp}}
\left(\frac{4}{3}\right)^{1/3}\left(\frac{L}{R}\right)^{2/3}
[3+(L/R)^{2}]\times\frac{1}{g_{\perp}}
\ena

We will use $\tau_{\rm H,\|,\perp}$ as a fiducial time scales for the
different sources of rotational damping.
Thus the rotational damping times due to process $j$ are
\beq
\tau_{j,\|,\perp}^{-1} = F_{j,\|,\perp} ~ \tH_{\|,\perp}^{-1}.
\eeq
The linear drag processes are additive:
\beq
F_{\|,\perp} = 
F_{\rm n,\|,\perp} + 
F_{\rm i,\|,\perp} + 
F_{\rm p,\|,\perp} + 
F_{\rm IR,\|,\perp},
\label{eq:fsum}
\eeq 
where $F_{\rm n}$, $F_{\rm i}$, $F_{\rm p}$, and $F_{\rm IR}$ are
the contributions from neutral impacts, ion impacts, plasma drag, and
thermal emission of infrared photons.

\subsection{B.1. Plasma Drag}

For a grain with an electric dipole moment $\bmu$, the interaction of
this electric dipole moment with passing ions in the plasma results in
a damping and excitation (Anderson \& Watson 1993). The damping due to
this process is obtained using the Fluctuation Dissipation theorem,
i.e., $F_{\rm p,\|,\perp}=G_{\rm p,\|,\perp}$ with $G_{\rm p,\|,\perp}$ 
are derived in Appendix C2.

\subsection{\label{app:rotdamp}
         B.2. Rotational Damping by IR Emission}

Let $\ba_{2}$, $\ba_{3}$, $\ba_{1}$ be
axes fixed in the grain.
Consider an oscillator with dipole moment 
\beqa
{\bf \mu}&=&\mu_2 \ba_{2} \cos\omega_0 t= \frac{\mu_2}{2}
\left[(\ba_{2}\cos\omega_0t+\ba_{3}\sin\omega_0t)
     +(\ba_{2}\cos\omega_0t-\ba_{3}\sin\omega_0t)
                 \right].
\eeqa
Now consider rotation of the grain around 
$\ba_{1}$ with angular velocity $\omega_r$:
\beqa
\ba_{2} 
&=& 
\hat{\bf x}\cos\omega_rt + \hat{\bf y}\sin\omega_rt,~~\ba_{3} 
= \hat{\bf y}\cos\omega_rt - \hat{\bf x}\sin\omega_rt,
\eeqa
where $\hat{\bf x}$, $\hat{\bf y}$, $\hat{\bf z}$ define an inertial
coordinate system.
The dipole moment is
\beqa
{\bmu} &=&
\frac{\mu_2}{2}\left[\hat{\bf x}\cos(\omega_r+\omega_0)t
                  +\hat{\bf y}\sin(\omega_r+\omega_0)t
                  +\hat{\bf x}\cos(\omega_r-\omega_0)t
                  +\hat{\bf y}\sin(\omega_r-\omega_0)t
             \right],
\eeqa
i.e., a dipole $(\mu_2/2)$ rotating with frequency $(\omega_0+\omega_r)$,
plus a dipole $(\mu_2/2)$ rotating retrograde 
with frequency $(\omega_0-\omega_r)$.
The net loss of angular momentum is
\beqa
\Gamma &=& -\left[\frac{2(\mu_2/2)^2}{3c^2}
                \frac{(\omega_r+\omega_0)^4}{\omega_r+\omega_0}
               -\frac{2(\mu_2/2)^2}{3c^2}
                \frac{(\omega_0-\omega_r)^4}{\omega_0-\omega_r}\right]
= -\frac{4\mu_2^2\omega_0^2}{c^3}\omega_r~.
\eeqa
The same result for the torque is obtained
for rotation around $\ba_{3}$, 
but rotation of this oscillator around $\ba_{2}$ would not result in
a torque.

Extending this to three incoherent oscillators:
\beq
{\bmu} = \mu_2 \ba_{2}\cos\omega_{02}t
+ \mu_3 \ba_{3}\cos\omega_{03}t
+ \mu_1 \ba_{1}\cos\omega_{01}t
\eeq
we find
\beqa
\frac{d}{dt}J_{i} &=& -\left[\frac{\mu_{j}^2\omega_0^2}{c^3} + 
                     \frac{\mu_{k}^2\omega_0^2}{c^3}\right]\omega_{i},
\eeqa
where $i,j$ and $k$ run from $1$ to $3$ so that if $i=2$ then $j=3$ and 
$k=1$ and so on; and replace $\omega_{01}\approx\omega_{02}\approx\omega_{03}$ by
$\omega_0$.

If we now consider a grain with symmetry axis $\ba_{1}$ and
$\mu_2=\mu_3$.
The power radiated by in-plane and out-of-plane vibrations is
\beqa
\dot{E}_{ip} = 
\dot{N}_{ip}\hbar\omega_0 = (\mu_2^2+\mu_3^2)\frac{\omega_0^4}{3c^3} ,~~
\dot{E}_{oop}= \dot{N}_{oop}\hbar\omega_0 = \mu_1^2\frac{\omega_0^4}{3c^3},
\eeqa
where $\dot{N}_{ip}$ and $\dot{N}_{oop}$ are the rates of photon emission
by in-plane and out-of-plane vibrations.
Then
\beqa
\frac{d}{dt}J_{\|} &=& -\left(\frac{3\hbar}{2\pi c I_{\|}}\right)
\left(\dot{N}_{ip}\lambda\right) J_{\|}~,
\\
\frac{d}{dt}J_\perp &=& -\left(\frac{3\hbar}{2\pi c I_{\perp}}\right)
\left(\frac{1}{2}\dot{N}_{ip}\lambda+ \dot{N}_{oop}\lambda\right) J_\perp~.
\eeqa
where $J_{\perp}=J_{2}$, $\lambda=2\pi c/\omega_0$ is the emitted wavelength.

\subsection{B3. Electric Dipole Damping}\label{dipole}

The spinning grain will radiate power, which results in rotational
damping of grain.  The associated rotational damping time depends on
$\bomega$.

In the grain body system, the dipole is given by 
\bea
\bmu=\mu_{1}\ba_{1}+\mu_{2}\ba_{2}+\mu_{3}\ba_{3}.\label{mueqa} 
\ena
For simplicity, we assume that $\mu_{1}^{2}\approx \mu_{2}^{2}\approx
\mu_{3}^{2}=\frac{1}{3}\mu^{2}$.  The angular velocity $\bomega$ in
general is not aligned with $\ba_{1}$. Thus, the rotation of the grain
consists of the rotation about $\ba_{1}$ with $\omega_{\|}$ and the
rotation with $\omega_{\perp}$ about axes perpendicular to $\ba_{1}$. 
The former rotation
of the dipole results in the damping for $\omega_{\|}$, and the later results
 in the damping for $\omega_{\perp}$.

Consider first the rotation with angular velocity $\omega_{\|}$ about $\ba_{1}$. 
After a time t, the dipole moment becomes
\bea
\bmu
&=&
\mu_{2}(\cos\omega_{\|}t\ba_{2}+\sin\omega_{\|}t\ba_{3})+
\mu_{3}(-\sin\omega_{\|}t\ba_{2}+\cos\omega_{\|}t\ba_{3}),\nonumber\\
&=&
\mu_{2}\sqrt{2}\left[-\ba_{2}\sin
\left(\frac{\pi}{4}+\omega_{\|}t\right)+
\ba_{3}\cos\left(\frac{\pi}{4}+\omega_{\|}t\right)\right],\label{bmu1}
\ena
The increase of angular momentum of the grain resulting from the rotation is
then
\bea
\frac{d{\bJ}}{dt}=-\frac{2}{3c^{3}}[\dot{\bmu}\times \ddot{\bmu}]=-
\frac{4\mu^{2}}{9c^{3}}\omega_{\|}^{3}\ba_{1},\label{dldt}
\ena
where we assumed $\mu_{2}^{2}=\frac{1}{3}\mu^{2}$.
Hence,
\bea
I_{\|}\frac{d\omega_{\|}}{dt}=
-\frac{4\mu^{2}}{9c^{3}}\omega_{\|}^{3}.\label{domepar}
\ena
Similarly, for the rotation with $\omega_{\perp}$ about 
$\ba_{2}$ and $\ba_{3}$ axes, we get
\bea
I_{\perp}\frac{d\omega_{\perp}}{dt}=
-\frac{4\mu^{2}}{9c^{3}}\omega_{\perp}^{3}.\label{domeper}
\ena

Since $\bomega$ is not parallel to $\ba_{1}$, let us define
characteristic damping times associated with $\omega_{\|}$ and
$\omega_{\perp}$, corresponding to the rotation along and
perpendicular to $\ba_{1}$.

As in DL98, we can define the damping times due to electric dipole emissions as 
\beq
\left({1 \over \omega_{\|,\perp}}{d\omega_{\|,\perp}\over dt}\right)_{\rm ed} = 
- {I_{\|,\perp}\omega_{\|,\perp}^2\over 3\kB T_{\rm gas}}
  {1\over \tau_{\rm ed,\|,\perp}}.\label{domepp}
\eeq
Then, combining equations (\ref{domepar}), (\ref{domeper}) and 
(\ref{domepp}), we obtain
\bea
\tau_{\rm ed,\|}\equiv \frac{3I_{\|}^{2}c^{3}}{4\mu^{2}\kB T_{\rm
gas}}
%=6.2\times10^{10}
%{\hat{\rho}^{2}R_{-7}^{10}\left(L/R\right)^{2}\over 
%{[(a_{\rm x}/a)^2 \langle Z^2\rangle+3.8(\beta/0.4\D)^2]a_{-7}^{2}T_2}}~\s,~~~~~~
=1.6\times10^{11}
\hat{\rho}^{2}a_{-7}^{8}\times\frac{(R/L)^{4/3}}
     {[(a_{\rm x}/a)^2 \langle Z^2\rangle+3.8(\beta/0.4~\D)^2 a_{-7}]T_2}~\s,~~~~~~
\label{tedpar}
\ena
and
\bea
\tau_{\rm ed,\perp}\equiv \frac{3I_{\perp}^{2}c^{3}}{4\mu^{2}\kB T_{\rm gas}}
%=1.55\times10^{10}{\hat{\rho}^{2}{R_{-7}^{10}
%\left(L/R\right)^{2}
%\left(1+\frac{1}{3}\left({L}/{R}\right)^{2}\right)^{2}}\over 
%{[(a_{\rm x}/a)^2 \langle Z^2\rangle+3.8(\beta/0.4\D)^2]a_{-7}^{2}T_2}}~\s,~~~~~~
=4.0\times10^{10}
\hat{\rho}^{2} a_{-7}^8\times\frac{(R/L)^{4/3}
     \left(1+\frac{1}{3}(L/R)^2\right)^2}
     {[(a_{\rm x}/a)^2 \langle
Z^2\rangle+3.8(\beta/0.4~\D)^2 a_{-7}]T_2}~\s,~~~~~~
\label{tedper}
\ena 
where $\mu^{2}$ from DL98b (eq. \ref{mu2}) has been used and $\langle
Z^{2}\rangle$ is the mean square grain charge.

\begin{figure}
\includegraphics[width=0.45\textwidth]{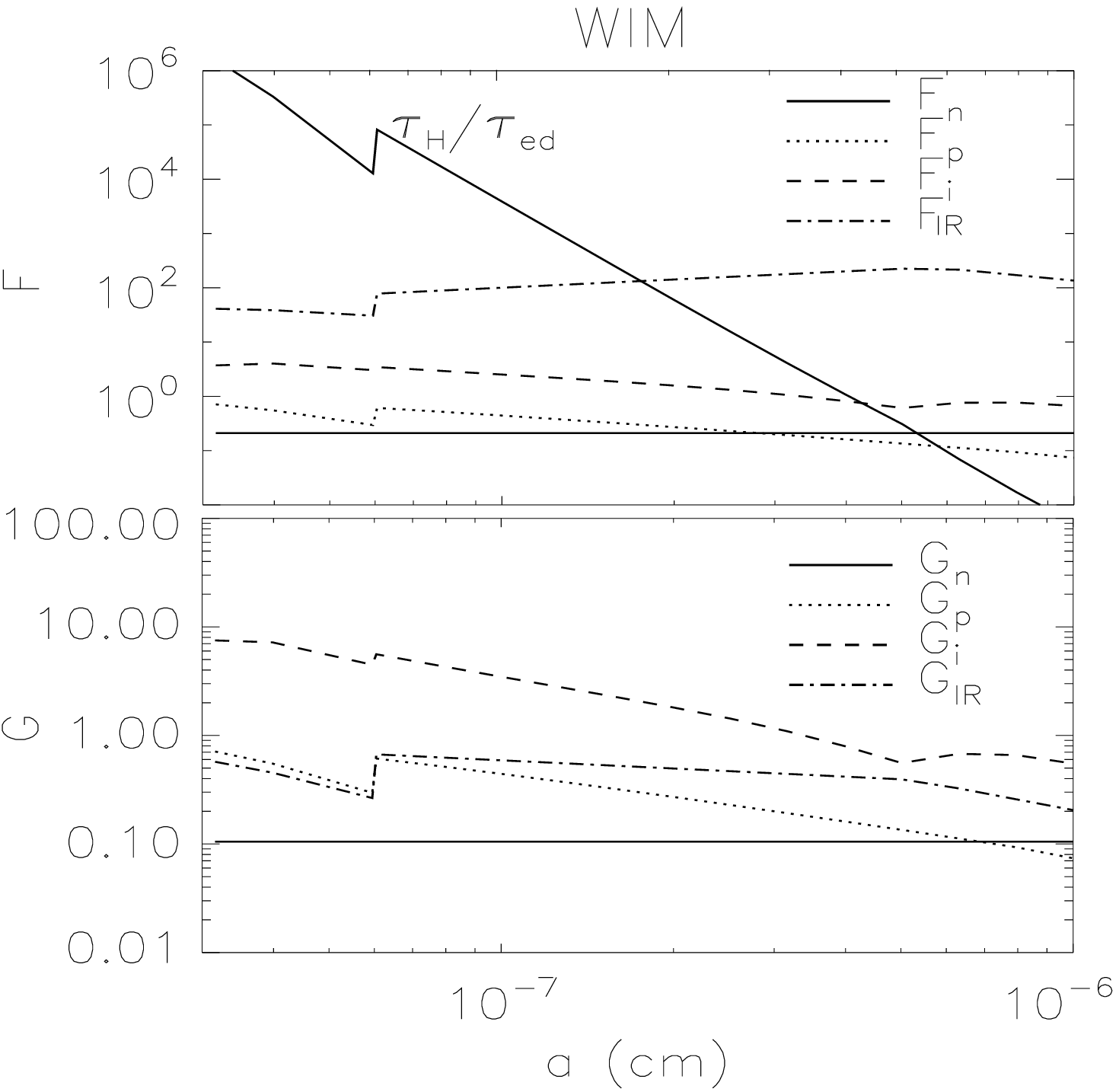}
\includegraphics[width=0.45\textwidth]{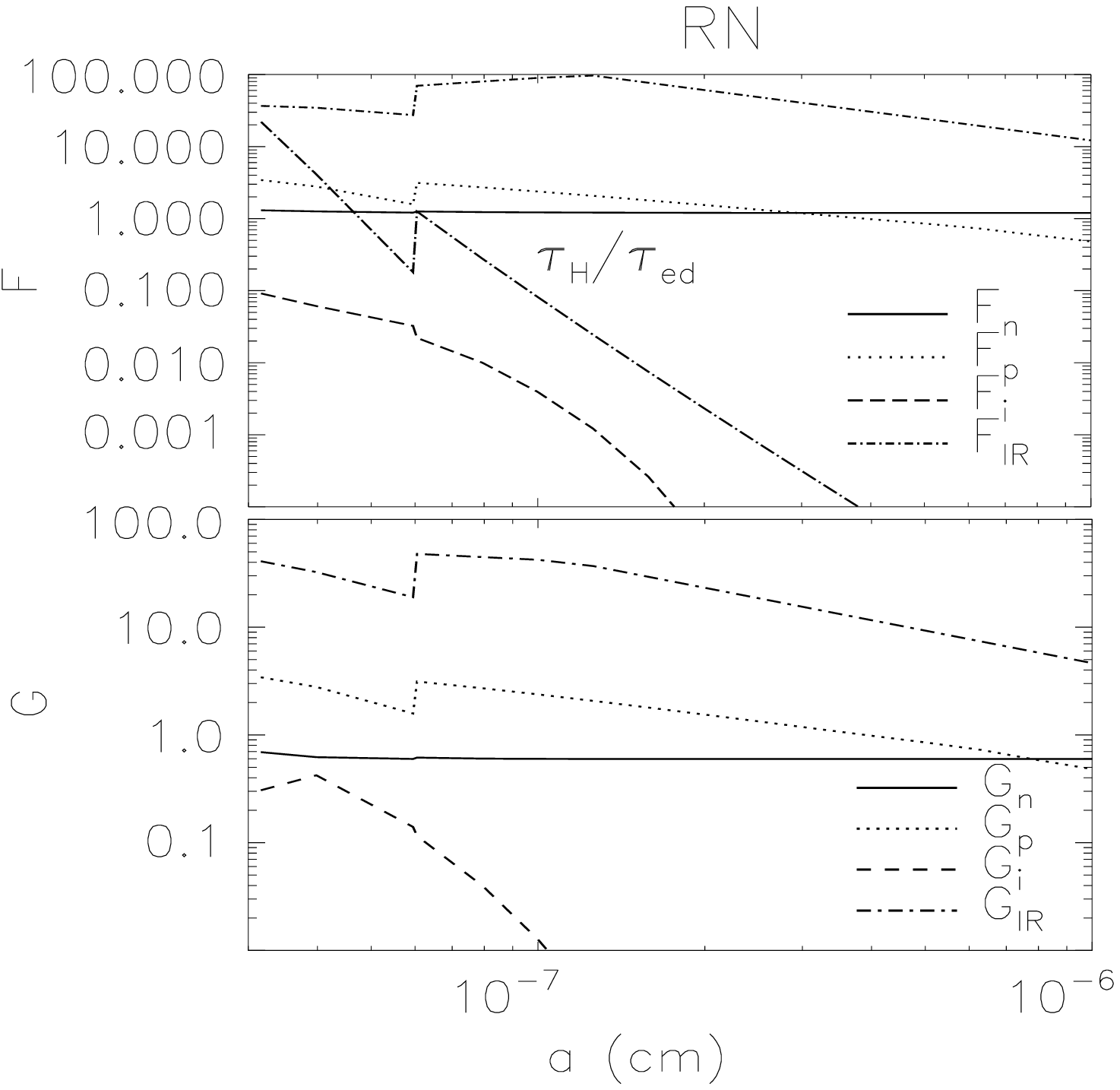}
\caption{{\it Left:} Damping coefficients $F_{\rm n}, F_{\rm i}, F_{\rm p}, F_{\rm
    IR}$ and $\tau_{\rm ed}/t_{\rm H}$ and excitation coefficients
  $G_{\rm n}, G_{\rm i},G_{\rm p}$ and $G_{\rm IR}$ from DL98b model as functions
  of grain size $a$ for WIM. {\it Right:} Similar to Figure
  \ref{fg_wim} but for the RN.}
\label{fg_wim}
\end{figure}

Using the distribution charge function $f(Z)$ and mean charge $\langle
Z\rangle$ obtained by solving equation of ionization equations (see
DL98b), we compute damping coefficients $F_{\rm n},F_{\rm i},F_{\rm p}, F_{\rm
  IR}$ and $\tau_{\rm ed}$. Our new results for $F_{\rm p}$, $F_{\rm IR}$ 
and $\tau_{\rm ed}$ are presented in Figures \ref{Gpar}, \ref{FG_IRWIM} and
 \ref{fir_new}, respectively.
For comparison, in Figures \ref{fg_wim} we present results from the DL98 model
for the WIM and RN.

\section{C. Rotational Excitation\label{sec:rotational_excitation}}

An initially stationary grain will have its rotational
kinetic energy increasing at an average rate
\bea
\label{eq:gdef}
{d\over dt}\left({1\over2}I_{\|,\perp}\omega_{\|,\perp}^2\right)
&=&
\frac{\kB T_{\rm gas}}{\tau_{\rm H,\|,\perp}}
\times(G_{{\rm n},\|,\perp}+G_{{\rm i},\|,\perp}+G_{{\rm p},\|,\perp}+G_{\rm IR,\|,\perp}),
\ena
where the normalized excitation rate
$G_{\rm n}$ is due to impacting neutrals,
$G_{\rm i}$ is due to impacting ions,
$G_{\rm p}$ is due to plasma drag,
and $G_\IR$ is due to recoil from infrared emission, 
and $\|, \perp$ denote excitation along and perpendicular to 
the grain symmetry axis. 

\subsection{C1. Recoil from Thermal Collisions and Evaporation}
For collisional excitations, similar to collisional drag, 
we have $G_{n,\|}=G_{n,\perp}$ and $G_{i,\|}=G_{i,\perp}$.

\subsection{C2. Excitation by the Plasma}\label{plasma}
Consider a disk-like grain with dipole moment $\bmu$, assumed to be parallel 
to $\ba_{2}$:
\bea
\bmu=|\mu|\ba_{2}.\label{mu_eq}
\ena 
Let us define the inertial coordinate system
$\xhat,\yhat,\zhat$. Grain is assumed to be rotating with angular velocity 
$\omega_{\|}$ and $\omega_{\perp}$ parallel and perpendicular to the symmetry axis.
Initially, $\ba_{2}\| \xhat$, $\ba_{3}\|\yhat$
and $\ba_{1}\| \zhat$. At the time t, the orientation of the grain in
the lab system can be determined by two Euler angles $\theta$ and
$\phi$ where $\theta$ and $\phi$ are the rotation angles of the grain
about $\ba_{1}$ and $\ba_{3}$, respectively.\footnote{Ali-Ha\"imoud et
  al. (2009) assumes that initial grain orientation is described by
  $\theta_{0}, \phi_{0}$, and later, they average over random
  $\theta_{0}, \phi_{0}$. Our assumption is special with
  $\theta_{0}=\phi_{0}=0$, but without loosing physical effects} They
are given by 
\bea
\ba_{1}=\cos\theta~\zhat+\sin\theta~\cos\phi~\xhat+\sin\theta~\sin\phi~\yhat,\\
\ba_{2}=-\sin\theta~\zhat+\cos\theta~\cos\phi~\xhat+\cos\theta~\sin\phi~\yhat,\\
\ba_{3}=\cos\phi~\yhat-\sin\phi~\xhat.
\ena

The trajectory plane of the incoming ion of charge $Z_{\rm i}e$ with
$e$ being the electron charge is $\yhat \zhat$ where $\zhat$ is
parallel to the initial direction of ion. The electric field produced
by the ion at the grain position is 
\bea
{\bf
  E}=\frac{Z_{\rm i}e}{r^{2}}{\bf
  r}=\frac{Z_{\rm i}e}{r^{2}}(\cos\alpha ~\yhat+\sin\alpha~\zhat).
\ena
The torque induced by the dipole-electric field interaction is given
by
\bea \frac{d{\bJ}}{dt}=[\bmu\times\bE]=
|\mu||E|\left(\xhat[\sin\theta~\cos\alpha+\cos\theta~\sin\phi~\sin\alpha]\right)~~~
\label{dLdt}
\\
 +|\mu||E|
\left(-\cos\theta~\cos\phi~\sin\alpha~\yhat\nonumber+
       \cos\theta~\cos\phi~\sin\alpha~\zhat
\right),
\ena
where equation (\ref{mueqa}) has been used.  By
projecting (\ref{dLdt}) onto two rotational axes $\ba_{1}$ and
$\ba_{3}$, we obtain
\bea
\frac{d\omega_{\|}}{d\alpha}=
\frac{-Z_{\rm i}e|\mu|}{I_{\|}bv}\cos\phi~\cos\alpha,~~
\frac{d\omega_{\perp}}{d\alpha}
=
\frac{-Z_{\rm i}e|\mu|}{I_{\perp}bv}
\left(\cos\theta~\sin\alpha+\sin\theta~\sin\phi~\cos\alpha\right),~~
\ena
where $bv=r^{2}d\alpha/dt$ has been taken.

The mean rotational excitation due to one ion is then given by
\bea
\langle (\delta \omega_{\|})^{2}\rangle=
\left(\frac{2Z_{\rm i}e|\mu|}{I_{\|}bv}\right)^{2}\mathcal I_{\|},~~
\langle (\delta \omega_{\perp})^{2}\rangle=
\left(\frac{2Z_{\rm i}e|\mu|}{I_{\perp}bv}\right)^{2}\mathcal I_{\perp},
\ena
where the factor of $2$ comes from the fact that 
$\alpha$ in the range from $0$ to $\pi/2$, 
instead from $-\pi/2$ to $\pi/2$, or time $t$ from 0 to $\infty$, and
\bea
\mathcal I_{\|}
=
\left(\int d\alpha \cos\phi~\cos\alpha\right)^{2},~~\mathcal I_{\perp}
=
\left(\int d\alpha (\cos\theta~\sin\alpha+\sin\theta~\sin\phi)\right)^{2},
\ena

For a neutral grain and the trajectory of ion is straight line, the
polar angle $\alpha$ and $t$ are related to each other as follows:
\bea
\sin\alpha=\frac{vt}{\sqrt{b^{2}+v^{2}t^{2}}},~~
\cos\alpha=\frac{b}{\sqrt{b^{2}+v^{2}t^{2}}},~~
d\alpha=\frac{1}{[1+(vt/b)^{2}]}\frac{vdt}{b},
\ena
where we assume
that $t=0$ corresponding to the time when the ion is closest to the
grain, i.e., $r=b$. We have, at the time t=0, the direction of the
grain is random, then at t, we have $\phi=\phi_{0}+\omega_{\|}t,
\theta=\theta_{0}+\omega_{\perp}t$ with $\phi_{0},\theta_{0}$ are
random variables.

Therefore, $\mathcal I_{\|}$ and $\mathcal I_{\perp}$ become
\bea
\mathcal I_{\|}(x)
&=&
\left(\int\cos\phi~\cos\alpha~
d\alpha\right)^{2}=\left(\int\cos\omega_{\|}t~\cos\alpha
~d\alpha\right)^{2}
=
\left(\int_{0}^{\infty}d\tau~\frac{\cos(x\tau)}{(1+\tau^{2})^{3/2}}\right)^{2}
=
x^{2}K_{1}(x)^{2},
\label{ipar}
\\
\mathcal I_{\perp}(y,y_{\pm})
&=&
\left(
\int [\cos\omega_{\perp}t~\sin\alpha+\sin\omega_{\perp}t~\sin\omega_{\|}t\cos\alpha]~
d\alpha\right)^{2},\nonumber
\\
&=&\left(\int_{0}^{\infty}\frac{\tau\cos(y\tau)
  d\tau}{(1+\tau^{2})^{3/2}}+
\frac{1}{2}(y_{+}K_{1}(y_{+})-y_{-}K_{1}(y_{-})\right)^{2}=
\left(H(y)+\frac{1}{2}[y_{+}K_{1}(y_{+})-y_{-}K_{1}(y_{-})\right)^{2},
\label{iper}
\ena
where $x=\omega_{\|}b/v, y=\omega_{\perp}b/v$ and
$y_{\pm}=(\omega_{\|}\pm\omega_{\perp})b/v$. $K_{n}(x)$ is the
modified Bessel function of the second kind, and $H(y)$ is given by
\bea
H(y)=\int_{0}^{\infty}\frac{\tau \cos (y\tau)~d\tau}{(1+\tau^{2})^{3/2}}.
\label{heq}
\ena
To calculate (\ref{heq}) numerically , we note that when the grain rotates much
  faster than the ion motion, i.e., $y=\omega_{\perp}b/v>>1$, the torque
  gets averaged out. And a truncation of (\ref{heq}) is $y=1.3$ is
  adopted without introducing large uncertainty.

The total excitation rate from all ions with density $n_{\rm i}$ following
normal distribution of velocity is given by 
\bea
\frac{d\langle
  \delta\omega_{\|})^{2}\rangle}{dt}
&=&
\int_{0}^{\infty} dv\left(4\pi v^{2} n_{\rm i}v 
(\frac{m_{\rm i}}{2\pi \kB T_{\rm gas}})^{3/2}e^{-m_{\rm i}v^{2}/
2\kB T_{\rm gas}}\right)\int_{b_{\rm max}}^{\infty}db~ 2\pi
b\left(\frac{Z_{\rm i}e|\mu|}{I_{\|}bv}\right)^{2}\mathcal
I_{\|}(x),\label{dome12}
\\
\frac{d\langle\delta\omega_{\perp})^{2}\rangle}{dt}&=&\int_{0}^{\infty}
dv\left(4\pi v^{2} n_{\rm i}v (\frac{m_{\rm i}}{2\pi \kB T_{\rm
    gas}})^{3/2}e^{-m_{\rm i}v^{2}/2\kB T_{\rm gas}}\right)
\int_{b_{\rm max}}^{\infty}db~ 2\pi
b\left(\frac{Z_{\rm i}e|\mu|}{I_{\perp}bv}\right)^{2}\mathcal
I_{\perp}(y,y_{\pm}),
\label{dome22}
\ena

In equations (\ref{dome12}) and (\ref{dome22}), the value $b_{\rm max}$ for
a neutral grain is given by
\bea b_{\rm max}
=a_{\rm cx}\sqrt{1+\frac{\Phi}{u}},~\Phi^{2}
=\frac{2Z_{\rm i}^{2}e^{2}}{a_{\rm cx}\kB T_{\rm gas}},~ u
=\frac{v}{v_{\rm T}} {\rm with }~ v_{\rm T}=(2\kB T_{\rm gas}/m_{\rm i})^{1/2},~~~~~\label{bmaxn}
\ena
and 
for a charged grain, $b_{\rm max}$ takes the form (see Spitzer 1941)
\bea
b_{\rm max} = 
\left\{
\begin{array}{ll}
0 
& 
\mbox{for ${m_{\rm i}v^2\over 2} < {Z_{\rm g} Z_{\rm i} e^2\over a_{\rm cx}}$}
\\
a_{\rm cx}\left(1-{2Z_{\rm g} Z_{\rm i} e^2\over m_{\rm i}a_{\rm cx}v^2 }\right)^{1/2}
&
\mbox{for ${m_{\rm i}v^2\over 2} > {Z_{\rm g} Z_{\rm i} e^2\over a_{\rm cx}}$},
\end{array}
\right.\label{bmaxc}
\ena
 and the upper limit of the integration over the impact factor is constrained
by the Debye length $\lambda_{\rm D}=\left(kT_{\rm gas}/(4\pi n_{e}e^{2})\right)^{1/2}$.

The dimensionless excitation coefficients for grain-plasma interactions are then
\bea
G_{\rm p,\|}
=
\frac{d\langle (\delta\omega_{\|})^{2}\rangle}{dt}
\frac{I_{\|}\tau_{\rm H,\|}}{2\kB T_{\rm gas}},~~G_{\rm p,\perp}
=
\frac{d\langle (\delta\omega_{\perp})^{2}\rangle}{dt}
\frac{I_{\perp}\tau_{H,\perp}}{2\kB T_{\rm gas}}.
\ena 
Plugging (\ref{dome12}) and (\ref{dome22}) into the above equations, we obtain
\bea
G_{\rm p,\|,\perp}
=
\frac{n_{\rm i}}{n_{\H}}
\left(\frac{m_{\rm i}}{m_{\H}}\right)^{1/2}
\left(\frac{Z_{\rm i}e\mu}{\kB T_{\rm gas}a_{\rm cx}^{2}}\right)^{2}
\left(3\over 2\right)\int_{0}^{\infty}ue^{-u^{2}}du\times g_{\|,\perp}
\ena
where
\bea
g_{\|,\perp}=
\int_{b_{\rm max}/a_{\rm cx}}^{\infty}\frac{dl}{l}
\mathcal I_{\|,\perp}\left(u\frac{\Omega_{\|,\perp} }{l}\right), 
{~\rm and~ }
\Omega_{\|,\perp}=\sqrt{\frac{a_{\rm cx}^{2}m_{\rm i}}{2\kB T_{\rm gas}}}\omega_{\|,\perp}.
\ena

Using the Fluctuation-Dissipation theorem, we obtain $F_{\rm
  p,\|}=G_{\rm p,\|}$, and $F_{\rm p,\perp}=G_{\rm p,\perp}$. In
Figure \ref{Gpar} we show our new results for $G_{\rm p,\|}$ and $G_{\rm p,\perp}$ as functions of 
grain size for neutral grains in the CNM.

Excitation coefficients from DL98b $G_{\rm n},G_{\rm i},G_{\rm p}$ and $G_{\rm
  IR}$ are also shown in the lower panels of Figure \ref{fg_wim} for the
WIM and RN, respectively.

\subsection{\label{app:rotexc}
         C3. Rotational Excitation by IR Photons}
Consider the case of a grain with dynamical symmetry,
with principal values of the moment of inertia tensor
$I_1 > I_2=I_3$, where $I_1$ is the moment of inertia for rotations around
the $\ba_{1}$ axis.
We assume the emission to come from optically active vibrations that are
either ``in-plane'' or ``out-of-plane'', and we assume the ``in plane''
oscillators to be symmetrically distributed around the $\ba_{1}$-axis (i.e.,
equal numbers oscillating in the $\ba_{2}$- and $\ba_{3}$-directions.

Let $P_{2}$, $P_3$, and $P_1$ be the power radiated by vibrational modes
with electric dipole moment oscillating in the $\ba_{2}$- $\ba_{3}$-, and 
$\ba_{1}$-directions.  
For electric dipole radiation, we can imagine that each oscillator radiates
50\% of its power in each of the two cardinal directions orthogonal to
the dipole.

For a nonrotating (or slowly rotating) grain, the angular momentum
will undergo a random walk with
\beqa
\langle \frac{d}{dt} J_{i}^2\rangle &=& \frac{1}{2}\frac{(P_{j} + P_{k})}{h\nu}
 2\hbar^{2},
\eeqa
where i, j, k run from 1 to 3 and the vector basis $\ba_{i}\ba_{j}\ba_{k}$ 
follow right-handed rule.

Thus, defining $J_\perp^2\equiv J_2^2= J_3^2$, and noting that $P_2$ and $P_3$
are in-plane vibrations, and $P_1$ is out-of-plane:
\beqa
\langle \frac{d}{dt} J_\perp^2\rangle = \frac{1}{2}\left[\frac{1}{2}\dot{N}_{ip}
                                  +\dot{N}_{oop}\right] 2\hbar^2,~~~
\langle \frac{d}{dt} J_{\|}^2\rangle =\frac{1}{2} \dot{N}_{ip} 2\hbar^2
\eeqa

\section{D. Collision of ions with grains}
The damping and excitation coefficients for the collision of ions with
grain were calculated in DL98b, and refined in Ali-Ha\"imoud et
al. (2009). The later study showed that the calculations by DL98b
where the effects of dipole were disregarded remain valid. Thus, below
we also disregard the effect of dipole in calculating the collision
rate for ion-grain collision.  The collision rate of ions with density
$n_{\rm i}$ with a grain of size $a$ is
\bea
R_{i}=n_{\rm i}\int_{0}^{\infty}v\pi b_{\rm max}^{2} 4\pi v^{2}f(v)dv
=
n_{\rm i}\int_{0}^{\infty}v \pi b_{\rm max}^{2} 4\pi
v^{2}A\mbox{exp}\left(\frac{-m_{\rm i}v^{2}}{2\kB T_{\rm gas}}\right)dv,
\label{ridef}
\ena
where $A=\left({2\kB T_{\rm gas}}/{m_{\rm i} \pi}\right)^{-3/2}$, and
$b_{\max}$ is the critical impact factor so that all incoming ions with
impact factor $b \le b_{\rm max}$ collide with the grain.

\subsection{D1. Neutral grain}
For neutral grain, $b_{\rm max}$ is given by equation (\ref{bmaxn}), thus, 
plugging (\ref{bmaxn}) into (\ref{ridef}) and performing the integral, we obtain
\bea
R_{i}(Z_{\rm g}=0)=
n_{\rm i}\pi a^{2}
\left(\frac{8\kB T_{\rm gas}}{m_{\rm i}\pi}\right)^{1/2} 
\left[1+\frac{\sqrt{\pi}}{2}\left(\frac{2Z_{\rm i}^{2}e^{2}}{a
\kB T_{\rm gas}}\right)^{1/2}\right].
\ena

The rms angular momentum per collision is obtained by dividing the rms angular
 momentum to the collision rate. For the ion-neutral collision, we have
\bea
\langle \delta J^{2}\rangle=m_{\rm i}\kB T_{\rm gas}a^{2}\frac{2 + 3\sqrt{\pi/2\tau} 
+ 2/\tau}{1+\sqrt{\pi/2\tau}},~~~\mbox{with~} 
\tau = \frac{a\kB T_{\rm gas}}{e^{2}} = 60\left(\frac{a}{10^{-7}{~\rm cm}}\right)
\left(\frac{T_{\rm gas}}{10^{4}{~\rm K}}\right).
\ena 

\subsection{D2. Charge grain}
For a charge grain, using the critical impact factor from (\ref{bmaxc}) 
for (\ref{ridef}) we obtain
\bea
R_{i}(Z_{\rm g}\ne 0)
=
n_{\rm i}\pi a^{2}\left(\frac{8k_{\rm B}T_{\rm gas}}{m_{\rm i}\pi}\right)^{1/2}g
\left(\frac{Z_{\rm g}Z_{\rm i}e^{2}}{a\kB T_{\rm gas}}\right),
\label{Ri}
\ena
where
$ g(x)= 1-x$ for $x<0$ and $g(x)=e^{-x}$ for  $x>0$

The collision of one ion with the negatively-charged grain has rms angular
 momentum per collision for this case given by
\bea
\langle \delta J^{2}\rangle=m_{\rm i}\kB T_{\rm gas}a^{2}\frac{2-2\phi+\phi^{2}}{1-\phi}, 
\mbox{~with~} \phi=\frac{Z_{\rm g}e^{2}}{a\kB T_{\rm gas}}.
\ena
Taking into account the grain charge distribution, the final form for
collision rate becomes 
\bea R_{i}
&=&
f(Z_{\rm g}=0) n_{\rm i}\pi a^{2}
\left(\frac{8\kB T_{\rm gas}}{m_{\rm i}\pi}\right)^{1/2}
\left[1+\frac{\sqrt{\pi}}{2}\left(\frac{2Z_{\rm i}^{2}e^{2}}{a\kB T_{\rm gas}}\right)^{
1/2}\right]+
\sum_{Z_{\rm g}\ne 0}f(Z_{\rm g}) n_{\rm i}\pi
a^{2}\left(\frac{8\kB T_{\rm gas}}{m_{\rm i}\pi}\right)^{1/2}g
\left(\frac{Z_{\rm g}Z_{\rm i}e^{2}}{a\kB T_{\rm gas}}\right),~~~~~\label{Ri}
\ena
where $f(Z_{\rm g})$ is the grain charge distribution function.

\section{E. Transformation of coordinate systems}
Damping coefficient $ A_{i}=\langle {\Delta J_{i}}/{\Delta t}\rangle$ 
and diffusion coefficients  
$B_{ij}=\langle {\Delta J_{i}\Delta J_{j}}/{\Delta t}\rangle$ are 
usually derived in the body coordinate system, while we are interested in 
the evolution of grain angular momentum in the inertial coordinate system. Let
us define an inertial coordinate system $\be_{1}\be_{2}\be_{3}$ in which 
the direction $\bJ$ is described by the angle $\xi$ between $\bJ$ with $\be_{1}$ , 
and the azimuthal angle $\chi$ (see Figure 2 in Roberge \& Lazarian 1999 in which
angles $\phi,~\beta$ are replaced by $\chi,~\xi$) . To
obtain these coefficients in the lab
coordinate system, we first transform the body system $\ba_{i}$ to the external 
system $\xhat \yhat \zhat$ where $\zhat\|\bJ$ and $\xhat,\yhat\perp \bJ$ 
(see Figure \ref{preces}).
Then, we perform the transformation from $\xhat\yhat\zhat$ system to
 the inertial system $\be_{1}\be_{2}\be_{3}$. 
In the body system, the damping coefficients are given by
\bea
A_{i}^{b}=\langle \frac{\Delta J_{i}^{b}}{\Delta t}\rangle=
-\frac{J_{i}}{\tau_{{\rm gas},i}}-\frac{J_{i}^{3}}{\tau_{{\rm ed},i}}
\left(\frac{1}{3I_{i}k_{\rm B}T_{\rm gas}}\right),
\ena
where $t_{{\rm gas},i}=F_{{\rm tot},i}/\tau_{\rm H,\|}$ and $i=x,y$ and $z$ with
 $z\| \ba_{1}$.
The diffusion coefficients, 
$B_{ij}^{b}=\langle {\Delta J_{i}^{b}\Delta J_{j}^{b}}/{\Delta t}\rangle$ 
with $B_{ij}^{b}=0$ for $i\ne j$ are related to the excitation coefficients as follows:
\bea
B_{zz}^{b}=B_{\|}=\frac{2I_{\|}\kB T_{\rm gas}}{\tau_{\rm H,\|}}G_{\rm tot,\|},{\rm ~and~~}
B_{xx}^{b}=B_{yy}^{b}=B_{\perp}=\frac{2I_{\perp}\kB T_{\rm gas}}{\tau_{\rm H,\perp}}G_{\rm tot,\perp}.
\ena
Transforming the vector $A_{i}^{b}$ and matrix $B_{ij}^{b}$ from the body
 system to the inertial system, and averaging over the fast precession of 
$\ba_{1}$ about $\bJ$, we obtain damping coefficient (see Lazarian \& Roberge 1997)
\bea
A_{i}=\langle \frac{\Delta J_{i}}{\Delta t}\rangle=-\frac{J_{i}}
{\tau_{\rm gas,\rm eff}}-\frac{J_{i}^{3}}{\tau_{\rm ed,\rm eff}}
\left(\frac{1}{3I_{\|}k_{\rm B}T_{\rm gas}}\right),\mbox{~for~} i=\mbox{~x,y,z~},
\ena
where the effective gas damping reads
\bea
\tau_{\rm gas,\rm eff}=\tau_{\rm H, \|}F_{\rm tot,\|}\frac{1}{\mcs\theta+
\gamma_{\rm H} \mss\theta},
\ena
and the effective time for electric dipole damping is given by
\bea
~~\tau_{\rm ed,\rm eff}=\tau_{\rm ed,\|}\frac{1}{\mbox{cos}^{4}\theta+
\gamma_{\rm ed}\mbox{sin}^{4}\theta}.\label{tau_eda}
\ena
where $\theta$ is the angle between $\ba_{1}$ and $\bJ$, and 
\bea
\gamma_{\rm H}=\frac{F_{\rm tot,\perp}\tau_{\rm H,\|}}{F_{\rm tot,\|}\tau_{\rm H,\perp}},
~~\gamma_{\rm ed}=\frac{I_{\|}\tau_{\rm ed,\|}}{I_{\perp}\tau_{\rm ed,\perp}}=h^{3}.
\ena
If the coupling of parallel and perpendicular rotation is accounted for, then the 
electric dipole damping is increased by a small amount, and equation (\ref{tau_eda}) 
is  rewritten
 as (communication with C Hirata)
\bea
\tau_{\rm ed,\rm eff}=\tau_{\rm ed,\|}\frac{1}{\mbox{cos}^{4}\theta+
\gamma_{\rm ed}\mbox{sin}^{4}\theta+(h^{3}+3h)\sin^{2}\theta\cos^{2}\theta/2}.
\ena

We can see that for disk-like grains, $\gamma_{\rm H}>1$, and
$\tau_{\rm gas,\rm eff}<\tau_{\rm H, \|}F_{\rm tot,\|}$,
which results in faster gas damping time. 

The diffusion coefficients in the inertial system $\be_{i}$, $B_{zz}$ and
$B_{xx},B_{yy}$ are given by
\bea
B_{zz}&=&B_{\|}\left(\frac{1}{2}
\mss\theta\mss\xi+\mcs\theta\mcs\xi\right)+B_{\perp}\left(\frac{1}{2}[
1+\mcs\theta]\mss\xi+\mss\theta \mcs\xi\right),~~~~~\\
B_{xx}&=&B_{\|}\left(\frac{1}{2}\mss\theta[\mcs\chi+\mss\chi\mcs\xi]
+\mcs\theta \mss\chi\mss\xi\right)+B_{\perp}\left(\frac{1}{2}[1+\mcs\theta]
[\mcs\chi+\mss\chi\mcs\xi]
+\mss\theta\mss\chi\mss\xi\right),~~~~~~\\
B_{yy}&=&B_{\|}\left(\frac{1}{2}\mss\theta[\mss\chi+\mcs\chi\mcs\xi]
+\mcs\theta \mss\chi\mss\xi\right)+B_{\perp}\left(\frac{1}{2}[1+\mcs\theta]
[\mss\chi+\mcs\chi\mcs\xi]
+\mss\theta\mss\chi\mss\xi\right),~~~~~~~
\ena
where $\xi$ is the angle between $\bJ$ and $\be_{1}$, and $\chi$ is the
azimuthal angle of $\bJ$ in the inertial system $\be_{i}$.

In the presence of fast internal fluctuations, we need to average the damping
 and diffusion coefficients over $\theta$. Therefore,
the terms containing $\theta$ in above equations are replaced by the averaged
 values, i.e., $\langle\mcs\theta\rangle=\int_{0}^{\pi} \mcs\theta f_{\rm LTE}(\theta) d\theta$,
$\langle\mss\theta\rangle=\int_{0}^{\pi} \mss\theta f_{\rm LTE}(\theta) d\theta$.
 Note that we do not average
the damping and diffusion coefficients over the precession angle $\chi$ here
 because the effect of magnetic field which results 
in the fast precession of $\bJ$ about the magnetic field is disregarded.

Then, the Langevin equations in the inertial system become
\bea
dJ_{i}=A_{i}dt+\sqrt{B_{ii}}dq_{i},{\rm where~} \langle dq_{i}^{2}\rangle=dt,{~\rm and~i=x,~y, ~z}.
\ena

\section{F. Dipole Emission of a Torque-Freely Rotating Grain}

Let us consider the simple case where the dipole is given by 
\bea
\bmu=\mu_{\|}\ba_{1}+\mu_{\perp}\ba_{2}
\ena
in the body system. In an external coordinate system $\xhat\yhat\zhat$ 
(see Figure \ref{preces}), $\ba_{1}, \ba_{2}$ and $\ba_{3}$ 
are described as
\bea
\ba_{1}&=&\sin\phi\sin\theta \xhat-\cos\phi\sin\theta \yhat+\cos\theta \zhat,\\
\ba_{2}&=&(\cos\phi\cos\psi-\sin\phi\sin\psi\cos\theta) \xhat+
(\sin\phi\cos\psi+\cos\phi\sin\psi\cos\theta)\yhat+ \sin\psi\sin\theta \zhat,\\
\ba_{3}&=&-(\cos\phi\sin\psi+\sin\phi\cos\psi\cos\theta) \xhat+
(-\sin\phi\sin\psi+\cos\phi\cos\psi\cos\theta)\yhat+ \cos\psi\sin\theta \zhat,~
\ena
where $\phi,\psi$ and $\theta$ are Euler angles (see Figure \ref{preces}).

Let us now consider the torque-free motion of an axisymmetric grain 
with the angular momentum $\bJ$ and the ratio of inertia moments $h$. 
For this case, the symmetry axis $\ba_{1}$ precesses about $\bJ$ with 
constant angle $\theta$ and the rate $\dot\phi$, and the grain rotates 
about the symmetry axis with the rate $\dot\psi$. They are given by (Landau \& Lifshitz 1976)
\bea
\dot\phi=\frac{J}{I_{\perp}},~~~\dot{\psi}=\frac{J}{I_{\|}}\cos\theta(1-h).~~\label{omepa}
\ena
Precession and rotation of the grain with respect to $\bJ$ results in an 
instantaneous acceleration for dipole moment:
\bea
\ddot{\bmu}=\mu_{\|}\ddot{\ba}_{1}+\mu_{\perp}\ddot{\ba}_{2},\label{dotmua}
\ena
where $\ddot{\ba}_{1}$ and $\ddot{\ba}_{2}$ are given by
\bea
\ddot{\ba}_{1}&=&-\dot{\phi}^{2}\sin\theta(\sin\phi \xhat-\cos\phi \yhat),
\label{dota1}\\
\ddot{\ba}_{2}&=&\left[-\dot{\phi}^{2}(\cos\phi\cos\psi-\sin\phi\sin\psi\cos\theta)-
2\dot\phi\dot\psi(-\sin\phi\sin\psi+\cos\phi\cos\psi\cos\theta)-\dot{\psi}^{2}
(\cos\phi\cos\psi-\sin\phi\sin\psi\cos\theta)\right]\xhat\nonumber\\
&&+\left[-\dot{\phi}^{2}(\sin\phi\cos\psi+\cos\phi\sin\psi\cos\theta)-
2\dot\phi\dot\psi(\cos\phi\sin\psi+\sin\phi\cos\psi\cos\theta)-{\dot\psi}^{2}
(\sin\phi\cos\psi+\cos\phi\sin\psi\cos\theta)\right]\yhat\nonumber\\
&&-{\dot\psi}^{2}\sin\psi\sin\theta \zhat,\label{dota2}\\
\ddot{\ba}_{3}&=&\left[\dot{\phi}^{2}(\cos\phi\sin\psi+\sin\phi\cos\psi\cos\theta)+
2\dot\phi\dot\psi(\sin\phi\cos\psi+\cos\phi\sin\psi\cos\theta)+\dot{\psi}^{2}
(\cos\phi\sin\psi+\sin\phi\cos\psi\cos\theta)\right]\xhat\nonumber\\
&&+\left[\dot{\phi}^{2}(\sin\phi\sin\psi-\cos\phi\cos\psi\cos\theta)-
2\dot\phi\dot\psi(\cos\phi\cos\psi-\sin\phi\sin\psi\cos\theta)+{\dot\psi}^{2}
(\sin\phi\sin\psi-\cos\phi\cos\psi\cos\theta)\right]\yhat\nonumber\\
&&-{\dot\psi}^{2}\cos\psi\sin\theta \zhat,~~~~~\label{dota2}
\ena

Solving equation (\ref{omepa}) for Euler angles, and plugging into equation 
(\ref{dotmua})  with the usage of equations (\ref{dota1}) and (\ref{dota2}) 
we obtain the dipole acceleration $\ddot\bmu$ as functions of time. Performing 
Fourier transform for the components of $\ddot\bmu$ gives us the spectrum  and frequency 
of electric dipole emission. 

The dipole emission power of this torque-freely rotating grain can be 
obtained by averaging the $\ddot\mu^{2}$ over Euler angles 
$\phi$ and $\psi$ in the range from $0$ to $2\pi$:
\bea
P_{\rm ed}(J,\theta)=\frac{2}{3c^{3}}\langle \ddot\bmu^{2}\rangle\equiv \int_{0}^{2\pi}
\int_{0}^{2\pi}\frac{d\phi}{2\pi}\frac{d\psi}{2\pi}\frac{2}{3c^{3}}\ddot\mu^{2}
\ena

\end{document}